\documentclass[11pt]{article}
\usepackage[top=0.8in, bottom=0.78in, left=0.87in, right=0.87in]{geometry}
\usepackage{setspace}
\usepackage[T1]{fontenc}
\usepackage{times}
\usepackage{booktabs}
\usepackage{rotating}
\usepackage{graphicx}
\usepackage{tikz}
\usepackage[section]{placeins} 
\usepackage[large, bf]{caption}
\usepackage[FIGTOPCAP]{subfigure}
\usepackage{palatino}
\usepackage{pdflscape}
\usepackage{textcomp}
\usepackage{longtable}
\usepackage{nicefrac}
\usepackage{adjustbox}	
\usepackage[hyphens]{url}

\usepackage{natbib}
\bibpunct{(}{)}{;}{a}{,}{,}

\usepackage{amsmath, amsfonts, amssymb, amsthm}

\usepackage{mathpazo} 
\usepackage{hyperref}
\hypersetup{colorlinks, citecolor=blue, filecolor=blue, linkcolor=blue, urlcolor=blue}

\def\urltilda{\kern -.15em\lower .7ex\hbox{\~{}}\kern .04em}

\usepackage{titlesec}

\titlespacing*{\section}{0pt}{1.5ex plus 1ex minus .2ex}{0.8ex plus .2ex}
\titlespacing*{\subsection}{0pt}{1.2ex plus 1ex minus .2ex}{0.8ex plus .2ex}

\usepackage{algorithm}
\usepackage{algpseudocode}
\usepackage{colortbl}
\usepackage{tabularray}
\usepackage{mathtools}
\begin{document}

\title{Effective and Scalable Programs to Facilitate Labor Market Transitions for Women in Technology\thanks{Athey: Stanford University; Palikot: Northeastern University. We thank Aleksandra Bis and Natalie Piling from Dare IT for support and collaboration. We thank Keshav Agrawal for excellent research assistance. We also thank Dean Karlan, Raviv Murciano-Goroff, Sagit Bar-Gill, Aruna Ranganathan, and seminar participants at HBS, NBER, Cornell, Purdue, Stanford, UT Dallas, South Florida, Berkeley, Michigan, Telecom Paris, Northeastern, and Booth for their helpful comments and suggestions. The Golub Capital Social Impact Lab at Stanford Graduate School of Business provided funding for this research. This research has been subject to review and approval by Research Compliance Office at Stanford University, protocol number IRB-62530 and registered at AEA RCT registry \citet{athey_palikot_rct, athey_palikot_rct_2}.}
}
\author{Susan Athey \and Emil Palikot}
\date{\today}
\maketitle

\begin{abstract}\noindent
We evaluate two interventions facilitating technology-sector transitions for women in Poland: \emph{Mentoring}, focused on expanding professional networks, and \emph{Challenges}, focused on building credible skill signals. Randomizing oversubscribed admissions, we find both programs substantially increase technology employment at twelve months---by 15 percentage points for \emph{Mentoring} and 11 p.p.\ for \emph{Challenges}. The distinct mechanisms through which the programs operate translate to heterogeneous treatment effects across geography, career stage, and baseline credentials. These differential effects create scope for improved allocation: algorithmic targeting across programs outperforms random assignment by 86\% and experts' selection into \emph{Mentoring} by 11\%. \textit{JEL:} J24, J08, J16, C93
\end{abstract}

\newpage

\section{Introduction}
Facilitating worker transitions into high-growth sectors is a core challenge for labor market policy, particularly for underrepresented groups. The technology sector is a prime example: despite rapid employment growth and attractive wages, women hold only 30\% of technology jobs worldwide.\footnote{\cite{unitednations2021} estimates the share of women in the technology sector at approximately 30\%. Many programs exist to support women in technology, including NGO initiatives (Women in Technology International, Girls Who Code), corporate programs (Women in Tech from Salesforce, The Women at Microsoft), and government efforts (UN's Girls in ICT Day, European Commission's Women in Digital Scoreboard).} While women face multiple barriers to entering technology, we focus on two labor market frictions that short-term interventions can directly address.\footnote{The gender gap in technology has been attributed to discrimination in hiring and promotion \citep{reuben2014gender, moss-racusin2012science}, work-family conflict and child penalties \citep{goldin2014grand, kleven2019children, bertrand2010dynamics}, and underrepresentation in STEM education pipelines \citep{bayer2016diversity, carrell2010sex}. We focus on network and information barriers because short-term labor market interventions can directly address these constraints, unlike structural factors requiring longer-term policy responses.} First, research across multiple settings has found that women have weaker professional networks than men, limiting access to job opportunities \citep{greguletz2019women, forret2004networking}. Second, gender stereotypes have been shown to lead employers to discount women's technical abilities, creating information frictions that credible skill signals could reduce \citep{cheryan2013stereotypical, del2022more, murciano2022missing}. Understanding which barriers pose the greatest impediment, and for whom, is critical for designing effective policy. If network barriers dominate, interventions should expand professional connections; if information frictions dominate, scalable credentialing programs may suffice. Alternatively, different populations may face different primary barriers, necessitating multiple program types and targeted assignment rules. We find evidence that both network deficits and information frictions matter, but for different populations: network deficits primarily constrain women outside major urban centers and those with caregiving responsibilities, while information frictions primarily constrain early-career women with limited work history.

We evaluate two interventions implemented in Poland by Dare IT, a social impact organization, each targeting one of these frictions. \emph{Mentoring} pairs participants with senior female technology professionals for four months of one-on-one sessions. While sessions are personalized, mentors consistently emphasize networking: facilitating introductions to hiring managers, expanding mentees' professional connections, and teaching participants how to effectively use these connections. \emph{Challenges} is a scalable online program (\$15 per participant) we designed during this research to address skill signaling barriers. The program helps participants build portfolios demonstrating technical proficiency through six bi-weekly industry-designed assignments, culminating in complete projects, such as mobile app designs or React applications.  Participants later reference these projects in job applications.\footnote{Program development began with structured interviews with hiring managers at twelve Polish technology firms, who uniformly cited the absence of credible practical experience signals in female applicants' profiles. We designed \emph{Challenges} to be delivered entirely online, centered on peer interaction rather than scarce mentors, and based on realistic business cases to ensure employer credibility. The program has been replicated across multiple specializations (software testing, cloud computing, UI design) and delivered in Ukrainian for refugees.} Both programs were substantially oversubscribed, allowing us to randomize access and measure employment outcomes via LinkedIn profiles over twelve months.\footnote{We define technology jobs as positions in technology firms (excluding finance, legal, HR, and accounting roles) or technical positions elsewhere (software development, IT support, data analytics). UX and front-end programming roles are classified as technology jobs regardless of firm type.}

Both programs generate large employment effects. \emph{Mentoring} increases technology employment by 15 percentage points (SE 5.7, from a 46\% control baseline); \emph{Challenges} by 11 percentage points (SE 4.8, from 27\%). Consistent with distinct mechanisms, \emph{Mentoring} increases LinkedIn connections by 52.7 (SE 20.8), a 23\% increase, while \emph{Challenges} shows no detectable network expansion. 

The programs benefit different populations. Professional experience reveals the starkest contrast: \emph{Challenges} generates 28 percentage points larger effects for participants with under five years of experience compared to those with longer careers, while \emph{Mentoring} shows the opposite pattern with 20 percentage points larger effects for experienced workers. This indicates that early-career women benefit from portfolio signals that compensate for limited work history, while experienced women benefit from network expansion that leverages their accumulated professional standing. Geographic heterogeneity supports the network mechanism: \emph{Mentoring} generates 26 percentage point effects for women in small towns compared to only 3 percentage points for Warsaw residents (difference: 23 percentage points, SE 14), while \emph{Challenges} shows no significant geographic heterogeneity (difference: -2 percentage points, SE 12). This pattern is consistent with network barriers binding tightly outside major urban centers where local technology ecosystems are sparse, while online portfolio signals function equally regardless of location. These differential patterns demonstrate that both network barriers and information frictions constrain technology employment, but for distinct populations defined by career stage or geography.

The differential treatment effects create an allocation problem when program capacity is limited. Both programs remain substantially oversubscribed (5:1 applicant-to-spot ratio for \emph{Mentoring}; all 300 spots filled in four hours for \emph{Challenges}), requiring selection among applicants. We evaluate counterfactual assignment policies using the methods of \cite{athey2021policy} and \cite{zhou2023offline}, estimating which applicants should receive which program to maximize aggregate outcomes. A targeted policy assigning 15\% of applicants to each program based on predicted treatment effects achieves treatment effects 86\% larger than randomly assigning from the pool of all applicants. This gain comes from two margins: selecting individuals with the highest treatment effects and matching individuals to their optimal program type. The matching component generates value even at scale: when capacity expands to 50\% per program (universal coverage), targeted allocation maintains outcomes 23\% larger than random assignment. This demonstrates that complementarity between individual characteristics and program mechanisms creates allocation gains beyond simple capacity expansion. Treatment effect heterogeneity enables improved matching even when all applicants can be served.

In \emph{Mentoring}, experienced female technology professionals select mentees from a pre-screened pool based on application materials: cover letters, resumes, pre-recorded videos, and registration surveys. Mentors vary systematically in whom they select and what outcomes they achieve. Mentors with managerial experience or above-median career tenure select applicants with weaker baseline employment prospects and have stronger treatment effects: a pattern consistent with identifying "high-potential, high-barrier" candidates rather than those already likely to succeed. Less experienced mentors favor applicants with stronger baseline employability, selecting candidates who would likely find technology jobs regardless of program participation. If effective selection heuristics can be learned through experience, data-driven tools may offer a path to improving targeting among less experienced selectors—a possibility we examine next.

We find that they can. Algorithmic allocation using only six observable characteristics (education, location, age, professional experience, degree level, domain expertise) achieves treatment effects 63\% larger than random assignment and 11\% than average mentor selections. The two approaches achieve substantial concordance (70\%): the algorithm largely validates expert choices while identifying additional high-benefit candidates. Despite mentors' rich soft information, systematic patterns in treatment response remain that simple demographic targeting can exploit. This result extends recent work showing algorithms can complement expert decisions in domains from bail \citep{kleinberg2018human} to medical diagnosis \citep{mullainathan2022diagnosing}, demonstrating that even sophisticated expert judgment leaves room for data-driven improvement when treatment effects vary systematically with observable characteristics.

Our focus on small, custom-designed programs limits external validity but enables deeper investigation of mechanisms. Mentoring programs are almost always small due to capacity constraints.\footnote{Our 150 participants is comparable to \cite{dennehy2017female} and exceeds the 30–40 per cohort in \cite{blau2010can}.} Working with each mentor, we implemented a paired randomization design that allows us to measure both selection and treatment effects—observing not only program impacts but how mentor choices vary with mentor experience. We developed \emph{Challenges} during this research and evaluate its first cohort in this paper; the program has since scaled to multiple technology domains (software testing, cloud computing, UI design) and languages of instruction, suggesting the model extends beyond our pilot context.

The remainder of the paper proceeds as follows. Section 2 reviews related literature. Section 3 describes the programs, theoretical framework, and testable predictions. Section 4 details the experimental design and data. Section 5 presents treatment effect estimates, mechanisms, and heterogeneity analysis. Section 6 evaluates counterfactual allocation policies. Section 7 concludes.

\section{Literature Review}

This paper contributes to the literature on treatment effect heterogeneity and optimal policy assignment. Building on recent advances in policy learning using machine learning methods \citep{athey2017econometrics, kitagawa2018should, athey2021policy}, we show that data-driven targeting can substantially improve program allocation decisions in active labor market policy. Theoretical work demonstrates that when treatment effects vary across individuals and program capacity is limited, policies that assign treatment based on predicted treatment effects can substantially improve welfare relative to random allocation or allocation based on baseline characteristics \citep{kitagawa2018should, bhattacharya2012inferring}. Empirical applications of these methods have demonstrated value across diverse domains: customer retention \citep{ascarza2018retention}, marketing \citep{hitsch2018heterogeneous}, development programs \citep{haushofer2022targeting}, agriculture \citep{kasy2021adaptive}, and medical treatment \citep{inoue2023machine}.

We extend this literature to active labor market policy. Optimal allocation based on predicted treatment effects achieves outcomes 86\% larger than random assignment at 15\% capacity and 23\% larger even at universal coverage. Crucially, targeting based on predicted treatment effects (CATE) dominates targeting based on baseline outcomes or predicted final outcomes, demonstrating that treatment response heterogeneity does not align with baseline heterogeneity. The persistent gains at near-universal coverage reflect complementarity between individual characteristics and program type, creating allocation value beyond capacity expansion.

We further contribute to the growing literature examining whether algorithmic decision-making can complement or improve upon expert human judgment. Recent work demonstrates that machine learning algorithms can match or exceed expert performance in domains from bail decisions \citep{kleinberg2018human} to medical diagnosis \citep{mullainathan2022diagnosing}, often by identifying systematic patterns that experts miss or cannot consistently apply. Similar to findings in online labor markets where revealing information about worker quality improves matching outcomes \citep{pallais2014inefficient}, we show that observable characteristics predict treatment effects systematically.

We demonstrate that data-driven targeting can improve upon expert human judgment. Mentor-based selection achieves treatment effects 48\% larger than random allocation. Yet algorithmic targeting achieves an additional 11\% improvement. The 70\% concordance between methods indicates complementarity: algorithms largely validate expert choices while identifying additional high-benefit candidates, demonstrating that systematic patterns in treatment response exist even after expert selection captures most available heterogeneity.

This paper contributes to the literature on mentoring effectiveness in improving career outcomes. Several influential papers focus on mentoring in academic contexts and show that mentorship increases sense of belonging among female engineering students \citep{dennehy2017female}, youth' learning outcomes \citep{resnjanskij2021can}, and tenure rates \citep{ginther2020can}, publications, and federal grants for women in research careers \citep{gardiner2007show, blau2010can}.\footnote{The impact of mentoring on career outcomes has also been studied using observational data. \cite{ragins1999mentor} survey over 600 individuals and find that mentored respondents have better career outcomes than non-mentored respondents, with effects varying by the gender composition of the mentoring pair. \cite{kammeyer2008quantitative} conduct a meta-analysis concluding that mentoring improves job and career satisfaction.} Mentoring programs may operate partly through expanding professional networks, which play a critical role in labor market outcomes \citep{granovetter1973strength, ioannides2004job, pallais2016referential}. \cite{athey2000mentoring} develop a formal theoretical model of the dynamics of mentoring and diversity. We provide randomized trial evidence on mentoring effectiveness in labor market transitions outside academic settings, demonstrating large effects on technology employment for women career-changers. Crucially, we provide experimental evidence that mentoring causally expands professional networks: \emph{Mentoring} increases participants' LinkedIn connections by 52.7 (SE 20.8), a 23\% increase from baseline, while the non-network-based \emph{Challenges} program shows no statistically significant effect. We further examine the evaluative selection mechanism inherent in mentor-based programs, where mentors exercise discretion in choosing mentees, and demonstrate that this selection generates substantial treatment effect heterogeneity with more experienced mentors systematically selecting mentees with higher treatment effects. 

To our knowledge, we are the first to evaluate mentoring effectiveness using a paired randomization design within mentor-selected candidates. This approach offers three advantages over standard randomization from an applicant pool. First, it allows us to estimate causal treatment effects while preserving the program's natural selection structure—mentors choose whom they want to work with, avoiding the forced assignments that standard randomization would impose and that mentors might resist. Second, by observing mentor choices, we can study how selection varies with mentor characteristics and whether experienced mentors develop effective selection heuristics. Third, because we observe outcomes for non-selected applicants alongside selected participants randomized to control, we can directly compare these groups, which is not possible in prior mentoring RCTs, which randomize only within pre-screened applicant pools.

We contribute to the extensive literature evaluating active labor market training programs. \cite{card2018works} conduct a comprehensive meta-analysis of over 200 studies, finding that traditional training programs aimed at unemployed workers exhibit near-zero short-term employment effects, with modest positive impacts emerging only 2-3 years post-completion.\footnote{See also \cite{heckman1999economics}, \cite{card2010active}, and \cite{crepon2016active} for reviews of ALMP effectiveness.} Programs incorporating on-the-job skill development tend to outperform classroom-based alternatives \citep{sianesi2008differential, lechner2000microeconometric}, though even intensive apprenticeships show mixed results \citep{biewen2014effectiveness}. \cite{fein2018bridging} show that YearUp, a sectoral training program targeting disadvantaged youth, generates 40\% earnings gains. However, recent evidence suggests that programs targeting skill signaling and information frictions generate substantially larger employment gains than traditional classroom training \citep{abebe2021anonymity, bassi2022screening, carranza2020job, athey2024moocs}. Building on the theoretical work of \cite{spence1978job}, this literature shows that credentials demonstrating capability can significantly boost earnings \citep{tyler2000estimating, hadavand2018can} and generate market-level benefits when employers face uncertainty about worker quality \citep{pallais2014inefficient}. Recent evidence further suggests that training effectiveness may depend on complementary interventions: \cite{batista2022closing} find that business training for female microentrepreneurs generates significant profit increases only when combined with mobile money access, highlighting that releasing multiple constraints simultaneously may be necessary for training programs to translate into economic gains.

We provide experimental evidence on training program effectiveness in technology sector transitions, a context involving higher-skilled, career-changing women rather than the unemployed workers typically studied. We document substantial short-term effects (visible within twelve months) for both programs, aligning with the signaling literature rather than the null findings in \cite{card2018works}' meta-analysis. We directly compare two program modalities: personalized one-on-one mentoring versus scalable online project-based training. This head-to-head comparison reveals complementarity rather than substitutability at a policy-portfolio level: the programs serve different populations and generate heterogeneous treatment effects that enable targeted assignment. Our subjects are women seeking technology roles who can utilize skill signaling to counter gender-based statistical discrimination.

Finally, this paper adds to the literature on the gender gap in the technology sector. The persistent gender wage gap reflects both occupational segregation and within-occupation differences \citep{juhn2017specialization}. Various reasons for low women's participation in the technology sector have been studied, including women's tendency to under-report programming skills \citep{murciano2022missing} and the impact of occupational stereotypes \citep{cheryan2013stereotypical, del2022more}. Strategies proposed to counter these barriers include increasing role model visibility \citep{correll2016succeed} and shifting perceptions of technology roles. We contribute to this literature by demonstrating how mentoring and credentialing programs can enhance women's success rates in obtaining technology jobs.
\section{Experimental Interventions and Theoretical Framework}\label{context_section}

Labor market transitions face two fundamental frictions: imperfect information 
about worker abilities and search costs in matching workers to opportunities. 
Workers can address these through skill signaling \citep{spence1978job} and professional 
networks \citep{granovetter1973strength, ioannides2004job, pallais2016referential}. For women entering 
male-dominated technology sectors, both mechanisms face gender-specific barriers. First, occupational stereotypes bias employer perceptions of female candidates' technical abilities, creating information frictions even for qualified women \citep{cheryan2013stereotypical, del2022more, murciano2022missing}. Second, network structures systematically exclude women from informal professional connections that facilitate hiring, with women reporting smaller networks and facing both structural barriers and personal hesitation in networking behaviors that benefit men \citep{greguletz2019women,forret2004networking}. Third, women in male-dominated fields face psychological barriers that may 
reduce job search effectiveness. Women systematically under-report their 
technical abilities on resumes compared to similarly skilled men 
\citep{murciano2022missing}, use less self-promoting language in professional 
contexts \citep{lerchenmueller2019gender}, and may experience reduced 
confidence when confronting occupational stereotypes \citep{spencer1999stereotype}. We study the relative importance of these barriers by evaluating two distinct interventions: a mentoring program that primarily focuses on network gaps and confidence building through connections to successful female tech professionals, and a scalable portfolio-building program designed to counter information frictions and stereotypes through credible skill signals.

\paragraph{Context.} Poland's technology sector employed 250,000 workers in 2021, with median monthly wages of USD 2,600, nearly triple the national median \citep{gus2022}. Despite strong labor demand driving 9\% annual wage growth, only 1.1\% of women in the labor force held positions in the technology sector, compared to 5.2\% of men \citep{EC2021, fluff2021}.  This persistent underrepresentation mirrors patterns across developed economies where women hold approximately 30\% of tech jobs and persists despite numerous initiatives supporting women in technology \citep{unitednations2021}. As a high-income economy and a EU member state with gender gaps in technology employment comparable to Western European averages, Poland provides an appropriate setting for studying barriers facing women entering male-dominated sectors.

\subsection{Program Designs and Mechanisms}
We partnered with Dare IT, a Polish social impact organization addressing gender barriers in technology employment, to evaluate two interventions targeting different constraints. Both programs serve women with baseline technical skills who have not yet secured technology jobs.

\paragraph{Mentoring:} Dare IT launched its \emph{Mentoring} program in 2018 to support women transitioning into technology careers. The program pairs participants with experienced female technology professionals for weekly one-on-one sessions over four months. Mentors, all women actively working in tech, volunteer their time and are selected based on technical expertise and professional standing. From its inception, demand has substantially exceeded capacity. The spring 2021 cohort received over 1,000 applications for 150 available spots, a ratio typical of recent cohorts. This persistent oversubscription reflects a fundamental scalability constraint: the program's reliance on volunteer mentors limits expansion, as the pool of qualified female tech professionals willing and able to dedicate four months of weekly mentoring remains finite.

Mentoring aims to alleviate gender-specific network barriers documented in prior research. Women in technology report smaller professional networks and face structural exclusions from informal networking opportunities as compared to their male colleagues \citep{greguletz2019women}. Female professionals may also hesitate to engage in networking behaviors that appear effective for men \citep{forret2004networking}. By connecting mentees to established female professionals, the program potentially overcomes these barriers and provides same-gender role models who can navigate gender-specific challenges in recruitment.

Our qualitative interviews with mentors conducted prior to the experiment revealed four operational channels of the \emph{Mentoring} program: \emph{(i.)} direct network expansion through mentor connections and introductions to hiring managers, \emph{(ii.)} transmission of tacit knowledge about navigating tech sector recruitment as a woman, including addressing potential bias in interviews, \emph{(iii.)} self-efficacy reinforcement to build confidence, and \emph{(iv.)} technical skill development through code review. Mentors consistently emphasize networking as the primary mechanism, actively facilitating introductions and teaching mentees how to leverage professional connections effectively.

\paragraph{Challenges:} To overcome \emph{Mentoring}'s capacity constraints, we designed \emph{Challenges} as a scalable alternative targeting information frictions exacerbated by gender stereotypes. Program development began with structured interviews of hiring managers at twelve prominent Polish technology firms. These conversations revealed a systematic pattern: hiring managers perceived female applicants as lacking practical technical experience, even when formal credentials suggested otherwise. Managers emphasized that portfolio projects demonstrating hands-on capability could substantially alter hiring decisions, but expressed skepticism about generic online certifications.
These perceptions align with documented gender bias in technical hiring: employers underestimate women's programming abilities \citep{murciano2022missing} and stereotypes about women's technical competence create additional screening burdens for female candidates \citep{cheryan2013stereotypical}. A credible skill signal may be particularly valuable for women if it counters these negative stereotypes by providing concrete, verifiable evidence of technical proficiency.

\begin{figure}
  \centering
  \caption{An Example of \emph{Challenges} Final Product.}\label{fig:examples_challenges}
    \includegraphics[width=0.75\textwidth]{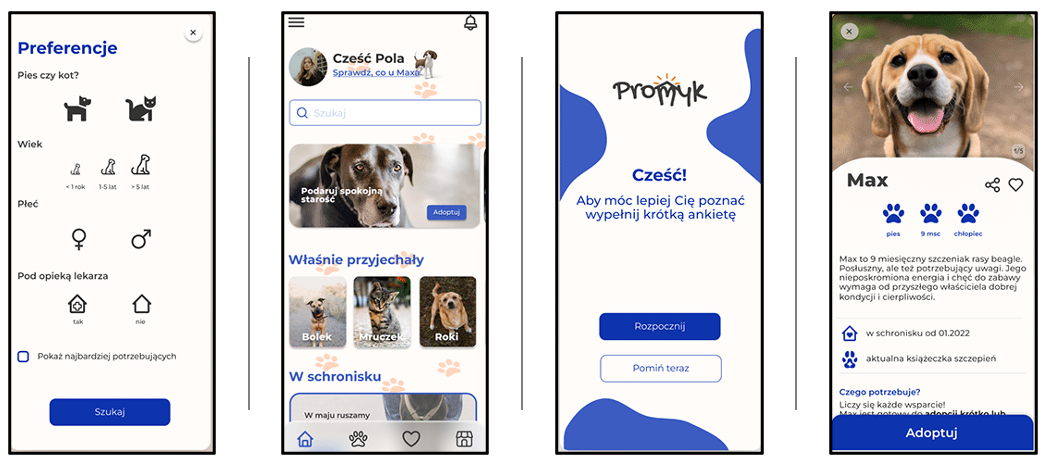}
  \caption*{\footnotesize{\textit{Note: Selected wireframes from a design of a mobile app for pets adoption prepared during Challenges. See: https://www.behance.net/gallery/142047393/Pet-Adoption-App-UXUI for the entire design.
}}}
\end{figure}

\emph{Challenges} provides six bi-weekly assignments in User Experience Design (UX) or front-end programing in React. All assignments are designed by industry practitioners, that culminate in a complete portfolio project (UX design or React application; see Figure \ref{fig:examples_challenges} for an example from the UX path). Each assignment replicates realistic business tasks and receives expert evaluation and feedback. Participants can add completed portfolios to resumes and reference them in interviews, creating a signal that hiring managers indicated they would find credible. The program operates online, accommodates cohorts of 300+ participants, and costs approximately \$15 per person.

 A more comprehensive discussion of the way the \emph{Challenges} program is organized and it's content is in Appendix \ref{challenges_appendix}.

\subsection{Theoretical Predictions and Empirical Strategy.} The programs' differential mechanisms generate testable predictions about the treatment effect heterogeneity. The patterns of heterogeneity reveal which populations benefit most from network expansion versus skill signaling, providing evidence on the economic importance of each mechanism.

If network barriers constrain technology employment for some women, \emph{Mentoring} should be especially effective for participants with weak existing professional connections: women in smaller cities and towns where local technology networks are sparse, those without family or friends working in technology, and those with longer gaps in employment who may have lost previous professional contacts. The network mechanism also predicts expansion of professional networks, which we measure through LinkedIn connection counts. Additionally, if experienced mentors' networks generate value, treatment effects should vary with mentor characteristics. Mentors with longer technology sector tenure and those in managerial positions likely possess richer professional networks and better access to hiring decision-makers, potentially generating larger effects for their mentees.

If imperfect information about technical abilities constrains hiring, particularly when exacerbated by gender stereotypes, \emph{Challenges} should disproportionately benefit women whose traditional credentials provide weak or ambiguous signals: for example, those early in their careers with limited work history. For these candidates, the portfolio complements weak existing signals. In contrast, women with strong observable credentials may gain less marginal value from an additional portfolio signal. 

We test these predictions by estimating treatment effects across participant subgroups defined by network strength, credential quality, and demographic characteristics. Differential patterns of heterogeneity provide evidence on whether these frictions meaningfully constrain employment and which populations each intervention serves most effectively. We further examine effects on LinkedIn connection counts to directly test the network expansion mechanism. 

Before proceeding, we note an important limitation: the two interventions do not isolate single mechanisms. \emph{Mentoring} may include technical feedback alongside network expansion, while completing \emph{Challenges} may affect confidence in addition to generating skill signals. Our empirical strategy therefore cannot definitively attribute effects to specific channels. Instead, we interpret differential treatment effects and heterogeneity patterns as evidence about the relative importance of network-based versus signaling-based approaches, recognizing that each program operates through multiple reinforcing channels.
\section{Experimental Design and Data}\label{design_data}

\paragraph{Mentoring.} We evaluate the cohort recruited in October 2021, with the program launching in January 2022. As in previous cohorts, applications substantially exceeded mentor availability. After Dare IT's eligibility screening, which excluded applicants currently employed in technology and those lacking junior-level technical skills, the applicant-to-mentor ratio remained 4:1, enabling a randomized evaluation.

We modified the standard selection procedure to generate exogenous variation in program access. Rather than each mentor selecting one mentee (the historical practice), we asked mentors to identify their two preferred candidates from eligible applicants. We then randomized within each pair in November 2021, assigning one applicant to treatment and one to control. This paired design controls for mentor selection preferences while ensuring balance on characteristics mentors value.

Of 160 mentors, 151 agreed to this modified process; 9 declined because they did not want to invest effort selecting candidates who might not become their mentees. This yielded 302 applicants (151 treatment, 151 control) across 151 mentor pairs. Between randomization and program launch, 4 treatment-assigned participants secured technology jobs and withdrew. To maintain program access for mentors who had invested time in selection, these slots were filled by the corresponding control-group applicants, yielding 151 treated participants and 147 controls.  We assess sensitivity to this replacement procedure in Appendix \ref{ate_bounds}, estimating effects under alternative sample definitions; results are substantively unchanged.

We also collected outcome data on 468 eligible applicants whom mentors did not select. These non-selected applicants provide a benchmark for assessing selection: control participants whom mentors chose exhibit substantially higher employment rates than non-selected applicants with similar eligibility, confirming that mentor selection predicts success and underscoring the importance of within-pair randomization for identification.

\paragraph{Challenges.} We launched recruitment in December 2021, with the program beginning in February 2022. To generate initial interest, we guaranteed admission to the first 100 applicants. Registration was capped at 500, but high demand produced 525 applications before we closed enrollment.

Applicants self-certified that they (i) were not currently employed in the technology sector, (ii) possessed English proficiency at B1 level or higher, and (iii) had advanced proficiency in either Figma (UX Design track) or React (Front-end Development track). Unlike \emph{Mentoring}, we did not manually verify technical skills prior to admission, reflecting the program's design for scalability.

We exclude the 100 guaranteed-admission applicants from our experimental analysis. Among the remaining 425 applicants, we randomly assigned 200 to treatment (program access) and 225 to control (no access), stratifying on track specialization, age category, residence type (Warsaw, other large city, or small town), and employment status (student, employed outside tech, or neither).

Between randomization and program launch, 17 treatment-assigned applicants secured technology employment and chose not to participate. Unlike \emph{Mentoring}, where the paired design required replacing such individuals, here we simply exclude them from the primary analysis. Our final analytic sample includes 183 treatment and 225 control participants. Appendix \ref{ate_bounds} demonstrates robustness to including these 17 cases with outcomes coded as 1.

Among treatment participants, 41 (22\%) submitted no assignments, though they retained access to program materials, weekly webinars with industry experts, and the Dare IT Slack community. Following intention-to-treat principles, we include all treatment-assigned participants regardless of engagement level; our estimates therefore capture the effect of program access rather than active participation. Under standard LATE assumptions, effects for active participants would be proportionally larger.

We also collected outcome data on 160 applicants who registered after enrollment closed. These late applicants contribute to the off-policy evaluation in Section \ref{section_offpolicy}, which estimates treatment effects for populations beyond the experimental sample. Appendix \ref{balance_appendix} reports covariate balance; of 9 baseline characteristics, only social science degree shows imbalance at the 5\% level, consistent with chance.

\subsection{Data}
We have four data sources: applicant registration surveys, mentor registration surveys, outcome surveys conducted four months post-program, and LinkedIn profile data.

\paragraph{Applicant Registration Survey.} Both programs collected basic information during registration: city of residence, age, employment situation, and time spent developing IT skills. The \emph{Mentoring} application additionally asked whether applicants have family or friends in IT and whether they have children. We also collected information on preferred domains: UX design, manager (project or product), or developer (front or back-end).

\paragraph{Mentors Registration Survey.} Mentors provided information enabling us to construct: first-time mentor (binary), career changer (whether the mentor worked outside tech before transitioning to the sector), managerial experience (binary), and years of experience in the tech sector.

\paragraph{LinkedIn Data.} Our primary outcome measure is whether participants secured technology-sector employment following program completion. We collected employment data from participants' LinkedIn profiles until twelve months post-program (May 2023 and June 2023, respectively).

LinkedIn provides standardized job title metadata for each employment entry. The research team and research assistants, blinded to treatment assignment, coded each profile according to our pre-registered definition. A technology job is:
\begin{enumerate}
\item Any position in a technology company: firms in software development, testing, and sales; data analytics; IT services; digital marketing; or online platforms; excluding finance, regulatory, legal, accounting, and HR roles.
\item Technical positions in non-technology companies involving software development and testing, IT support, or data analytics (e.g., software engineers at banks, data analysts at consulting firms).
\end{enumerate}

Self-employment, freelancing, and volunteer positions were not coded as new employment. Appendix \ref{glassdoor} validates this classification by comparing average salaries for categories (1) and (2) using Glassdoor data; we find no statistically significant difference between (1) and (2) (p=0.9).

We obtained complete LinkedIn data for all Mentoring participants. For \emph{Challenges}, we collected data on 384 of 408 subjects (94\%). Missing data resulted from participants changing LinkedIn profile URLs or privacy settings between recruitment and follow-up waves. Appendix \ref{ate_bounds} provides bounds and estimates under extreme attrition assumptions; main estimates remain statistically significant under all scenarios.

\paragraph{Outcome variables.} We define three binary measures: new job (any employment starting January 2022 or later for \emph{Mentoring}, February 2022 or later for \emph{Challenges}), tech job (technology-sector position per above criteria), and non-tech job (new employment outside technology). In a few cases, subjects reported only the start year of new employment on their LinkedIn profiles, we conservatively assign December 2022 or May 2023 as the start date.

We also collected from LinkedIn: highest education level (bachelor's, master's, postgraduate, non-traditional credential, or other), college major (constructing binary variables for STEM and social science/business), age (combining registration survey data with degree completion dates), and professional experience length (from first job start date to 2022). Table \ref{sum_stats} summarizes key variables.

\begin{table}
     \caption{Summary Statistics of the Main Variables: Treatment, Control, and Applicants}
     \label{sum_stats}
      \centering
          \resizebox{0.9\textwidth}{!}{%
      \begin{tabular}{ l c c *{8}{cc}}
        \toprule
        \toprule
      Variables & \multicolumn{4}{c}{Treated group} & \multicolumn{4}{c}{Control group} \\ 
      \cmidrule(l){2-6} \cmidrule(l){7-10}
         & N & Mean & St. Dev. & Min & Max&  N &Mean & St. Dev. & Min & Max\\ 
      \midrule
      \multicolumn{9}{@{}l}{\textit{Mentoring: mentor characteristics}}\\ 
      \addlinespace
      First time mentor & 151 & 0.47 & 0.50 & 0 & 1 & 145 & 0.43 & 0.50 & 0 & 1\\
Managerial experience & 151 & 0.52 & 0.50 & 0 & 1 & 145 & 0.51 & 0.50 & 0 & 1\\ 
Years of tech experience & 151 & 5.86 & 3.79 & 2 & 35 & 145 & 6.02 & 3.79 & 2 & 35\\
Career changer & 151 & 0.70 & 0.46 & 0 & 1 & 145 & 0.71 & 0.46 & 0 & 1\\
        \midrule
            \multicolumn{9}{@{}l}{\textit{Mentoring: experimental groups}}\\ 
      \addlinespace
New job & 151 & 0.65 & 0.48 & 0 & 1 & 147 & 0.57 & 0.50 & 0 & 1\\
Non-tech job & 151 & 0.04 & 0.20 & 0 & 1 & 147 & 0.12 & 0.32 & 0 & 1\\
Tech job & 151 & 0.61 & 0.49 & 0 & 1 & 147 & 0.46 & 0.50 & 0 & 1\\
LinkedIn connections & 151 & 277.45 & 173.47 & 0 & 500 & 147 & 223.20 & 185.60 & 0 & 500\\
STEM & 151 & 0.49 & 0.50 & 0 & 1 & 147 & 0.53 & 0.50 & 0 & 1\\
Student & 151 & 0.14 & 0.35 & 0 & 1 & 147 & 0.16 & 0.37 & 0 & 1\\
Over 30 & 151 & 0.47 & 0.50 & 0 & 1 & 147 & 0.50 & 0.50 & 0 & 1\\
Small town & 151 & 0.25 & 0.44 & 0 & 1 & 147 & 0.24 & 0.43 & 0 & 1\\
Applied before & 151 & 0.24 & 0.43 & 0 & 1 & 147 & 0.23 & 0.42 & 0 & 1\\
Mother & 151 & 0.26 & 0.44 & 0 & 1 & 147 & 0.26 & 0.44 & 0 & 1\\
High proficiency & 151 & 0.54 & 0.50 & 0 & 1 & 147 & 0.57 & 0.50 & 0 & 1\\
Family or friends in IT & 151 & 0.70 & 0.46 & 0 & 1 & 147 & 0.70 & 0.46 & 0 & 1\\
UX & 151 & 0.42 & 0.50 & 0 & 1 & 147 & 0.44 & 0.50 & 0 & 1\\
Masters & 151 & 0.48 & 0.50 & 0 & 1 & 147 & 0.41 & 0.49 & 0 & 1\\
Years of professional experience & 151 & 6.99 & 5.01 & 0 & 31 & 147 & 7.20 & 4.82 & 0 & 23\\
      \midrule
      \multicolumn{9}{@{}l}{\textit{Challenges: experimental groups}}\\
      \addlinespace 
New job & 166 & 0.57 & 0.50 & 0 & 1 & 218 & 0.50 & 0.50 & 0 & 1\\
Non tech job & 166 & 0.19 & 0.40 & 0 & 1 & 218 & 0.23 & 0.42 & 0 & 1\\
Tech job & 166 & 0.38 & 0.49 & 0 & 1 & 218 & 0.27 & 0.44 & 0 & 1\\
LinkedIn connections & 166 & 221.00 & 180.74 & 0 & 500 & 218 & 232.40 & 186.32 & 0 & 500\\
STEM & 183 & 0.42 & 0.50 & 0 & 1 & 225 & 0.39 & 0.49 & 0 & 1\\
Social science & 183 & 0.21 & 0.41 & 0 & 1 & 225 & 0.27 & 0.44 & 0 & 1\\
Warsaw & 183 & 0.20 & 0.40 & 0 & 1 & 225 & 0.20 & 0.40 & 0 & 1\\
Years of professional experience& 183 & 7.64 & 4.98 & 0 & 31 & 225 & 7.67 & 6.24 & 0 & 49\\
UX & 183 & 0.49 & 0.50 & 0 & 1 & 225 & 0.54 & 0.50 & 0 & 1\\
Over 30 & 183 & 0.47 & 0.50 & 0 & 1 & 225 & 0.44 & 0.50 & 0 & 1\\
Small town & 183 & 0.61 & 0.49 & 0 & 1 & 225 & 0.63 & 0.48 & 0 & 1\\
      \midrule
      \multicolumn{9}{@{}l}{\textit{Applicants both programs (Non-experimental comparison group)}}\\
      \addlinespace 
Tech job 12 months  & - & - & - & - & -&  628 & 0.28 & 0.45 & 0 & 1\\
STEM  & - & - & - & - & -&  628 & 0.20 & 0.40 & 0 & 1\\
Warsaw  & - & - & - & - & -&  628 & 0.31 & 0.46 & 0 & 1\\
Small town  & - & - & - & - & -&  628 & 0.39 & 0.49 & 0 & 1\\
Years of professional experience & - & - & - & - & -&  606 & 7.47 & 6.32 & 0 & 37\\
Over 30  & - & - & - & - & -&  628 & 0.14 & 0.35 & 0 & 1\\
Masters degree  & - & - & - & - & -&  628 & 0.31 & 0.46 & 0 & 1\\
Mentoring applicant  & - & - & - & - & -&  628 & 0.75 & 0.44 & 0 & 1\\
     \bottomrule
     \bottomrule
       \end{tabular}
       }
       \caption*{\footnotesize{\textit{Note: Summary statistics of selected variables. The top panel characteristics of mentors. The second panel characteristics of treatment and control groups in the Mentoring experiment. The third panel characteristics of treatment and control groups in the Challenges experiment. The bottom panel, applicants to both programs that are not in experimental groups; this group includes subsets of applicants to Mentoring that were not selected by mentors and applicants to Challenges that submitted their application past registration closure. Career changer takes the value of one when the mentor reported in the registration survey having changed occupations. The variable years of tech experience is reported by mentors in the survey. STEM excludes social sciences and architecture. Social science includes business, management, finance, and accounting. The variable years of professional experience is measured based on LinkedIn profiles. Appendix \ref{balance_appendix} shows the covariate balance between treatment and control.}}}
     \end{table}

\paragraph{Outcomes Survey.} We surveyed participants and control group members four months post-program, asking about new employment, salary ranges, number of job offers, and negotiation behavior. We conducted three surveys: the main \emph{Mentoring} survey (162 responses, 111 from treatment), an initial \emph{Mentoring} survey (194 responses), and the \emph{Challenges} survey (68 control, 61 treatment). Appendix \ref{survey} analyzes survey outcomes.
\section{Effectiveness of Dare IT Programs}\label{main_results}

\subsection*{Analytical Framework}

Let each applicant $i = 1, \ldots, N$ be characterized by observable characteristics $X_i \in \mathcal{X}$, unobservable characteristics $Z_i \in \mathcal{Z}$, treatment assignment $W_i^P \in \{0,1\}$ for program $P \in \{M, C\}$, and observed outcome $Y_i \in \mathcal{Y}$. Let $Y_i(P)$ denote the potential outcome under participation in program $P$, where $M$ denotes \emph{Mentoring} and $C$ denotes \emph{Challenges}, and let $Y_i(0)$ denote the potential outcome absent any program participation. For a subpopulation $\mathcal{G} \subseteq \{1, \ldots, N\}$, we define the average treatment effect as
\begin{align}
    \tau^P_{\mathcal{G}} \coloneqq \mathbb{E}\left[Y_i(P) - Y_i(0) \mid i \in \mathcal{G}\right].
\end{align}

\paragraph{Selection into Treatment.} The two programs employ different selection mechanisms, which affects the interpretation of our estimates. In the \emph{Challenges} program, participants were randomly selected from all eligible applicants, where eligibility was determined by observable characteristics. Let $\mathcal{G}^C$ denote this eligible population. Random selection implies that our estimate of $\tau^C_{\mathcal{G}^C}$ identifies the average treatment effect for the eligible population.

In the \emph{Mentoring} program, applicants were first screened by mentors who selected preferred mentees based on application materials, including short introductory videos, some of which we do not observe in the data. Let $\mathcal{G}^S \subseteq \{1, \ldots, N\}$ denote the subpopulation of applicants selected by mentors; only selected applicants entered the randomization, and thus we do not observe treated outcomes for non-selected applicants. This selection process implies that our estimand for the mentoring program is:
\begin{align}
    \tau^M_{\mathcal{G}^S} \coloneqq \mathbb{E}\left[Y_i(M) - Y_i(0) \mid i \in \mathcal{G}^S\right].
\end{align}
This quantity represents the average treatment effect for applicants whom mentors chose to work with, a population that may differ systematically from the broader applicant pool on both observable and unobservable dimensions. To the extent that mentors select applicants with higher potential returns to mentoring, $\tau^M(\mathcal{G}^S)$ may exceed the average treatment effect that would obtain under random assignment from the full applicant pool.

\subsection{Treatment Effects}

\begin{figure}
  \centering
  \caption{Share of Applicants with a Tech Job per Group Over Time.}\label{fig:outcomes_over_time}
  \begin{minipage}[b]{0.49\textwidth}
    \includegraphics[width=\textwidth]{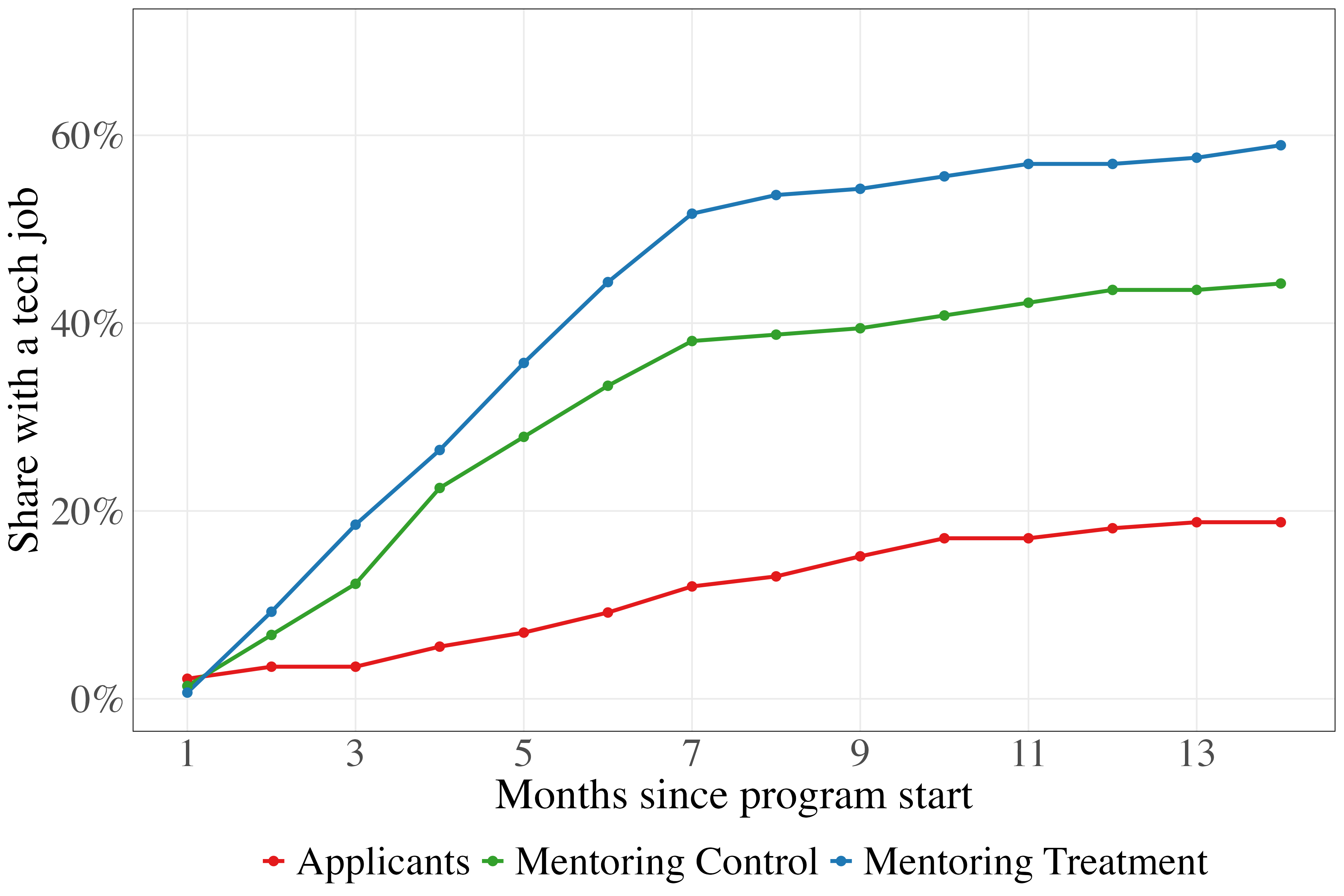}
  \end{minipage}
  \hfill
  \begin{minipage}[b]{0.49\textwidth}
    \includegraphics[width=\textwidth]{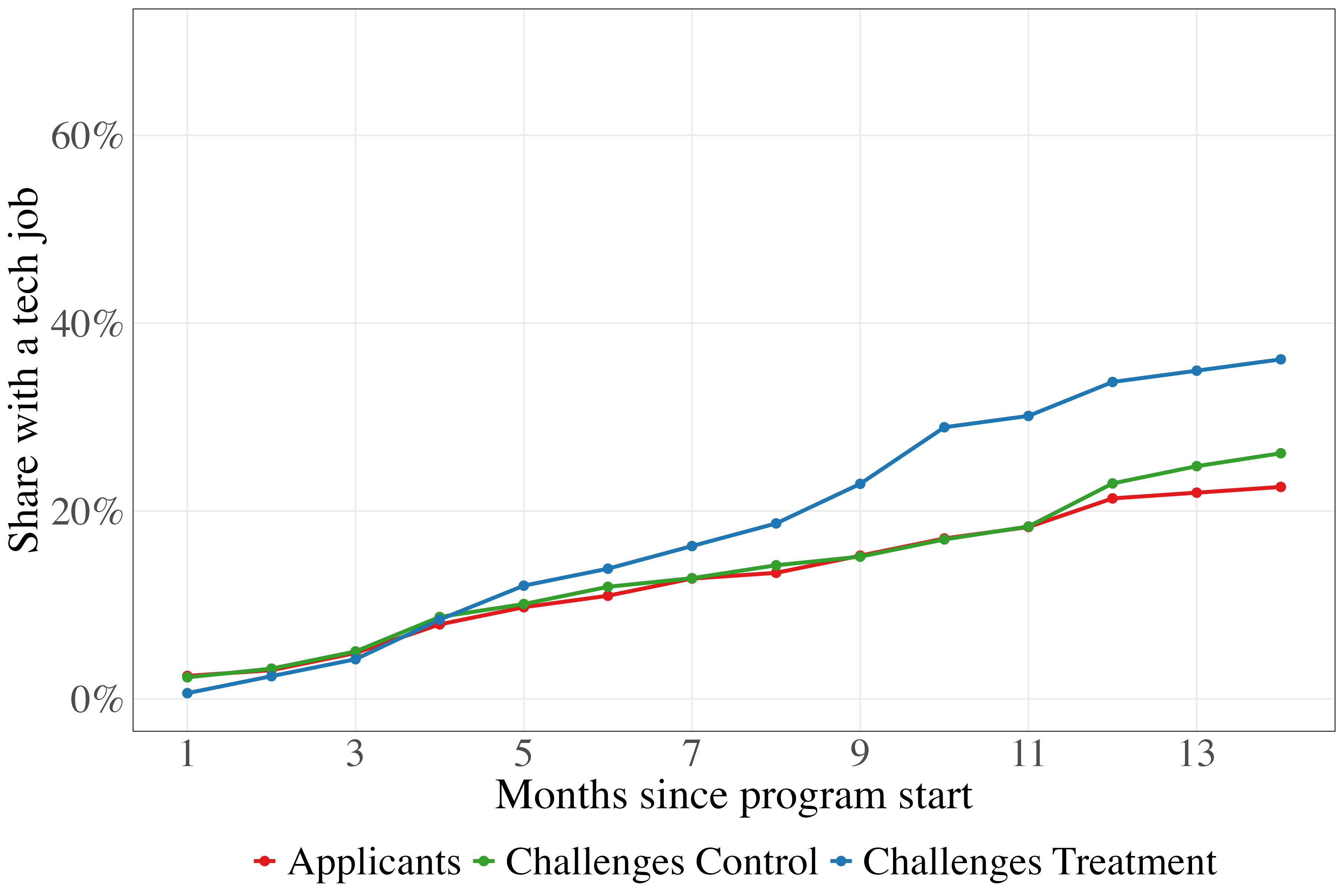}
  \end{minipage}
  \caption*{\footnotesize{\textit{Note: Share of subjects with a tech job per group across time. The month zero is the month of application to Dare IT. In the left figure, we present outcomes of three Mentoring groups: the treatment group (in blue), the control group (in green), and the applicants who were not selected by mentors (in red). In the right figure, we group focus on the Challenges program groups. The average outcome in the treated group (in blue), the control group (in green) and the late applicants to the program who were not included in the randomization and the main analysis (in red).
}}}
\end{figure}

Figure \ref{fig:outcomes_over_time} displays technology employment rates over fourteen months following program application, separately for each program.\footnote{These are simple unweighted averages. We drop from these plots subjects who reported only a year in which they started a new job. Appendix \ref{KM_curves} presents the plots with these subjects included; the only group with noticeable difference are non-selected applicants to \emph{Mentoring}. In the analysis, we imput month 14 for cases when the month is missing.}

The left panel presents results for the \emph{Mentoring} program, comparing three groups: treated applicants (selected by mentors and assigned to receive mentoring), control applicants (selected by mentors but assigned to the control group), and non-selected applicants (not chosen by any mentor and therefore excluded from randomization). All groups begin at zero as no applicants held technology jobs at baseline. Employment rates rise over time across all groups, but at markedly different rates. The treatment group exhibits rapid gains beginning immediately after program start, reaching 52\% employment by month eight. The control group follows a similar but attenuated trajectory, stabilizing at 43\%. Non-selected applicants show substantially lower employment throughout, reaching only 19\% by the end of the observation period.

Two comparisons merit emphasis. First, the gap between selected-control and non-selected applicants (43\% versus 19\% at month thirteen) confirms that mentors identify promising candidates, whether through observable signals or information revealed in application videos. Second, the treatment effect among selected applicants (52\% minus 43\% = 9 percentage points) is considerably smaller than this selection effect (24 percentage points), underscoring the importance of randomization for isolating program impacts from mentor selection.

The right panel presents results for the \emph{Challenges} program, comparing three groups: treated applicants (randomly assigned to participate in the \emph{Challenges} program), control applicants (randomly assigned to the control group), and late applicants (who applied after randomization closed and therefore received no program exposure). The control group and late applicants exhibit nearly identical employment trajectories, with both reaching approximately 20\% by month thirteen. This rate mirrors that of non-selected applicants in the \emph{Mentoring} program, suggesting a common counterfactual employment rate for women seeking to enter technology absent intervention. The treatment group, by contrast, achieves 39\% employment by month fourteen, nearly double the control rate.

The timing of employment gains differs notably across programs. \emph{Mentoring} participants begin finding technology jobs during the program itself, consistent with mentors actively facilitating job placement through referrals or direct assistance. \emph{Challenges} participants, by contrast, show employment gains concentrated after program completion, consistent with the program's focus on building portable credentials that participants subsequently deploy in job search.

Table \ref{ate_main} presents estimated average treatment effects using two approaches: difference-in-means estimation (columns 1--3 and 5--7) and Cox proportional hazard models (columns 4 and 8). We report effects on three outcomes: any new job, new technology job, and new non-technology job. Appendix \ref{appendix_prop_hazard} verifies the proportional hazards assumption; Appendix \ref{KM_curves} presents Kaplan-Meier survival curves confirming the timing patterns visible in Figure \ref{fig:outcomes_over_time}; Appendix \ref{ate_surv} estimates treatment effects using causal survival forests, the method employed in the counterfactual policy analysis of Section \ref{section_offpolicy}; Appendix \ref{app:stratified} reweighs the estimates for \emph{Challenges} by stratum used for randomization.

Both programs generate substantial and statistically significant increases in technology employment. \emph{Mentoring} raises the probability of technology employment by 15.3 percentage points (SE 5.7), representing a 34\% increase relative to the control group mean of 45.6\%. \emph{Challenges} produces an 11.3 percentage point effect (SE 4.8), a 42\% increase relative to the control mean of 26.6\%. Cox proportional hazard estimates, which exploit the full monthly panel structure, yield nearly identical results: 14.5 percentage points for \emph{Mentoring} and 11.2 percentage points for \emph{Challenges}.

\begin{table}[!htbp] \centering 
  \caption{Average Treatment Effects on Employment Outcomes} 
  \label{ate_main} 
  \resizebox{\textwidth}{!}{%
    \begin{tabular}{lcccccccc}
      \toprule      \toprule

      & \multicolumn{4}{c}{\emph{Mentoring}} & \multicolumn{4}{c}{\emph{Challenges}} \\
      \cmidrule(lr){2-5} \cmidrule(lr){6-9}
      & \multicolumn{3}{c}{Difference-in-means} & Cox PH & \multicolumn{3}{c}{Difference-in-means} & Cox PH \\
      \cmidrule(lr){2-4} \cmidrule(lr){5-5} \cmidrule(lr){6-8} \cmidrule(lr){9-9}
      & New job & Tech job & Non-tech job & Tech job & New job & Tech job & Non-tech job & Tech job \\
      \midrule
      ATE & 0.078 & 0.153*** & $-$0.076** & 0.145*** & 0.072 & 0.113** & $-$0.041 & 0.112** \\
      & (0.057) & (0.057) & (0.031) & (0.055) & (0.051) & (0.048) & (0.042) & (0.051) \\[0.5em]
      ATE / Control mean (\%) & 13.6 & 33.7** & $-$65.6** & 31.7*** & 14.5 & 42.6** & $-$17.6 & 42.2** \\
      & (9.9) & (14.8) & (26.7) & (12.0) & (10.3) & (18.1) & (18.0) & (19.3) \\
      \midrule
      Treated & 151 & 151 & 151 & 2,265 & 166 & 166 & 166 & 2,656 \\
      Control & 147 & 147 & 147 & 2,205 & 218 & 218 & 218 & 3,488 \\
      \bottomrule      \bottomrule

    \end{tabular}%
  }
  \caption*{\footnotesize\textit{Notes:} Average treatment effects on employment outcomes measured at month 12. Columns 1--3 and 5--7 report difference-in-means estimates; columns 4 and 8 report Cox proportional hazard estimates with the full covariate set. Cox models use person-month observations. ``ATE / Control mean'' expresses the treatment effect as a percentage of the control group mean. Standard errors in parentheses. *** $p<0.01$, ** $p<0.05$, * $p<0.1$.}
\end{table}

\paragraph{Job Search Selectivity.} Estimates in Column 3 of Table \ref{ate_main} suggest that \emph{Mentoring} operates partly through increased job search selectivity. While \emph{Mentoring} increases overall employment by only 7.8 percentage points (SE 5.7), statistically insignificant, it increases technology employment by 15.3 percentage points and decreases non-technology employment by 7.6 percentage points (SE 3.1). This implies that approximately half of \emph{Mentoring}'s impact reflects participants becoming more selective, holding out for technology positions rather than accepting readily available employment in other sectors. In contrast, \emph{Challenges} shows no significant effect on non-technology employment (-4.1pp, SE 4.2), consistent with this program operating through skill signaling rather than changing search behavior or persistence. The differential patterns support distinct mechanisms: \emph{Mentoring} increases both participants' capability to access technology jobs (through networks and tacit knowledge) and their willingness to persist in pursuing them (for example, through confidence-building), while \emph{Challenges} primarily improves application success conditional on applying. 

In Appendix \ref{survey}, we show the results based on the outcomes survey. Response rates were modest, which is why we rely on LinkedIn data for our primary analysis. Among respondents, both treatment groups report a higher number of job offers and higher salaries than the control group. The differences between treatment and control are greater in the case of \emph{Mentoring} than \emph{Challenges}.

\subsection{Impact of Program Participation on LinkedIn Connections}

We directly test the network mechanism by examining whether program participation increases professional network size, measured by LinkedIn connection counts. This analysis distinguishes network expansion from other channels: if \emph{Mentoring} operates solely through skill development or signaling, we should observe no differential effect on connections relative to \emph{Challenges}. Conversely, if networking is a primary mechanism, \emph{Mentoring} should increase connections while \emph{Challenges}, which operates entirely online without mentor introductions, should not.

The data support the network mechanism. \emph{Mentoring} increases LinkedIn connections by 52.7 (SE 20.8) from a control group mean of 225, a 23\% increase. \emph{Challenges}, by contrast, produces a statistically insignificant change of $-11.4$ connections (SE 17.8) from a control mean of 232. The differential effect between programs is 64 connections (SE 28), providing direct evidence that \emph{Mentoring} expands professional networks while \emph{Challenges} does not.

Table \ref{tab:li_connections} presents quantile regression estimates to examine where these effects appear in the connection distribution. \emph{Mentoring} generates large, statistically significant increases at the 25th percentile (62 connections, SE 27) and median (92 connections, SE 35), but no detectable effect at the 75th percentile. The program thus particularly benefits participants with initially sparse networks. \emph{Challenges} shows no significant effects at any quantile, confirming that this program does not operate through network channels.

\begin{table}[htbp]
\centering
\caption{Treatment Effects on LinkedIn Professional Networks} 
\label{tab:li_connections} 
\resizebox{0.55\textwidth}{!}{%
\begin{tabular}{lccc}
\toprule\toprule
& \emph{Mentoring} & \emph{Challenges} & Difference \\
\midrule
25th percentile & 62** & 5 & 57* \\
& (27.2) & (13.4) & (30.3) \\[0.5em]
Median & 92*** & 5 & 87* \\
& (35.4) & (29.5) & (46.1) \\[0.5em]
75th percentile & 5 & $-$8 & 13 \\
& (47.0) & (50.7) & (69.1) \\
\bottomrule\bottomrule
\end{tabular}%
}
\caption*{\footnotesize\textit{Notes:} Quantile regression estimates of treatment effects on LinkedIn connection counts. The difference column reports \emph{Mentoring} minus \emph{Challenges} treatment effects. Standard errors in parentheses. *** $p<0.01$, ** $p<0.05$, * $p<0.1$.}
\end{table}

The magnitude of network expansion is economically meaningful. A 92-connection increase at the median represents substantial growth, comparable to network size differences between mid-career and senior professionals in our sample. Moreover, the measured effects likely understate true network expansion due to LinkedIn's 500-connection display cap: 25\% of \emph{Mentoring} treatment participants hit this ceiling compared to 23\% of controls, attenuating our estimates toward zero.

This network expansion is not driven by new employment. Both programs increase technology employment at substantial rates, yet only \emph{Mentoring} participants expand their networks. If job acquisition were the primary driver of connection growth, through adding new colleagues and industry contacts post-employment, we should observe comparable network increases in both programs. The small and statistically insignificant estimate for \emph{Challenges}, despite a 38\% technology employment rate among participants, indicates that \emph{Mentoring}'s network effects operate independently of job acquisition. This pattern is consistent with mentors actively facilitating introductions during the program, as reported in qualitative interviews, rather than networks expanding passively after job placement.

\subsection{Heterogeneous Treatment Effects}\label{hte}

The contrasting mechanisms underlying \emph{Mentoring} and \emph{Challenges} generate testable predictions about which participants benefit most from each program. If \emph{Mentoring} operates primarily through network expansion, as suggested by the LinkedIn analysis, effects should concentrate among participants with weak existing networks. If \emph{Challenges} works through skill signaling, effects should be largest for participants with weaker observable credentials. We estimate heterogeneous treatment effects using Cox proportional hazard models with the full covariate set, estimating separate models for subgroups defined by pre-treatment characteristics.\footnote{The characteristics we examine are motivated by our theoretical framework. However, one might be concerned that these subgroup analyses reflect specification search. Section \ref{section_offpolicy} addresses this concern using data-driven methods: we estimate optimal treatment assignment policies via causal survival forests with sample splitting, allowing the algorithm to identify predictive heterogeneity without researcher input on which variables matter. The resulting policies generate substantial gains over random assignment, confirming that policy-relevant heterogeneity exists in the data.}

Table \ref{hte_main} presents results across three participant characteristics that map to credential strength and network access. The clearest differential pattern emerges from professional experience. \emph{Challenges} generates 28 percentage points larger effects for participants with under five years of experience compared to those with longer careers (SE 10), whereas \emph{Mentoring} shows the opposite pattern, with 20 percentage points larger effects for experienced participants (SE 11). The cross-program differential is 49 percentage points (SE 11). This contrast directly supports our theoretical predictions: early-career women lack established work history to credibly signal technical abilities, making the \emph{Challenges} portfolio particularly valuable as a complement to sparse employment records. Experienced workers, by contrast, have credentials but may lack networks in technology; mentoring addresses this constraint.

Educational background reveals an asymmetric pattern. Among \emph{Mentoring} participants, those without STEM degrees show 19 percentage points larger effects than those with STEM backgrounds (SE 11). A STEM degree might bundle both network access, through connections formed during technical education, and credential signaling value; we cannot cleanly separate these channels. However, the fact that \emph{Challenges} shows no educational heterogeneity (difference: 0.3pp, SE 12) is informative. One explanation is that STEM education, as measured here, is too broad to function as a meaningful signal for specific technology roles: a degree in physics or electrical engineering may not credibly signal competence as a front-end developer. If so, the \emph{Challenges} portfolio provides equally valuable skill demonstration for STEM and non-STEM participants alike.

Geography provides additional evidence on network mechanisms. \emph{Mentoring} generates 26 percentage point effects for participants currently residing in small towns (SE 10) compared to only 3 percentage points for Warsaw residents (difference: 23pp, SE 14, $p = 0.11$). Network barriers bind more tightly outside major urban centers, where local technology ecosystems are sparse. Mentor networks based in Warsaw and other major cities provide greater marginal value when participants lack local access to technology professionals. \emph{Challenges} shows no significant geographic heterogeneity (difference: $-2$pp, SE 12), consistent with online portfolio signals functioning independently of applicant location.

\begin{table}[!htbp] \centering 
  \caption{Heterogeneous Treatment Effects: \emph{Mentoring} vs.\ \emph{Challenges}} 
  \label{hte_main} 
  \resizebox{0.9\textwidth}{!}{%
    \begin{tabular}{lccccc}
      \toprule      \toprule

      & \multicolumn{2}{c}{\emph{Mentoring}} & \multicolumn{2}{c}{\emph{Challenges}} & \\
      \cmidrule(lr){2-3} \cmidrule(lr){4-5}
      & Control mean & CATE & Control mean & CATE & Difference \\
      \midrule
      \multicolumn{6}{@{}l}{\textit{STEM degree}}\\
      \addlinespace
      Yes & 0.538 (0.057) & 0.038 (0.079) & 0.283 (0.059) & 0.120 (0.099) & $-$0.081 (0.082) \\
      No & 0.362 (0.058) & 0.228*** (0.070) & 0.259 (0.035) & 0.116** (0.057) & 0.111 (0.068) \\
      Difference & 0.176** (0.081) & $-$0.189* (0.106) & 0.024 (0.068) & 0.003 (0.114) & $-$0.193* (0.106) \\
      \midrule
      \multicolumn{6}{@{}l}{\textit{Professional experience}}\\
      \addlinespace
      More than 5 years & 0.412 (0.049) & 0.202*** (0.066) & 0.310 (0.039) & $-$0.003 (0.065) & 0.205*** (0.063) \\
      Less than 5 years & 0.556 (0.075) & 0.000 (0.087) & 0.190 (0.042) & 0.281*** (0.075) & $-$0.281*** (0.086) \\
      Difference & $-$0.144 (0.089) & 0.203* (0.109) & 0.120** (0.057) & $-$0.284*** (0.099) & 0.486*** (0.106) \\
      \midrule
      \multicolumn{6}{@{}l}{\textit{Location}}\\
      \addlinespace
      Small town & 0.429 (0.085) & 0.260** (0.101) & 0.256 (0.067) & 0.149 (0.130) & 0.111 (0.108) \\
      Warsaw & 0.447 (0.082) & 0.033 (0.103) & 0.182 (0.044) & 0.166* (0.089) & $-$0.133 (0.093) \\
      Difference & $-$0.019 (0.118) & 0.227 (0.144) & 0.074 (0.157) & $-$0.019 (0.118) & 0.246 (0.197) \\
      \bottomrule      \bottomrule

    \end{tabular}%
  }
  \caption*{\footnotesize\textit{Notes:} Heterogeneous treatment effects on technology employment by participant characteristics. Estimates from Cox proportional hazard models with the full covariate set, estimated separately by subgroup. ``Control mean'' is the employment rate in the control group for each subgroup. ``CATE'' is the average treatment effect conditional on being in a subgroup. ``Difference'' column reports \emph{Mentoring} CATE minus \emph{Challenges} CATE. Standard errors in parentheses. *** $p<0.01$, ** $p<0.05$, * $p<0.1$.}
\end{table}

\paragraph{Additional Heterogeneity in Mentoring.} The \emph{Mentoring} registration survey collected more extensive information than the \emph{Challenges} survey, and we observe mentor characteristics. Table \ref{hte_add} presents heterogeneity along these additional dimensions.

Mentor characteristics strongly predict treatment effects. Mentors with managerial experience generate 26 percentage points larger effects than those without (SE 10). Similarly, mentors with over five years of technology sector experience produce 26 percentage points larger effects than less experienced mentors (SE 11). These patterns are consistent with the network channel: managers and senior professionals possess richer networks, better access to hiring decision-makers, and greater ability to facilitate job placements directly.

Interestingly, mentor selection patterns reveal that more experienced mentors choose less promising candidates. Control group employment among participants selected by mentors with managerial experience is 38\%, compared to 55\% for those matched to non-managerial mentors. Yet treated participants of experienced mentors achieve substantially higher treatment effects (26pp versus 0pp). This pattern suggests that experienced mentors either deliberately select candidates who need more assistance, recognizing their capacity to help difficult cases, or value candidate attributes that differ from baseline employment probability.

First-time mentors are as effective as repeat mentors (difference: 6pp, SE 11). This finding indicates that mentoring effectiveness stems from mentors' accumulated professional networks and industry standing rather than mentoring skill developed through practice. 

\begin{table}[!htbp] \centering 
  \caption{Heterogeneous Treatment Effects in \emph{Mentoring}: Additional Characteristics} 
  \label{hte_add} 
  \resizebox{0.8\textwidth}{!}{%
    \begin{tabular}{lccccc}
      \toprule      \toprule

      & \multicolumn{2}{c}{Yes} & \multicolumn{2}{c}{No} & \\
      \cmidrule(lr){2-3} \cmidrule(lr){4-5}
      & Control mean & CATE & Control mean & CATE & Difference \\
      \midrule
      \multicolumn{6}{@{}l}{\textit{Mentee characteristics}}\\
      \addlinespace
      Mother & 0.237 (0.070) & 0.348*** (0.097) & 0.532 (0.048) & 0.071 (0.062) & 0.277** (0.115) \\
      Applied before & 0.500 (0.087) & 0.139 (0.109) & 0.442 (0.047) & 0.143** (0.060) & $-$0.004 (0.124) \\
      \midrule
      \multicolumn{6}{@{}l}{\textit{Mentor characteristics}}\\
      \addlinespace
      Managerial experience & 0.378 (0.056) & 0.259*** (0.073) & 0.549 (0.059) & 0.002 (0.074) & 0.257** (0.104) \\
      Long experience in tech & 0.390 (0.063) & 0.277*** (0.085) & 0.512 (0.054) & 0.021 (0.066) & 0.257** (0.107) \\
      First-time mentor & 0.444 (0.062) & 0.155* (0.080) & 0.476 (0.055) & 0.100 (0.073) & 0.056 (0.108) \\
      \bottomrule      \bottomrule

    \end{tabular}%
  }
  \caption*{\footnotesize\textit{Notes:} Heterogeneous treatment effects in the \emph{Mentoring} program by mentee and mentor characteristics. Estimates from Cox proportional hazard models with the full covariate set, estimated separately by subgroup. ``Long experience in tech'' indicates mentor has worked over 5 years in the technology sector (median value). Standard errors in parentheses. *** $p<0.01$, ** $p<0.05$, * $p<0.1$.}
\end{table}

\paragraph{Mothers and Work-Family Barriers.} The largest heterogeneity we observe is by parental status. Among non-mothers, \emph{Mentoring} generates a 7 percentage point treatment effect (SE 6); among mothers, the effect is 35 percentage points (SE 10)---a difference of 28 percentage points (SE 12). Mothers in the control group face substantial disadvantage, with 24\% technology employment compared to 53\% for non-mothers. \emph{Mentoring} eliminates this gap: 59\% of treated mothers secure technology employment.

This large effect for mothers is consistent with \emph{Mentoring} addressing multiple barriers simultaneously. Mothers may have weaker professional networks if career interruptions reduced opportunities to maintain connections. Work-family conflict concerns may also reduce mothers' confidence in pursuing demanding technology careers; mentor encouragement and role-modeling by successful women in technology may counter these concerns.

This finding contributes to understanding gender gaps in technology beyond representation. Even among women who have acquired technical skills and are actively seeking technology employment, motherhood creates additional barriers. Traditional explanations emphasize mothers' labor supply decisions or employer discrimination; our results suggest that network deficits and confidence also play important roles, and mentoring interventions can substantially mitigate these barriers.

The contrasting heterogeneity patterns across programs align with our theoretical framework. Expanding professional networks creates the greatest opportunity for geographically isolated women and mothers, the groups showing largest \emph{Mentoring} effects. Building credible skill signals offers the greatest opportunity for early-career women with limited work history, those showing largest \emph{Challenges} effects. These differential impacts indicate that effective policy must target multiple, distinct barriers rather than assuming a single mechanism operates uniformly. Moreover, the patterns demonstrate that even within the selected population of women with baseline technical skills, substantial heterogeneity exists in where the greatest opportunities for improvement lie. We explore implications for optimal program assignment in Section \ref{section_offpolicy}.
\section{Counterfactual Program Assignment Policies}\label{section_offpolicy}

Program capacity is limited. Mentor availability restricts \emph{Mentoring} capacity, while we capped \emph{Challenges} at 300 participants to manage operational complexity. Given the substantial treatment effects documented above, relaxing these constraints would increase Dare IT's impact. However, capacity constraints are likely to persist, particularly for \emph{Mentoring}. Dare IT therefore faces an ongoing allocation problem: which applicants should be admitted to each program? The heterogeneous treatment effects documented in Section \ref{hte} suggest that admission policies favoring applicants with high predicted treatment effects could enhance program impact. This section quantifies potential gains from such targeting, characterizes which applicant profiles benefit most from each program, and examines whether the patterns align with the theoretical mechanisms identified above.

We perform three analyses: (i) we estimate gains from an assignment policy that allocates applicants to programs to maximize the sum of predicted treatment effects, given capacity constraints; (ii) we evaluate gains from relaxing capacity constraints with this targeting policy in place; and (iii) we compare overall \emph{Mentoring} impact under mentor-based selection to alternative assignment strategies.

\paragraph{Policy framework.} An assignment policy maps applicant characteristics $X \in \mathcal{X}$ and program capacities $Q = (Q^M, Q^C) \in \mathcal{Q}$ to assignments: $\pi: \mathcal{X} \times \mathcal{Q} \rightarrow \mathcal{P}$, where $\mathcal{P} \equiv \{M, C, O\}$ denotes \emph{Mentoring}, \emph{Challenges}, and \emph{Out of Dare IT}, respectively. $\mathcal{Q}$ is the grid of capacity levels under consideration.

We perform policy estimation and evaluation in separate samples. We randomly split the data at the applicant level with equal probability into training and test sets ($\mathcal{I}_{\text{train}}$ and $\mathcal{I}_{\text{test}}$). Using the training set, we estimate the policy $\hat{\pi}$. To evaluate the policy, we apply $\hat{\pi}$ to the test set, obtaining counterfactual assignments $\hat{\pi}(X_i, Q) \in \mathcal{P}$ for each applicant $i$ and capacity level $Q$. We then construct augmented inverse propensity weighted (AIPW) estimates of the policy value---the expected outcome when treatments are assigned according to $\hat{\pi}$. We implement this using cross-fitting at the applicant level in the test sample: we estimate outcome and propensity models and obtain AIPW estimates in held-out folds, using 100 repetitions with 5 folds and reporting the median across repetitions \citep{chernozhukov2017double}.

\paragraph{Sample and estimation.} We combine the two experimental samples with non-selected applicants to \emph{Mentoring} and late applicants to \emph{Challenges}, yielding 1,314 total applicants (682 in experimental samples, 632 non-participants) and 19,000 applicant-month observations. We use survival forests to estimate outcome models and regression forests for propensity models. The propensity model adjusts for selection into \emph{Mentoring} and for differences between the \emph{Mentoring} and \emph{Challenges} applicant populations. Appendix \ref{appendix_algorithms} details the estimation procedure; Appendix \ref{eval_models} describes the evaluation models and cross-fitting procedure, and validates the approach by comparing off-policy estimates to experimental estimates.

\paragraph{Outcome measure.} The preceding analyses measured treatment effects as changes in the probability of technology employment. For policy evaluation, we adopt the restricted mean survival time (RMST), which measures the average time until technology job acquisition over our observation period. RMST incorporates both whether participants find jobs and how quickly they do so: a policy that accelerates job-finding generates economic value even if it does not change ultimate employment probability.

We report RMST effects in months, with negative values indicating faster job acquisition. An RMST effect of $-3$ months means treated participants find technology jobs three months earlier on average than non-participants. We estimate treatment effects using causal survival forests \citep{athey2019generalized, cui2023estimating}. Appendix \ref{ate_surv} presents average treatment effect estimates in the experimental samples using this method.

\subsection{Targeted Assignment Policies}\label{optim_policies}

We formulate optimal program assignment as an integer linear program. Let $z_{ip} \in \{0,1\}$ indicate whether applicant $i$ is assigned to program $p \in \mathcal{P}$, where $\mathcal{P} = \{M, C, O\}$ denotes \emph{Mentoring}, \emph{Challenges}, and no program (Out), respectively. The optimal assignment solves:
\begin{align}\label{eq_opt}
    \max_{z_{ip}} \sum_{i=1}^{N} \sum_{p \in \mathcal{P}} z_{ip} \tau_i^p \quad \text{s.t.} \quad \sum_{i=1}^{N} z_{ip} \leq Q^p \; \forall p, \quad \sum_{p \in \mathcal{P}} z_{ip} = 1 \; \forall i,
\end{align}
where $\tau_i^p = \mathbb{E}[Y_i(p) - Y_i(O)]$ is applicant $i$'s treatment effect from program $p$ relative to no program, and $Q^p$ is the capacity of program $p$ (with $Q^O = \infty$). The first constraint ensures capacity limits are respected; the second ensures each applicant receives exactly one assignment.

We call the policy inducing the solution $z_{ip}^*$ the \emph{targeted} policy. Individual treatment effects $\tau_i^p$ are estimated using AIPW with models trained on the training set. Algorithm \ref{alg:policy} in Appendix \ref{appendix_algorithms} provides implementation details.

To evaluate policies, we estimate the \emph{policy value}: the average treatment effect achieved when applicants are assigned according to policy $\pi$ rather than receiving no program. This focus on program impact---the causal effect of participation---rather than raw employment rates among participants aligns with Dare IT's goal of maximizing benefit to participants. Below, we compare targeting based on predicted treatment effects to alternative approaches: selecting applicants with the lowest baseline employment probability or those with the highest baseline probability.

\subsection{Value of Targeted Policies}

Given limited program capacity, which applicants should be assigned to which program to maximize overall impact? We estimate targeted policies at three capacity levels, where capacity is expressed as the percentage of applicants assigned to \emph{Mentoring}/\emph{Challenges}/Out (no program): 15/15/70 approximates existing capacity; 33/33/34 expands both programs moderately; and 50/50/0 ensures every applicant receives a program slot. As a benchmark, we compare these targeted policies against random assignment, which allocates applicants to \emph{Mentoring} or \emph{Challenges} with equal probability. We evaluate the random policy at full capacity (50/50/0) because lower capacity levels yield the same expected treatment effects but noisier estimates due to fewer treated subjects.

\begin{figure}[!htbp]
\caption{Characteristics of Groups Under Targeted Policy}
\label{fig:opt_cat}
\centering
\includegraphics[width=0.8\textwidth]{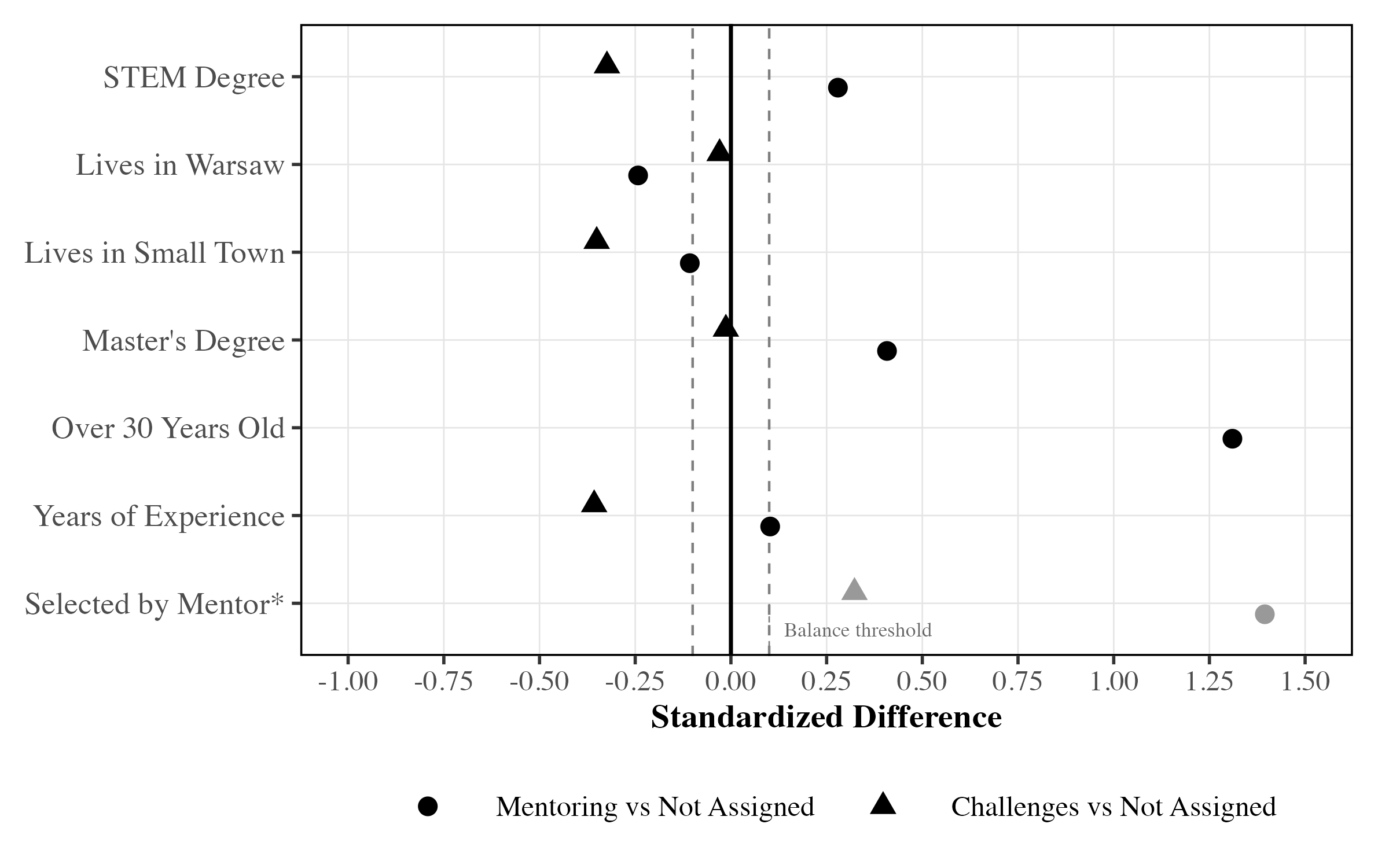}
\caption*{\footnotesize\textit{Notes:} Standardized differences in baseline characteristics between applicants assigned to each program and those assigned to no program under the 15/15/70 targeted policy. Circles show \emph{Mentoring} versus Out; triangles show \emph{Challenges} versus Out. Dashed lines mark the $\pm 0.1$ threshold. Standardized differences are calculated as the difference in means divided by the pooled standard deviation; positive values indicate the assigned group has higher values on that characteristic. \emph{Selected by Mentor} is not used in the policy assignment, presented here for comparison.}
\end{figure}

To understand which applicant types the algorithm assigns to each program, Figure~\ref{fig:opt_cat} presents standardized differences in baseline characteristics between optimally assigned applicants and those assigned to no program under the 15/15/70 policy. Circles represent \emph{Mentoring}; triangles represent \emph{Challenges}. Dashed vertical lines mark the $\pm 0.1$ threshold conventionally used to indicate meaningful imbalance.

The algorithm exhibits clear differentiation in targeting across programs. Applicants assigned to \emph{Mentoring} are more likely to hold STEM degrees, live outside Warsaw, have Master's degrees, and be over 30 years old. These patterns align with the heterogeneous treatment effects documented in Section \ref{hte}, where \emph{Mentoring} showed larger effects for participants outside Warsaw and those without STEM backgrounds. Although the algorithm does not observe prior mentor selection status, applicants whom mentors selected in the experiment are disproportionately assigned to \emph{Mentoring}, suggesting the algorithm captures similar signals to those mentors used when choosing mentees.

For \emph{Challenges}, targeting is more balanced across characteristics. Assigned applicants are moderately more likely to have been mentor-selected but are less likely to hold STEM degrees, live in small towns, or have extensive professional experience. This pattern is consistent with \emph{Challenges} benefiting early-career women who lack credentials to signal their skills, as documented in the HTE analysis.

\paragraph{Main results.}

Table~\ref{tab:optimal_allocation} presents treatment effects under data-driven targeting at varying capacity levels, estimated over the pooled population of experimental participants from both programs and all other eligible applicants (non-selected \emph{Mentoring} applicants and late \emph{Challenges} applicants). Recall that negative values indicate faster job finding (months saved). Under the most selective policy (15/15/70), the overall treatment effect reaches $-4.34$ months, an 86\% improvement over random allocation's $-2.33$ months. These gains exhibit a capacity-selectivity tradeoff: as capacity expands from 15\% to 50\% per program, average treatment effects decline from $-4.34$ to $-2.86$ months as the algorithm must include applicants with lower predicted benefits. Even at universal coverage (50/50/0), however, targeted assignment achieves effects 23\% larger than random allocation.\footnote{Appendix~\ref{app:crossfitting} presents an alternative specification using repeated cross-fitting following \citet{fava2025}, which uses all observations for both policy estimation and evaluation rather than separate training and test samples. Point estimates are nearly identical (e.g., $-4.60$ versus $-4.34$ months for the 15/15/70 policy), while standard errors are smaller due to efficiency gains from the larger effective sample size. The qualitative conclusions are unchanged.}

Targeting gains differ substantially across programs. \emph{Mentoring} exhibits the largest effects under selective policies: $-5.55$ months at 15\% capacity versus $-2.31$ under random assignment. \emph{Challenges} shows more modest gains from targeting ($-3.13$ versus $-2.35$ months, a 33\% improvement). Even at 50\% capacity, targeted allocation achieves \emph{Mentoring} treatment effects of
-3.55 months compared to -2.31 months under random allocation (54\% larger), demonstrating that observable characteristics contain substantial predictive power for treatment response.

\begin{table}[!htbp]
\caption{Treatment Effects Under Targeted Allocation Policies}
\label{tab:optimal_allocation}
\centering
\resizebox{0.6\textwidth}{!}{%
\begin{tabular}{lccc}
\toprule\toprule

& \emph{Mentoring} & \emph{Challenges} & Overall \\
\midrule
\multicolumn{4}{l}{\textit{Panel A: Benchmark}} \\
Random allocation & $-$2.31 (0.02) & $-$2.35 (0.02) & $-$2.33 (0.02) \\
\midrule
\multicolumn{4}{l}{\textit{Panel B: Targeted allocation policies}} \\
15\% / 15\% / 70\% & $-$5.55 (0.09) & $-$3.13 (0.07) & $-$4.34 (0.06) \\
33\% / 33\% / 34\% & $-$4.33 (0.05) & $-$2.68 (0.04) & $-$3.50 (0.03) \\
50\% / 50\% / 0\% & $-$3.55 (0.04) & $-$2.17 (0.03) & $-$2.86 (0.02) \\
\bottomrule\bottomrule

\end{tabular}%
}
\caption*{\footnotesize\textit{Notes:} Treatment effects measured as change in restricted mean survival time (months); negative values indicate faster job finding. Capacity allocations specify the share of applicants assigned to \emph{Mentoring}/\emph{Challenges}/Out. Random allocation assigns each applicant to either program with equal probability. Standard errors in parentheses.}
\end{table}

Table~\ref{targetting} examines whether targeting correctly matches applicants to their highest-benefit program. We estimate counterfactual treatment effects for each assignment group under the 15/15/70 policy. In other words, we estimate, what would applicants assigned to \emph{Mentoring} gain if instead assigned to \emph{Challenges}, and vice versa?

The results confirm effective assignment. Applicants assigned to \emph{Mentoring} gain $-5.55$ months from their assignment but would gain only $-3.12$ months from \emph{Challenges}---a difference of 2.43 months (SE 0.11). Applicants assigned to \emph{Challenges} similarly benefit more from their assigned program ($-3.13$ months) than from \emph{Mentoring} ($-1.87$ months), a difference of 1.26 months (SE 0.09). The algorithm thus identifies not only who benefits most from treatment, but which treatment benefits each applicant most.

Applicants assigned to no program (70\% of the pool) exhibit smaller predicted effects under either counterfactual: $-1.72$ months for \emph{Mentoring} and $-2.02$ months for \emph{Challenges}. While these applicants would benefit from participation, their predicted gains are smaller than those of optimally assigned applicants. Interestingly, this group would benefit relatively more from \emph{Challenges} than \emph{Mentoring}, suggesting that among lower-priority applicants, the lower-cost program delivers comparable or greater value.

\begin{table}[!htbp]
\caption{Counterfactual Treatment Effects by Assignment Group}
\label{targetting}
\centering
\resizebox{0.8\textwidth}{!}{%
\begin{tabular}{lccc}
\toprule\toprule

& Assigned to \emph{Mentoring} & Assigned to \emph{Challenges} & Assigned to Out \\
\midrule
Share of applicants & 0.15 & 0.15 & 0.70 \\
\midrule
\multicolumn{4}{@{}l}{\textit{Treatment effect if assigned to:}}\\
\addlinespace
\emph{Mentoring} & $-$5.55 (0.09) & $-$1.87 (0.05) & $-$1.72 (0.02) \\
\emph{Challenges} & $-$3.12 (0.06) & $-$3.13 (0.07) & $-$2.02 (0.02) \\
\midrule
\multicolumn{4}{@{}l}{\textit{Difference:}}\\
\addlinespace
\emph{Mentoring} $-$ \emph{Challenges} & $-$2.43 (0.11) & 1.26 (0.09) & 0.30 (0.03) \\
\bottomrule\bottomrule

\end{tabular}%
}
\caption*{\footnotesize\textit{Notes:} Treatment effects measured as change in restricted mean survival time (months); negative values indicate faster job finding. Each column represents applicants assigned to that program under the 15/15/70 targeted policy. Rows show estimated treatment effects if that group were counterfactually assigned to each program. Standard errors in parentheses.}
\end{table}

\paragraph{Policy value across capacity levels.}

\begin{figure}[!htbp]
\caption{Policy Value Across Capacity Levels}
\label{fig:capacity_value}
\centering
\includegraphics[width=0.75\textwidth]{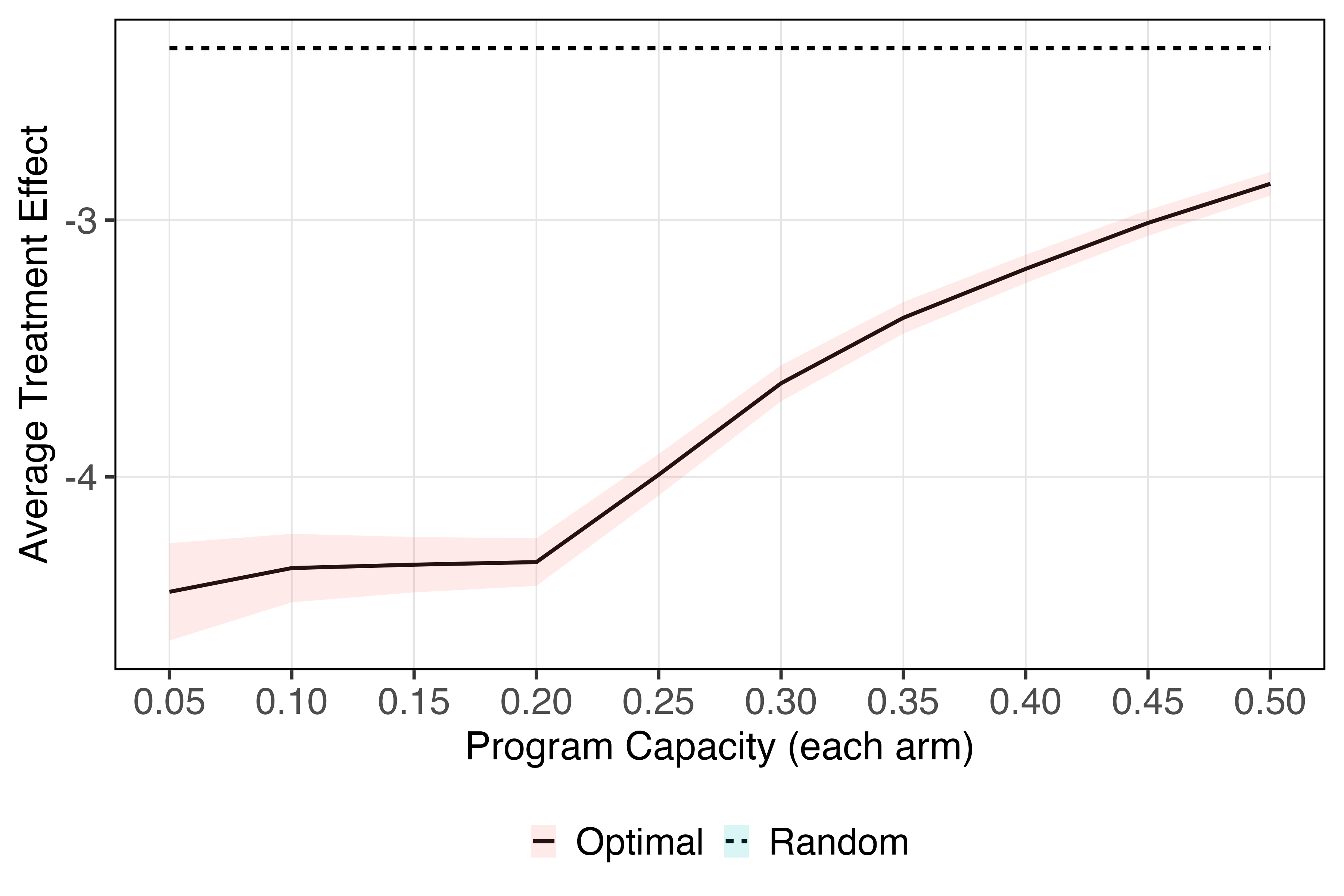}
\caption*{\footnotesize\textit{Notes:} Average treatment effects (RMST, in months) under targeted allocation as capacity varies from 5\% to 50\% per program. Shaded region indicates 95\% confidence interval. Policy value evaluated using cross-fitted AIPW estimates on the test set. At each capacity level, targeted allocation solves the assignment problem in Equation~\ref{eq_opt}. Dashed line shows the average treatment effect under random assignment with equal probability across programs.}
\end{figure}

The preceding results examine discrete capacity levels. To characterize the full capacity-selectivity tradeoff, Figure~\ref{fig:capacity_value} plots average treatment effects as program capacity varies continuously from 5\% to 50\% per program. The solid curve shows targeted allocation; the dashed horizontal line represents random assignment.

Three patterns emerge. First, targeted allocation outperforms random assignment at every capacity level, with the largest gains when capacity is most constrained. At 5\% capacity, targeting achieves effects of approximately $-4.5$ months compared to random allocation's $-2.3$ months. Second, returns to targeting diminish as capacity expands: per-person effects decline steadily as the algorithm exhausts high-benefit applicants and must include those with smaller predicted gains. Third, and most notably, targeted allocation maintains a meaningful advantage even at universal coverage (50\% per program). The 0.6-month improvement at full capacity reflects gains from matching applicants to their best-suited program, not just from selecting high-benefit individuals. This complementarity between applicant characteristics and program type generates value independent of capacity constraints.

\subsection{Counterfactual Mentors' Selection Policies}\label{mentors_selection}

Now, we examine selection within the \emph{Mentoring} program: how do mentors choose mentees, and could alternative selection strategies improve program impact? Understanding selection matters for two reasons. First, econometrically, mentor selection determines the population for which our treatment effects are identified. Our estimates of $\tau^M(\mathcal{G}^S)$ apply to the mentor-selected population, which may differ from the broader applicant pool. Second, mentoring occurs outside experimental settings, and mentor preferences may not align with social efficiency: if mentors favor applicants who would succeed regardless of intervention, or if less experienced mentors lack the judgment to identify high-potential candidates, there is scope for improved targeting. Mentors do demonstrably identify promising applicants: as shown in Figure~\ref{fig:outcomes_over_time}, applicants selected by mentors but randomly assigned to the control group achieved substantially higher employment rates than non-selected applicants. But whether they identify applicants with high \emph{treatment effects}---rather than high baseline employability---is an empirical question. Table~\ref{tab:selection_mentor_characteristics} characterizes these selection patterns by comparing selected versus non-selected applicants (Panel A) and by comparing applicants chosen by different types of mentors (Panels B and C).

\begin{table}[!ht]
  \caption{Differences in Applicant Characteristics: Selection and Mentor Experience}
  \label{tab:selection_mentor_characteristics}
  \centering
  \resizebox{\textwidth}{!}{%
  \begin{tabular}{lccccccccccc}
    \toprule
    \toprule
     & Baseline & Small town & Warsaw & Over 30 & Prof. exp. & Social science & STEM & Applied before & Mother & High skill & Family/friends in IT \\
    \midrule

    \multicolumn{12}{l}{\textit{Panel A: Selected vs. Not Selected}}\\
    Selected & 0.456 & 0.245 & 0.268 & 0.487 & 7.097 & 0.134 & 0.510 & -- & -- & -- & -- \\
     & (0.041) & (0.025) & (0.026) & (0.029) & (0.284) & (0.020) & (0.029) & (--) & (--) & (--) & (--) \\
    \addlinespace
    Not & 0.286 & 0.476 & 0.282 & 0.147 & 7.489 & 0.207 & 0.182 & -- & -- & -- & -- \\
     & (0.021) & (0.023) & (0.021) & (0.016) & (0.288) & (0.019) & (0.018) & (--) & (--) & (--) & (--) \\
    \midrule

    \multicolumn{12}{l}{\textit{Panel B: Mentor Managerial Experience}}\\
    High & 0.378 & 0.229 & 0.281 & 0.588 & 7.595 & 0.137 & 0.497 & 0.261 & 0.288 & 0.614 & 0.699 \\
     & (0.057) & (0.034) & (0.036) & (0.040) & (0.417) & (0.028) & (0.041) & (0.036) & (0.037) & (0.039) & (0.037) \\
    \addlinespace
    Low & 0.549 & 0.266 & 0.245 & 0.378 & 6.503 & 0.133 & 0.531 & 0.210 & 0.231 & 0.497 & 0.699 \\
     & (0.059) & (0.037) & (0.036) & (0.041) & (0.384) & (0.028) & (0.042) & (0.034) & (0.035) & (0.042) & (0.038) \\
    \midrule

    \multicolumn{12}{l}{\textit{Panel C: Mentor Total Experience (above vs. below median)}}\\
    Above & 0.390 & 0.239 & 0.299 & 0.590 & 8.197 & 0.171 & 0.564 & 0.282 & 0.308 & 0.573 & 0.718 \\
     & (0.064) & (0.040) & (0.043) & (0.046) & (0.442) & (0.035) & (0.046) & (0.042) & (0.043) & (0.046) & (0.042) \\
    \addlinespace
    Below & 0.512 & 0.251 & 0.240 & 0.419 & 6.330 & 0.112 & 0.480 & 0.207 & 0.229 & 0.547 & 0.687 \\
     & (0.054) & (0.033) & (0.032) & (0.037) & (0.364) & (0.024) & (0.037) & (0.030) & (0.031) & (0.037) & (0.035) \\
    \bottomrule
    \bottomrule
  \end{tabular}
  }
  \caption*{\footnotesize{\textit{Notes: Means shown with standard errors in parentheses. Baseline refers to the average outcome when not participating in Mentoring. Panel A compares applicants who were selected vs. not selected. Panel B compares applicants selected by mentors with vs. without managerial experience. Panel C compares applicants selected by mentors above vs. below the median years of experience.}}}
\end{table}

Panel A reveals that mentors select applicants with substantially stronger baseline employment prospects: 46\% of selected applicants in the control group find technology jobs, compared to 29\% of non-selected applicants. Selected applicants are more likely to be over 30 (49\% vs.\ 15\%), hold STEM degrees (51\% vs.\ 18\%), and come from urban areas outside small towns (25\% vs.\ 48\% from small towns). This selection pattern presents a puzzle: Section~\ref{hte} documented larger treatment effects for applicants from small towns and those without STEM degrees. Mentors appear to select on baseline employability rather than on treatment effect magnitude.

Panels B and C suggest that experienced mentors employ a different selection philosophy. Mentors with managerial experience select applicants with \emph{weaker} baseline employment prospects (38\% vs.\ 55\%) but who are older (59\% vs.\ 38\% over 30) and have more professional experience (7.6 vs.\ 6.5 years). Panel C shows similar patterns for mentors with above-median career experience. These experienced mentors appear to identify applicants with high potential who face significant barriers; perhaps recognizing that such applicants offer greater scope for mentoring impact. Less experienced mentors select more conventionally, favoring applicants with stronger baseline employability.

\paragraph{Alternative selection strategies.} We use our policy evaluation framework to examine how alternative selection strategies would affect \emph{Mentoring} effectiveness. We consider four algorithmic approaches, each selecting 15\% of applicants (approximating Dare IT's typical capacity):

\begin{itemize}
    \item \emph{Highest Treatment Effect}: select applicants with the largest predicted treatment effects $\hat{\tau}_i^M$
    \item \emph{Most Promising}: select applicants with the highest predicted outcomes \emph{absent} treatment, $\hat{Y}_i(0)$
    \item \emph{Least Promising}: select applicants with the lowest predicted outcomes absent treatment
    \item \emph{Best if Treated}: select applicants with the highest predicted outcomes \emph{under} treatment, $\hat{Y}_i(M)$
\end{itemize}

\noindent We benchmark these against random selection and the current mentor-based allocation.

\paragraph{Overlap between selection strategies.}

Table~\ref{tab:policy_overlap} examines the extent to which different selection strategies choose the same individuals. Each cell shows the percentage of individuals selected under the row policy who also appear in the column policy. Importantly, the algorithmic policies do not use information on whether an applicant was actually selected by mentors; any overlap reflects convergence on similar applicant characteristics through independent criteria.

\begin{table}[!htbp]
\caption{Overlap Between Selection Strategies (\%)}
\label{tab:policy_overlap}
\centering
\begin{tabular}{lccccc}
\toprule\toprule
& Highest TE & Most Prom. & Least Prom. & Best if Treated & Current \\
\midrule
Highest Treatment Effect & \textbf{100.0} & & & & \\
Most Promising & 28.5 & \textbf{100.0} & & & \\
Least Promising & 18.2 & 0.0 & \textbf{100.0} & & \\
Best if Treated & 27.9 & 26.7 & 31.5 & \textbf{100.0} & \\
Current (Mentor) & 70.4 & 29.6 & 18.3 & 19.7 & \textbf{100.0} \\
\bottomrule\bottomrule
\end{tabular}
\caption*{\footnotesize\textit{Notes:} Each cell shows the percentage of row-policy selections that also appear in the column policy. Lower triangle only; upper triangle symmetric. All policies select 15\% of applicants. Under random selection, expected overlap would be 15\%.}
\end{table}

The current mentor-based selection exhibits substantial overlap with the \emph{Highest Treatment Effect} policy: 70\% of mentor-selected applicants would also be selected by the treatment-effect-maximizing algorithm. This overlap far exceeds the 15\% expected under random selection, indicating that mentors, through whatever criteria they apply, tend to choose applicants whom the algorithm also identifies as high-benefit. The overlap with \emph{Most Promising} is considerably lower (30\%), and overlap with \emph{Least Promising} is lower still (18\%), suggesting mentors do not primarily select based on baseline employability in either direction.

These patterns are consistent with two interpretations. First, mentors may implicitly target treatment effects, recognizing which applicants would benefit most from mentoring. Second, the characteristics that predict high treatment effects (such as facing network barriers) may correlate with other attributes mentors find compelling in application materials. We cannot distinguish these explanations, but the high concordance between mentor judgment and algorithmic targeting suggests that scaling the \emph{Mentoring} program with algorithmic selection would likely preserve much of the effectiveness achieved under mentor-based selection.

\paragraph{Comparing selection strategies.}

Table~\ref{tab:mentoring_policies} compares treatment effects across six selection strategies for the \emph{Mentoring} program. The \emph{Highest Treatment Effect} policy achieves the largest impact: $-3.59$ months (SE 0.03), meaning selected participants find technology jobs 3.6 months faster than they would without the program. \emph{Random} selection serves as a benchmark, generating effects of $-2.20$ months (SE 0.04). The \emph{Current} mentor-based selection performs well, achieving $-3.25$ months (SE 0.10), a 48\% improvement over random allocation. This validates that mentors successfully identify high-benefit applicants through their selection process.

\begin{table}[!htbp]
\caption{Treatment Effects Under Alternative Selection Strategies: \emph{Mentoring}}
\label{tab:mentoring_policies}
\centering
\begin{tabular}{lcc}
\toprule\toprule

Strategy & Baseline & Treatment Effect \\
& (months) & (months) \\
\midrule
Current (Mentor) & 11.09 (0.13) & $-$3.25 (0.10) \\
Random & 11.36 (0.06) & $-$2.20 (0.04) \\
\midrule
\multicolumn{3}{l}{\textit{Algorithmic strategies:}} \\
Highest Treatment Effect & 11.08 (0.09) & $-$3.59 (0.03) \\
Most Promising & 9.71 (0.04) & $-$2.27 (0.08) \\
Least Promising & 13.44 (0.09) & $-$1.94 (0.07) \\
Best if Treated & 11.77 (0.15) & $-$2.14 (0.08) \\
\bottomrule\bottomrule

\end{tabular}
\caption*{\footnotesize\textit{Notes:} Comparison of selection strategies for \emph{Mentoring}, each selecting 15\% of applicants. ``Baseline'' is predicted months to employment without treatment (lower = faster job finding absent intervention). ``Treatment Effect'' is the change in months from treatment (negative = faster job finding due to program). Standard errors in parentheses.}
\end{table}

The \emph{Highest Treatment Effect} policy achieves a 10\% improvement over mentor selection ($-3.59$ versus $-3.25$ months). Notably, this algorithmic policy uses only observable application characteristics, while mentors access richer information including video introductions and holistic assessment. The substantial overlap between mentor selections and algorithmic targeting (70\%, from Table~\ref{tab:policy_overlap}) suggests that mentors and the algorithm identify similar applicant profiles through different information sets. Data-driven selection could complement mentor judgment by providing a systematic check or by scaling selection when mentor capacity is limited.

Policies targeting objectives other than treatment effects perform substantially worse. \emph{Most Promising} selects applicants with the strongest baseline prospects---those predicted to find jobs in 9.7 months even without intervention---yet achieves treatment effects of only $-2.27$ months, barely exceeding random allocation. Strong baseline prospects do not predict strong treatment response. Conversely, \emph{Least Promising} targets applicants with the weakest baselines (13.4 months to employment without intervention) but generates the smallest treatment effects ($-1.94$ months). This pattern reveals a tension between equity and efficiency: applicants with the greatest baseline disadvantage benefit less from the program than those with moderate baselines. However, even the \emph{Least Promising} group experiences meaningful gains: nearly two months of accelerated job finding, suggesting that equity-focused selection remains beneficial in absolute terms, if not maximal in efficiency terms.

\emph{Best if Treated}, which selects applicants predicted to have the best outcomes under treatment rather than the largest treatment \emph{effects}, performs similarly to random selection ($-2.14$ versus $-2.20$ months). This underscores a key distinction: applicants who perform best after treatment are not the same as applicants who benefit most from treatment. Program impact depends on the counterfactual, what would have happened absent intervention, not just realized outcomes.
\section{Conclusion}

Changing labor markets push workers to want to transition into high-growth sectors, and facilitating these transitions is an important goal for labor policy. However, effectively designing such policies requires principled evaluation of various training programs in terms of their overall impact and their heterogeneity. In this paper, we provide evidence of the high effectiveness of two programs supporting women transitioning into the technology sector, identifying a class of interventions that can effectively narrow the gender gap in technology.

Both programs generate large, statistically significant effects on technology employment. \emph{Mentoring} increases technology employment by 15 percentage points (from 46\% to 61\%), while \emph{Challenges} increases it by 11 percentage points (from 27\% to 38\%), with effects persisting throughout our observation period. The \emph{Challenges} program operates at approximately \$15 per participant, implying a cost of roughly \$150 per additional technology job—a remarkably low figure relative to traditional training programs.

The two programs address distinct barriers to labor market entry. \emph{Mentoring} expands professional networks: participants gain 53 LinkedIn connections on average, a 23\% increase, while \emph{Challenges} participants show no network growth. \emph{Challenges} instead provides credible skill signals through portfolio projects that demonstrate technical competence to prospective employers. These distinct mechanisms generate predictable heterogeneity in treatment effects. Expanding professional networks creates the greatest opportunity for women from small towns, those without STEM degrees, and mothers. Building credible skill signals offers the greatest opportunity for early-career women with limited work history. The 28 percentage point larger effect for mothers in \emph{Mentoring} is particularly striking, suggesting that mentoring can substantially reduce motherhood penalties in technology employment.

These heterogeneous effects have implications for program design. We develop an assignment policy that allocates applicants to their highest-benefit program based on predicted treatment effects. At current capacity levels, targeted assignment achieves effects 86\% larger than random allocation. Even at expanded capacity where every applicant receives a program slot, targeting generates 23\% larger effects than random assignment by matching individuals to their best-suited program. This demonstrates that the value of targeting comes not only from selecting high-benefit individuals but also from exploiting complementarities between participant characteristics and program mechanisms.

We also examine mentor-based selection within the \emph{Mentoring} program. Mentors successfully identify promising candidates: their selections achieve effects 48\% larger than random allocation. Algorithmic selection based on observable characteristics achieves a further 10\% improvement. The 70\% overlap between mentor and algorithmic selections suggests both approaches identify similar applicant profiles. Data-driven tools could complement mentor judgment, particularly for scaling selection when mentor capacity is limited.

Several limitations merit acknowledgment. First, moderate sample sizes yield wide confidence intervals for some estimates, particularly for subgroup analyses. Second, while our evidence is consistent with network and signaling mechanisms, we cannot definitively isolate the causal channels through which each program operates. Understanding these mechanisms more precisely could inform efforts to scale \emph{Mentoring} by identifying its essential components. Third, for \emph{Challenges}, our estimates come from the first program cohort; subsequent cohorts may differ in composition or in treatment response as the programs mature and selection patterns evolve. Yet studying small, custom-designed programs enabled the access and stakeholder cooperation necessary for our research design, including the novel paired randomization within mentor-selected candidates that allows us to separately identify selection and treatment effects, a contribution absent from prior mentoring RCTs.

More broadly, having two randomized interventions targeting distinct mechanisms within the same population enables comparisons rarely possible in program evaluation. We observe not only employment outcomes but also professional network expansion, allowing us to directly test whether mentoring operates through networks. We examine mentor selection behavior—how mentors choose, whether experienced mentors select differently, and how mentor choices compare to algorithmic efficiency. These analyses were possible only because our design preserves mentor discretion while enabling randomization.

Our findings demonstrate that women's underrepresentation in technology reflects multiple distinct barriers---network access, credentialing, confidence---that require different interventions. Effective policy should provide multiple pathways tailored to different populations rather than assuming a single intervention serves all. The substantial heterogeneity we document, and the gains from exploiting it through targeted assignment, suggest that personalized approaches to workforce development may substantially outperform one-size-fits-all programs.
\newpage
\bibliographystyle{chicago}
\bibliography{refs}

@article{zhou2023offline,
  title={Offline multi-action policy learning: Generalization and optimization},
  author={Zhou, Zhengyuan and Athey, Susan and Wager, Stefan},
  journal={Operations Research},
  volume={71},
  number={1},
  pages={148--183},
  year={2023},
  publisher={INFORMS}
}

@incollection{athey2017econometrics,
  title={The econometrics of randomized experiments},
  author={Athey, Susan and Imbens, Guido W},
  booktitle={Handbook of Economic Field Experiments},
  volume={1},
  pages={73--140},
  year={2017},
  publisher={Elsevier}
}

@article{athey2019generalized,
  title={Generalized random forests},
  author={Athey, Susan and Tibshirani, Julie and Wager, Stefan},
  journal={The Annals of Statistics},
  volume={47},
  number={2},
  pages={1148--1178},
  year={2019},
  publisher={Institute of Mathematical Statistics}
}

@article{dennehy2017female,
  title={Female peer mentors early in college increase women’s positive academic experiences and retention in engineering},
  author={Dennehy, Tara C and Dasgupta, Nilanjana},
  journal={Proceedings of the National Academy of Sciences},
  volume={114},
  number={23},
  pages={5964--5969},
  year={2017},
  publisher={National Acad Sciences}
}

@article{resnjanskij2021can,
  title={Can mentoring alleviate family disadvantage in adolescence? A field experiment to improve labor market prospects},
  author={Resnjanskij, Sven and Ruhose, Jens and Wiederhold, Simon and Woessmann, Ludger and Wedel, Katharina},
  journal={Journal of Political Economy},
  volume={132},
  number={3},
  pages={1013--1062},
  year={2024},
  publisher={The University of Chicago Press Chicago, IL}
}

@article{gardiner2007show,
  title={Show me the money! {A}n empirical analysis of mentoring outcomes for women in academia},
  author={Gardiner, Maria and Tiggemann, Marika and Kearns, Hugh and Marshall, Kelly},
  journal={Higher Education Research \& Development},
  volume={26},
  number={4},
  pages={425--442},
  year={2007},
  publisher={Taylor \& Francis}
}

@inproceedings{ginther2020can,
  title={Can mentoring help female assistant professors in economics? {A}n evaluation by randomized trial},
  author={Ginther, Donna K and Currie, Janet M and Blau, Francine D and Croson, Rachel TA},
  booktitle={AEA Papers and Proceedings},
  volume={110},
  pages={205--09},
  year={2020}
}

@article{blau2010can,
  title={Can mentoring help female assistant professors? {I}nterim results from a randomized trial},
  author={Blau, Francine D and Currie, Janet M and Croson, Rachel TA and Ginther, Donna K},
  journal={American Economic Review},
  volume={100},
  number={2},
  pages={348--52},
  year={2010}
}

@article{correll2016succeed,
  title={To succeed in tech, women need more visibility},
  author={Correll, Shelley and Mackenzie, Lori},
  journal={Harvard Business Review},
  pages={2--6},
  year={2016}
}

@article{murciano2022missing,
  title={Missing women in tech: The labor market for highly skilled software engineers},
  author={Murciano-Goroff, Raviv},
  journal={Management Science},
  volume={68},
  number={5},
  pages={3262--3281},
  year={2022},
  publisher={INFORMS}
}

@article{cheryan2013stereotypical,
  title={The stereotypical computer scientist: Gendered media representations as a barrier to inclusion for women},
  author={Cheryan, Sapna and Plaut, Victoria C and Handron, Caitlin and Hudson, Lauren},
  journal={Sex roles},
  volume={69},
  number={1},
  pages={58--71},
  year={2013},
  publisher={Springer}
}

@article{del2022more,
  title={More women in tech? {E}vidence from a field experiment addressing social identity},
  author={Del Carpio, Lucia and Guadalupe, Maria},
  journal={Management Science},
  volume={68},
  number={5},
  pages={3196--3218},
  year={2022},
  publisher={INFORMS}
}

@article{athey2021policy,
  title={Policy learning with observational data},
  author={Athey, Susan and Wager, Stefan},
  journal={Econometrica},
  volume={89},
  number={1},
  pages={133--161},
  year={2021},
  publisher={Wiley Online Library}
}

@article{card2018works,
  title={What works? {A} meta analysis of recent active labor market program evaluations},
  author={Card, David and Kluve, Jochen and Weber, Andrea},
  journal={Journal of the European Economic Association},
  volume={16},
  number={3},
  pages={894--931},
  year={2018},
  publisher={Oxford University Press}
}

@article{sianesi2008differential,
  title={Differential effects of active labour market programs for the unemployed},
  author={Sianesi, Barbara},
  journal={Labour Economics},
  volume={15},
  number={3},
  pages={370--399},
  year={2008},
  publisher={Elsevier}
}

@article{lechner2000microeconometric,
  title={Microeconometric Evaluation of the Active Labour Market Policy in {S}witzerland},
  author={Lechner, Michael and Gerfin, Michael},
  journal={Available at SSRN 233906},
  year={2000}
}

@article{biewen2014effectiveness,
  title={The effectiveness of public-sponsored training revisited: The importance of data and methodological choices},
  author={Biewen, Martin and Fitzenberger, Bernd and Osikominu, Aderonke and Paul, Marie},
  journal={Journal of Labor Economics},
  volume={32},
  number={4},
  pages={837--897},
  year={2014},
  publisher={University of Chicago Press Chicago, IL}
}

@article{fein2018bridging,
  title={Bridging the Opportunity Divide for Low-Income Youth: Implementation and Early Impacts of the {Y}ear {U}p Program. {P}athways for {A}dvancing {C}areers and {E}ducation. {OPRE} {R}eport 2018-65.},
  author={Fein, David and Hamadyk, Jill},
  journal={Office of Planning, Research and Evaluation},
  year={2018},
  publisher={ERIC}
}

@incollection{spence1978job,
  title={Job market signaling},
  author={Spence, Michael},
  booktitle={Uncertainty in Economics},
  pages={281--306},
  year={1978},
  publisher={Elsevier}
}

@article{tyler2000estimating,
  title={Estimating the labor market signaling value of the {GED}},
  author={Tyler, John H and Murnane, Richard J and Willett, John B},
  journal={The Quarterly Journal of Economics},
  volume={115},
  number={2},
  pages={431--468},
  year={2000},
  publisher={MIT Press}
}

@article{hadavand2018can,
  title={Can {MOOC} Programs Improve Student Employment Prospects?},
  author={Hadavand, Aboozar and Gooding, Ira and Leek, Jeffrey T},
  journal={Available at SSRN 3260695},
  year={2018}
}

@incollection{heckman1999economics,
  title={The economics and econometrics of active labor market programs},
  author={Heckman, James J and LaLonde, Robert J and Smith, Jeffrey A},
  booktitle={Handbook of Labor Economics},
  volume={3},
  pages={1865--2097},
  year={1999},
  publisher={Elsevier}
}

@report{fluff2021,
author = {Gawlowska-Bujok, Magdalena and Bujok, Tomasz},
title = {{IT} Job Market in {P}oland in 2021. {S}alaries, technologies and requirements in job ads},
institution = {No Fluff Jobs},
year = {2021},
}

@report{gus2022,
author = {GUS},
title = {Structure of wages and salaries by occupations in {O}ctober 2020},
institution = {Polish Statistics Office},
year = {2022},
}

@report{EC2021,
 author = {{European Commission}},
 year = {2021},
 title = {Women in Digital Scoreboard 2021},
 journal = {European Commission},
 url = {https://digital-strategy.ec.europa.eu/en/news/women-digital-scoreboard-2021},
 urldate = {2022-10-13}
}

@article{ragins1999mentor,
  title={Mentor functions and outcomes: a comparison of men and women in formal and informal mentoring relationships.},
  author={Ragins, Belle Rose and Cotton, John L},
  journal={Journal of applied psychology},
  volume={84},
  number={4},
  pages={529},
  year={1999},
  publisher={American Psychological Association}
}

@article{kammeyer2008quantitative,
  title={A quantitative review of mentoring research: Test of a model},
  author={Kammeyer-Mueller, John D and Judge, Timothy A},
  journal={Journal of Vocational Behavior},
  volume={72},
  number={3},
  pages={269--283},
  year={2008},
  publisher={Elsevier}
}

@techreport{chernozhukov2017double,
  title={Double/debiased machine learning for treatment and causal parameters},
  author={Chernozhukov, Victor and Chetverikov, Denis and Demirer, Mert and Duflo, Esther and Hansen, Christian and Newey, Whitney and Robins, James and others},
  year={2017}
}

@article{ascarza2018retention,
  title={Retention futility: Targeting high-risk customers might be ineffective},
  author={Ascarza, Eva},
  journal={Journal of Marketing Research},
  volume={55},
  number={1},
  pages={80--98},
  year={2018},
  publisher={SAGE Publications Sage CA: Los Angeles, CA}
}

@article{haushofer2022targeting,
  title={Targeting impact versus deprivation},
  author={Haushofer, Johannes and Niehaus, Paul and Paramo, Carlos and Miguel, Edward and Walker, Michael},
  journal={American Economic Review},
  volume={115},
  number={6},
  pages={1936--1974},
  year={2025},
  publisher={American Economic Association 2014 Broadway, Suite 305, Nashville, TN 37203}
}

@article{inoue2023machine,
  title={Machine-learning-based high-benefit approach versus conventional high-risk approach in blood pressure management},
  author={Inoue, Kosuke and Athey, Susan and Tsugawa, Yusuke},
  journal={International Journal of Epidemiology},
  pages={dyad037},
  year={2023},
  publisher={Oxford University Press}
}

@article{hitsch2018heterogeneous,
  title={Heterogeneous treatment effects and optimal targeting policy evaluation},
  author={Hitsch, G{\"u}nter J and Misra, Sanjog and Zhang, Walter W},
  journal={Quantitative Marketing and Economics},
  volume={22},
  number={2},
  pages={115--168},
  year={2024},
  publisher={Springer}
}

@article{pallais2014inefficient,
  title={Inefficient hiring in entry-level labor markets},
  author={Pallais, Amanda},
  journal={American Economic Review},
  volume={104},
  number={11},
  pages={3565--3599},
  year={2014},
  publisher={American Economic Association 2014 Broadway, Suite 305, Nashville, TN 37203}
}

@article{pallais2016referential,
  title={Why the referential treatment? Evidence from field experiments on referrals},
  author={Pallais, Amanda and Sands, Emily Glassberg},
  journal={Journal of Political Economy},
  volume={124},
  number={6},
  pages={1793--1828},
  year={2016},
  publisher={University of Chicago Press Chicago, IL}
}

@article{athey_palikot_rct,
  author    = {Athey, Susan and Palikot, Emil},
  title     = {Mentoring program experiment},
  journal   = {AEA RCT Registry},
  year      = {2022},
  url       = {https://www.socialscienceregistry.org/trials/10045}
}

@article{athey_palikot_rct_2,
  author    = {Athey, Susan and Palikot, Emil},
  title     = {Portfolio challenges},
  journal   = {AEA RCT Registry},
  year      = {2022},
  url       = {https://www.socialscienceregistry.org/trials/10044}
}

@article{lee2009training,
  title={Training, Wages, and Sample Selection: Estimating Sharp Bounds on Treatment Effects},
  author={Lee, David S},
  journal={The Review of Economic Studies},
  volume={76},
  number={3},
  pages={1071--1102},
  year={2009},
  publisher={Review of Economic Studies Ltd}
}

@article{forret2004networking,
  title={Networking behaviors and career outcomes: differences for men and women?},
  author={Forret, Monica L and Dougherty, Thomas W},
  journal={Journal of Organizational Behavior: The International Journal of Industrial, Occupational and Organizational Psychology and Behavior},
  volume={25},
  number={3},
  pages={419--437},
  year={2004},
  publisher={Wiley Online Library}
}

@article{greguletz2019women,
  title={Why women build less effective networks than men: The role of structural exclusion and personal hesitation},
  author={Greguletz, Elena and Diehl, Marjo-Riitta and Kreutzer, Karin},
  journal={Human Relations},
  volume={72},
  number={7},
  pages={1234--1261},
  year={2019},
  publisher={Sage Publications Sage UK: London, England}
}

@article{lee2021bounding,
  title={Bounding treatment effects by pooling limited information across observations},
  author={Lee, Sokbae and Weidner, Martin},
  journal={arXiv preprint arXiv:2111.05243},
  year={2021}
}

@article{granovetter1973strength,
  title={The strength of weak ties},
  author={Granovetter, Mark S},
  journal={American journal of sociology},
  volume={78},
  number={6},
  pages={1360--1380},
  year={1973},
  publisher={University of Chicago Press}
}

@article{ioannides2004job,
  title={Job information networks, neighborhood effects, and inequality},
  author={Ioannides, Yannis M and Loury, Linda Datcher},
  journal={Journal of economic literature},
  volume={42},
  number={4},
  pages={1056--1093},
  year={2004},
  publisher={American Economic Association}
}

@article{lerchenmueller2019gender,
  title={Gender differences in how scientists present the importance of their research: observational study},
  author={Lerchenmueller, Marc J and Sorenson, Olav and Jena, Anupam B},
  journal={bmj},
  volume={367},
  year={2019},
  publisher={British Medical Journal Publishing Group}
}

@article{spencer1999stereotype,
  title={Stereotype threat and women's math performance},
  author={Spencer, Steven J and Steele, Claude M and Quinn, Diane M},
  journal={Journal of experimental social psychology},
  volume={35},
  number={1},
  pages={4--28},
  year={1999},
  publisher={Elsevier}
}

@techreport{unitednations2021,
  title={Measuring digital development: Facts and figures},
  author={{United Nations}},
  year={2021},
  institution={International Telecommunication Union},
  address={Geneva, Switzerland}
}

@article{cui2023estimating,
  title={Estimating heterogeneous treatment effects with right-censored data via causal survival forests},
  author={Cui, Yifan and Kosorok, Michael R and Sverdrup, Erik},
  journal={Journal of the Royal Statistical Society Series B: Statistical Methodology},
  volume={85},
  number={2},
  pages={179--211},
  year={2023},
  publisher={Oxford University Press}
}

@manual{tibshirani2024grf,
  title={grf: Generalized Random Forests},
  author={Tibshirani, Julie and Athey, Susan and Sverdrup, Erik and Wager, Stefan},
  year={2024},
  note={R package version 2.3.2},
  url={https://CRAN.R-project.org/package=grf}
}

@article{kleinberg2018human,
  title={Human decisions and machine predictions},
  author={Kleinberg, Jon and Lakkaraju, Himabindu and Leskovec, Jure and Ludwig, Jens and Mullainathan, Sendhil},
  journal={The Quarterly Journal of Economics},
  volume={133},
  number={1},
  pages={237--293},
  year={2018},
  publisher={Oxford University Press}
}

@article{card2010active,
  title={Active labour market policy evaluations: A meta-analysis},
  author={Card, David and Kluve, Jochen and Weber, Andrea},
  journal={The Economic Journal},
  volume={120},
  number={548},
  pages={F452--F477},
  year={2010},
  publisher={Wiley Online Library}
}

@article{crepon2016active,
  title={Active labor market policies},
  author={Cr{\'e}pon, Bruno and van den Berg, Gerard J},
  journal={Annual Review of Economics},
  volume={8},
  number={1},
  pages={521--546},
  year={2016},
  publisher={Annual Reviews}
}

@article{kitagawa2018should,
  title={Who should be treated? empirical welfare maximization methods for treatment choice},
  author={Kitagawa, Toru and Tetenov, Aleksey},
  journal={Econometrica},
  volume={86},
  number={2},
  pages={591--616},
  year={2018},
  publisher={Wiley Online Library}
}

@article{bhattacharya2012inferring,
  title={Inferring welfare maximizing treatment assignment under budget constraints},
  author={Bhattacharya, Debopam and Dupas, Pascaline},
  journal={Journal of Econometrics},
  volume={167},
  number={1},
  pages={168--196},
  year={2012},
  publisher={Elsevier}
}

@article{kasy2021adaptive,
  title={Adaptive treatment assignment in experiments for policy choice},
  author={Kasy, Maximilian and Sautmann, Anja},
  journal={Econometrica},
  volume={89},
  number={1},
  pages={113--132},
  year={2021},
  publisher={Wiley Online Library}
}

@article{mullainathan2022diagnosing,
  title={Diagnosing physician error: A machine learning approach to low-value health care},
  author={Mullainathan, Sendhil and Obermeyer, Ziad},
  journal={The Quarterly Journal of Economics},
  volume={137},
  number={2},
  pages={679--727},
  year={2022},
  publisher={Oxford University Press}
}

@article{juhn2017specialization,
  title={Specialization then and now: Marriage, children, and the gender earnings gap across cohorts},
  author={Juhn, Chinhui and McCue, Kristin},
  journal={Journal of Economic Perspectives},
  volume={31},
  number={1},
  pages={183--204},
  year={2017},
  publisher={American Economic Association 2014 Broadway, Suite 305, Nashville, TN 37203-2418}
}

@article{reuben2014gender,
  title={How stereotypes impair women's careers in science},
  author={Reuben, Ernesto and Sapienza, Paola and Zingales, Luigi},
  journal={Proceedings of the National Academy of Sciences},
  volume={111},
  number={12},
  pages={4403--4408},
  year={2014},
  publisher={National Acad Sciences}
}

@article{moss-racusin2012science,
  title={Science faculty's subtle gender biases favor male students},
  author={Moss-Racusin, Corinne A and Dovidio, John F and Brescoll, Victoria L and Graham, Mark J and Handelsman, Jo},
  journal={Proceedings of the National Academy of Sciences},
  volume={109},
  number={41},
  pages={16474--16479},
  year={2012},
  publisher={National Acad Sciences}
}

@article{goldin2014grand,
  title={A grand gender convergence: Its last chapter},
  author={Goldin, Claudia},
  journal={American Economic Review},
  volume={104},
  number={4},
  pages={1091--1119},
  year={2014}
}

@article{kleven2019children,
  title={Children and gender inequality: Evidence from Denmark},
  author={Kleven, Henrik and Landais, Camille and S{\o}gaard, Jakob Egholt},
  journal={American Economic Journal: Applied Economics},
  volume={11},
  number={4},
  pages={181--209},
  year={2019}
}

@article{bertrand2010dynamics,
  title={Dynamics of the gender gap for young professionals in the financial and corporate sectors},
  author={Bertrand, Marianne and Goldin, Claudia and Katz, Lawrence F},
  journal={American Economic Journal: Applied Economics},
  volume={2},
  number={3},
  pages={228--255},
  year={2010}
}

@article{bayer2016diversity,
  title={Diversity in the economics profession: A new attack on an old problem},
  author={Bayer, Amanda and Rouse, Cecilia Elena},
  journal={Journal of Economic Perspectives},
  volume={30},
  number={4},
  pages={221--242},
  year={2016}
}

@article{carrell2010sex,
  title={Sex and science: How professor gender perpetuates the gender gap},
  author={Carrell, Scott E and Page, Marianne E and West, James E},
  journal={The Quarterly Journal of Economics},
  volume={125},
  number={3},
  pages={1101--1144},
  year={2010},
  publisher={MIT Press}
}

@article{athey2024moocs,
  title={The value of non-traditional credentials in the labor market},
  author={Athey, Susan and Palikot, Emil},
  journal={arXiv preprint arXiv:2405.00247},
  year={2024}
}

@article{abebe2021anonymity,
  title={Anonymity or distance? Job search and labour market exclusion in a growing African city},
  author={Abebe, Girum and Caria, A Stefano and Fafchamps, Marcel and Falco, Paolo and Franklin, Simon and Quinn, Simon},
  journal={The Review of Economic Studies},
  volume={88},
  number={3},
  pages={1279--1310},
  year={2021},
  publisher={Oxford University Press}
}

@article{bassi2022screening,
  title={Screening and signalling non-cognitive skills: experimental evidence from Uganda},
  author={Bassi, Vittorio and Nansamba, Aisha},
  journal={The Economic Journal},
  volume={132},
  number={642},
  pages={471--511},
  year={2022},
  publisher={Oxford University Press}
}

@article{carranza2020job,
  title={Job search and hiring with limited information about workseekers’ skills},
  author={Carranza, Eliana and Garlick, Robert and Orkin, Kate and Rankin, Neil},
  journal={American Economic Review},
  volume={112},
  number={11},
  pages={3547--3583},
  year={2022},
  publisher={American Economic Association 2014 Broadway, Suite 305, Nashville, TN 37203}
}

@article{batista2022closing,
  title={Closing the gender profit gap?},
  author={Batista, Catia and Sequeira, Sandra and Vicente, Pedro C},
  journal={Management Science},
  volume={68},
  number={12},
  pages={8553--8567},
  year={2022},
  publisher={INFORMS}
}

@article{athey2000mentoring,
  title={Mentoring and diversity},
  author={Athey, Susan and Avery, Christopher and Zemsky, Peter},
  journal={American Economic Review},
  volume={90},
  number={4},
  pages={765--786},
  year={2000},
  publisher={American Economic Association}
}

@misc{fava2025,
      title={Training and Testing with Multiple Splits: A Central Limit Theorem for Split-Sample Estimators}, 
      author={Bruno Fava},
      year={2025},
      eprint={2511.04957},
      archivePrefix={arXiv},
      primaryClass={econ.EM},
      url={https://arxiv.org/abs/2511.04957}, 
}

@techreport{chernozhukov2018generic,
  title={Generic machine learning inference on heterogeneous treatment effects in randomized experiments, with an application to immunization in India},
  author={Chernozhukov, Victor and Demirer, Mert and Duflo, Esther and Fernandez-Val, Ivan},
  year={2018},
  institution={National Bureau of Economic Research}
}
\newpage{}
\setcounter{page}{1}
\gdef\thepage{A\arabic{page}}
\appendix
\section*{Appendix}

\section{Challenges program}\label{challenges_appendix}

The \emph{Challenges} program is designed for the same population as the \emph{Mentoring} program: individuals who possess technical competencies but have not yet secured employment in the technology sector. The primary aim of the \emph{Challenges} program is to assist these individuals in developing credible credentials that can effectively signal their skill set. In its first edition, the \emph{Challenges} program provided two distinct tracks: User Experience Design (UX) and Front-end Development in React (front-end). The selection of these specific tracks was influenced by several determinants. Primarily, there was a significant volume of applications for the \emph{Mentoring} program related to these two domains. Furthermore, Dare IT possessed both internal expertise and established partnerships in these areas. Lastly, a pronounced market demand existed for professionals in these specializations in Poland.

To achieve the program's scalability goal and streamline the application process, unlike the \emph{Mentoring} program, the review of the minimum technical skills during application was based solely on applicants' declaration. The registration website explicitly mentioned the necessity for an advanced understanding of Figma for those interested in UX design, and React for those leaning towards front-end engineering as prerequisites for the \emph{Challenges} program. The applicants had to confirm that they indeed have the minimum skill level. Additionally, English proficiency was mandated due to the inclusion of course content delivered in English.

The \emph{Challenges} program spanned 16 weeks. Initially, the first two weeks focused on an introductory event and a refresher course in Figma and React. The concluding two weeks were allocated for the refinement of the final projects, HR workshops, and a closing event. Central to the program, the intermediate 12 weeks were primarily aimed at the progression of an individual project. Within both specializations, experts from partner organizations, specifically Flying Bisons and Grover -- Polish tech companies, were asked to prepare six tasks. These tasks, inspired by their recent undertakings, encapsulated the comprehensive procedure of designing a mobile application for the UX domain and detailed the facets of front-end engineering for the respective specialization.

The program is based on two-week cycles culminating on Thursdays when submissions are expected and news assignments are posted. Every subsequent Friday, post-submission, an evaluative session with a subject-matter specialist takes place. During this session, a selection of submissions undergoes review, identifying prevalent errors and underscoring industry best practices. This grants participants the opportunity to amend and refine their submissions from the previous week. The domain experts, either mentors from Dare IT or other professionals with an appropriate level of seniority, carry out the reviews. Each assignment has been designed to take approximately eight hours of time to finalize. As the program approaches its conclusion, participants engage with a Human Resources specialist. This guidance is primarily aimed at instructing participants on leveraging their newly-acquired qualifications, both as skill indicators in job applications and as a resource during subsequent interviews.

In the first week of the UX path, participants conceptualize their project. They are given the flexibility to either devise their own application idea or choose from two provided project themes, followed by a detailed analysis of market competitors and benchmarks. The subsequent phases guide participants through a structured journey from ideation to delivery. They familiarize themselves with tools like Personas, User Journey Maps, and User Stories, ensuring a systematic organization of the information gathered. As the program advances, participants delve deeper into content hierarchy, Information Architecture, and user pathways, culminating in the creation of UX wireframes. The latter weeks emphasize usability testing, where participants craft interactive prototypes, refining them based on feedback. Crucially, they also submit project presentations and receive feedback on them. 

In the case of the front-end path, in the first weeks, participants created reusable UI components, often termed Atoms. This initial phase emphasized the foundational importance of efficient design. In the subsequent weeks, attendees delved into topics of server communication: creating APIs, handling different requests, and DELETE queries. Following these modules, the middle segment of the program trained participants in creating models and form-based data collection. Later, the curriculum shifted towards data visualization using charts and the integration of real-time user feedback systems, e.g., notifications. These modules were designed to underscore the advantages of adept data presentation and heightened user experience. Finally, the program introduced participants to TypeScript, showcasing its emerging prevalence in modern web applications.

In addition to the assignments and review sessions, participants were granted access to the exclusive Dare IT community on the Slack platform. During the kick-off meeting, the participants were recommended to collaborate in teams to tackle the assignments. A dedicated process was created to support collaboration; participants were asked whether they wanted to be matched with a group. If they were interested, they received access to a group's channel.\footnote{The emphasis on collaborative efforts and peer support was further amplified in the ensuing editions of the \emph{Challenges} program, where matching events were organized.} Within each path's Slack channel, distinct discussion forums were established, allowing participants to pose queries, circulate resources, and engage in discourse about assignment components. Activity within this Slack community was consistently high, with frequent discussions across multiple channels.

The first edition of \emph{Challenges} has attracted a lot of popularity amongst the participants, applicants that we could not accommodate due to capacity limits, and the Polish tech community more broadly, thereby galvanizing its further development. Currently, the program is instituted on a trimestral cycle, extending its offerings to encompass Manual Testing, Automated Testing in Python, Cloud Computing, and User Interface Design. Furthermore, in the summer of 2022, the program was adapted to cater to Ukrainian refugees in Poland. This edition was delivered in Ukrainian, incorporating bespoke modules tailored to guide foreigners in navigating the Polish job market.

\section{Covariate balance}\label{balance_appendix}
\begin{figure}[htp]
\caption{Covariate Balance Mentoring.}\label{fig:balance_1}
\centering
\includegraphics[width=.8\textwidth]{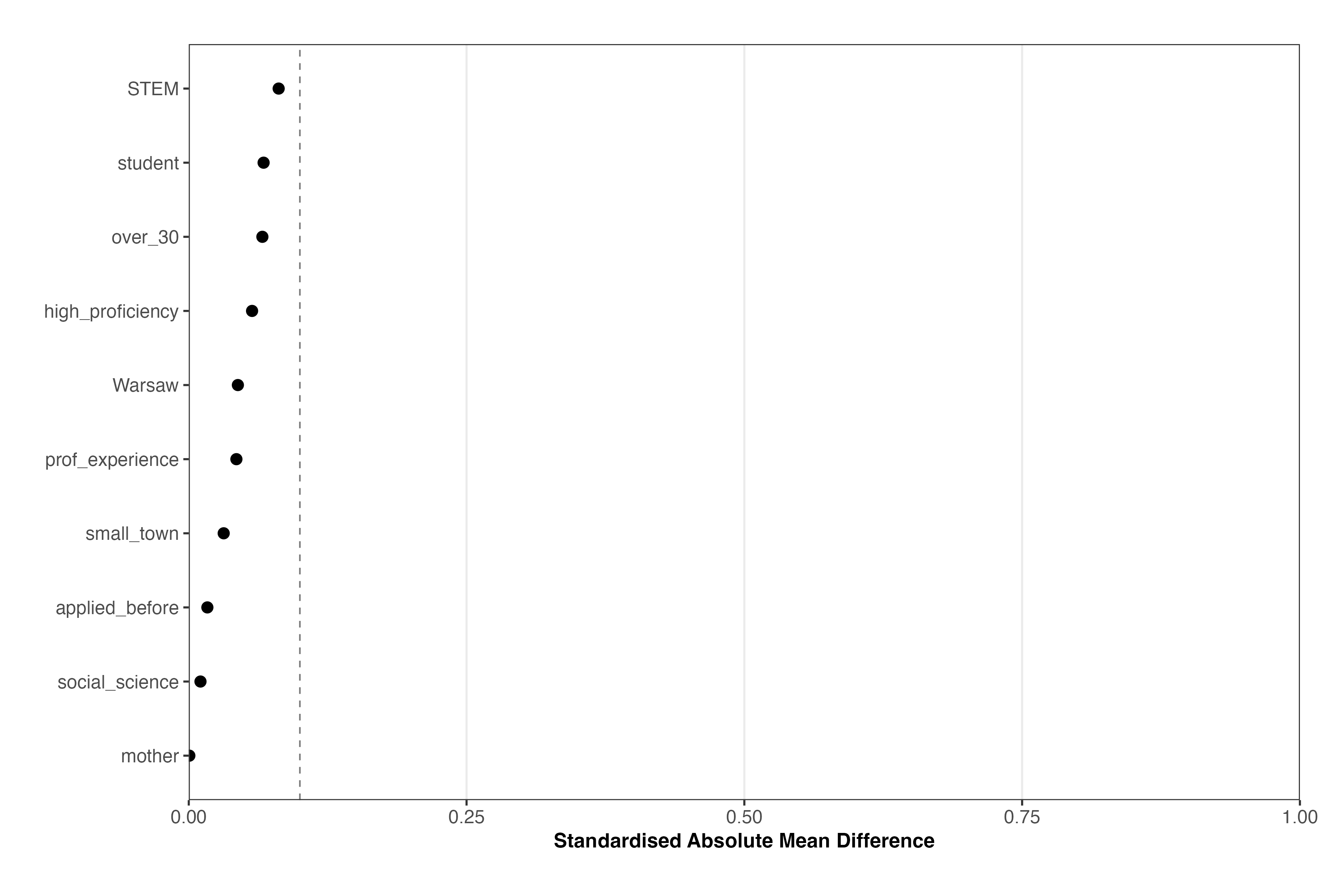}
\caption*{\footnotesize{\textit{Note: Balance across treatment and control of the main variables. Points correspond to the absolute difference in means between treatment and control divided by the square root of the sum of variances in treatment and control.}}}
\end{figure}

In Figure \ref{fig:balance_1} we show the balance of covariates in the mentoring experiment: each dot corresponds to the absolute difference of means in treatment and control divided by the square root of the sum of variances.

\begin{figure}[htp]
\caption{Covariate Balance Challenges.}\label{fig:balance_2}
\centering
\includegraphics[width=.8\textwidth]{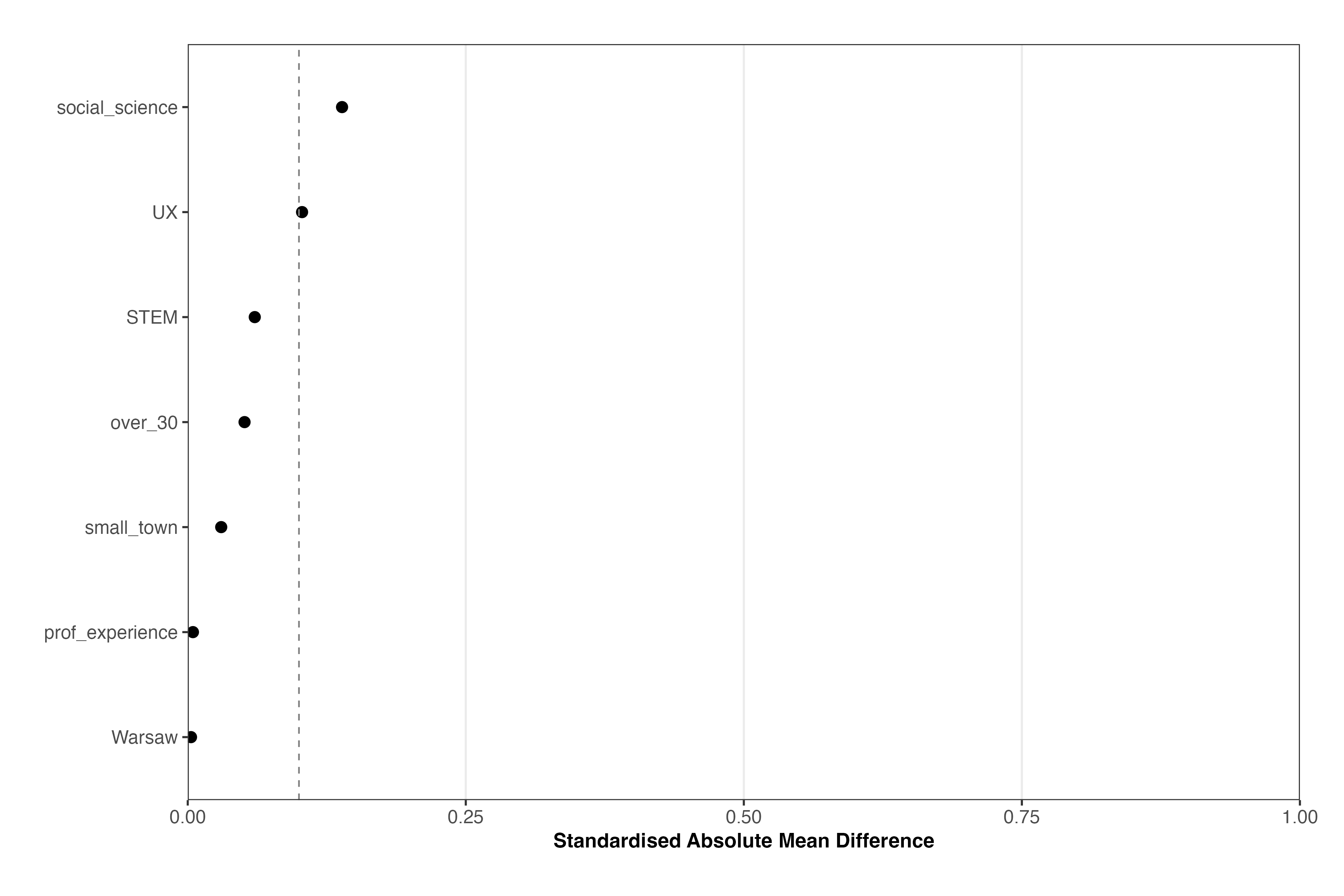}
\caption*{\footnotesize{\textit{Note: Balance across treatment and control of the main variables. Points correspond to the absolute difference in means between treatment and control divided by the square root of the sum of variances in treatment and control.}}}
\end{figure}

In Figure \ref{fig:balance_2} we show the balance of covariates in the Challenges experiment: each dot corresponds to the absolute difference of means in treatment and control divided by the square root of the sum of variances. The only variable with a standardized absolute mean difference above the 0.1 threshold is  is \emph{social science}.

\section{Monthly Outcomes Plots}\label{KM_curves}

\begin{figure}
  \centering
  \caption{Share of Applicants with a Tech Job per Group Over Time.}\label{fig:outcomes_over_time_corrected}
  \begin{minipage}[b]{0.49\textwidth}
    \includegraphics[width=\textwidth]{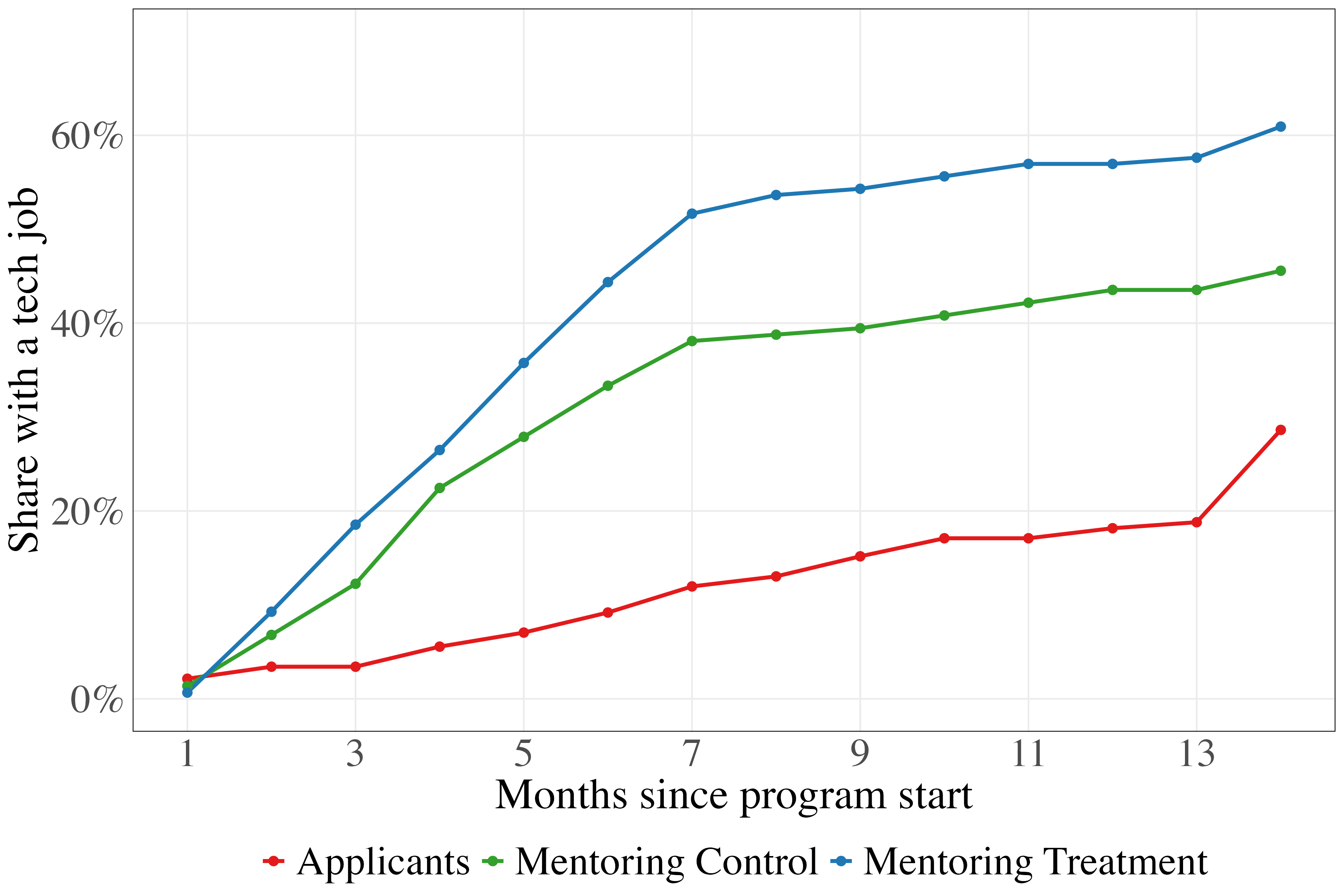}
  \end{minipage}
  \hfill
  \begin{minipage}[b]{0.49\textwidth}
    \includegraphics[width=\textwidth]{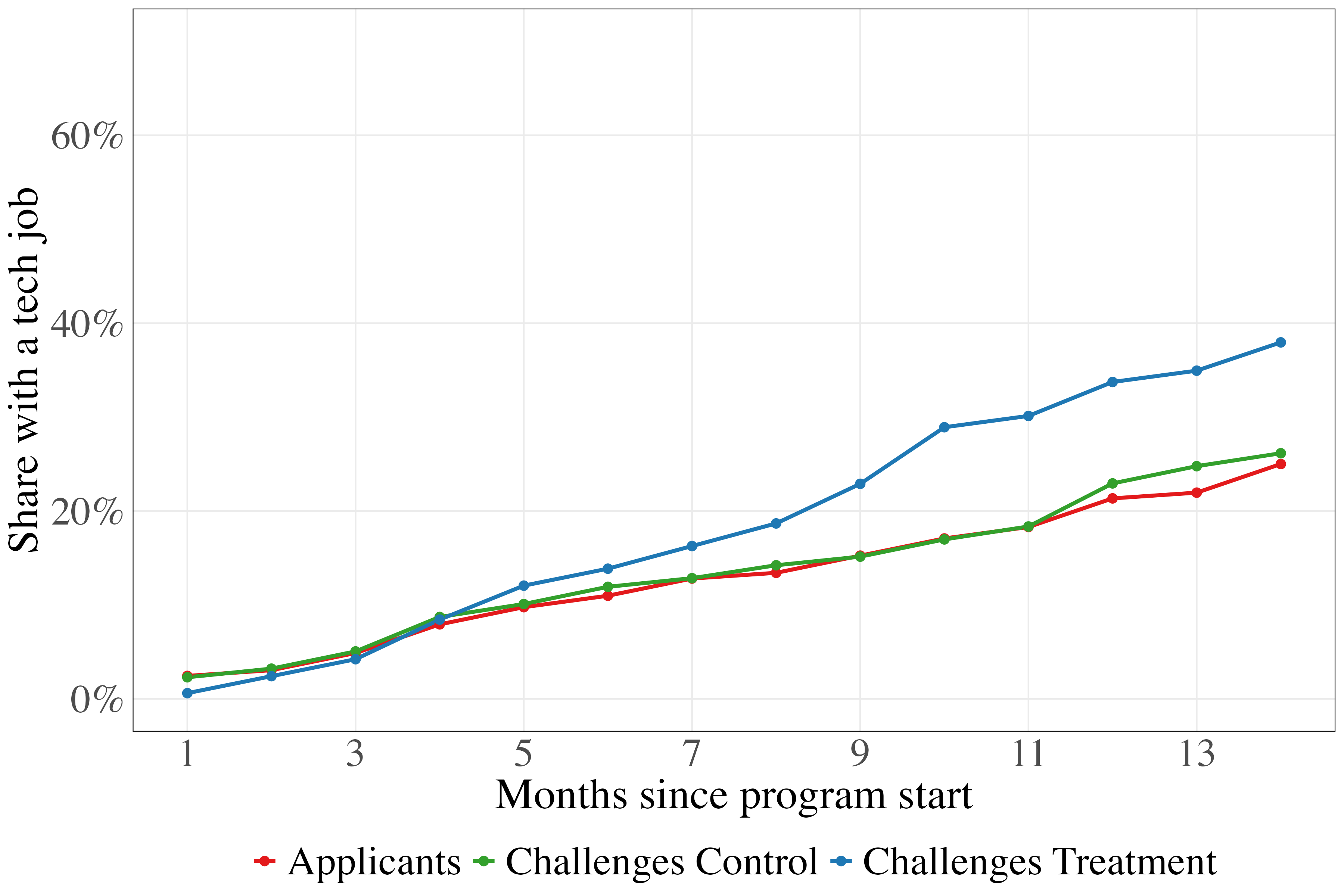}
  \end{minipage}
  \caption*{\footnotesize{\textit{Note: Share of subjects with a tech job per group across time. The month zero is the month of application to Dare IT. In the left figure, we present outcomes of three Mentoring groups: the treatment group (in blue), the control group (in green), and the applicants who were not selected by mentors (in red). In the right figure, we group focus on the Challenges program groups. The average outcome in the treated group (in blue), the control group (in green) and the late applicants to the program who were not included in the randomization and the main analysis (in red).
}}}
\end{figure}

\paragraph{Monthly Employment Outcomes.} Figure \ref{fig:outcomes_over_time_corrected} presents proportions of subjects across programs and treatment groups who reported a job in the technology sector with the inclusion of subjects who reported a job without a month in which they started. When a subject reported a job without a month, but only with the year in which they started we assign the last month in the data.

We notice that there is an increase in the shares that are employed in the last month and this is particularly pronounced for non-selected \emph{Mentoring} applicants.

\paragraph{Kaplan Meier Survival Curves.} We present Kaplan Meier estimates of the survival function $S(t)$, 
defined as the probability of not yet obtaining a technology job by month $t$.

\begin{figure}[H]
    \centering
    \caption{Survival Curves Mentoring Program.}
    \includegraphics[width=0.8\textwidth]{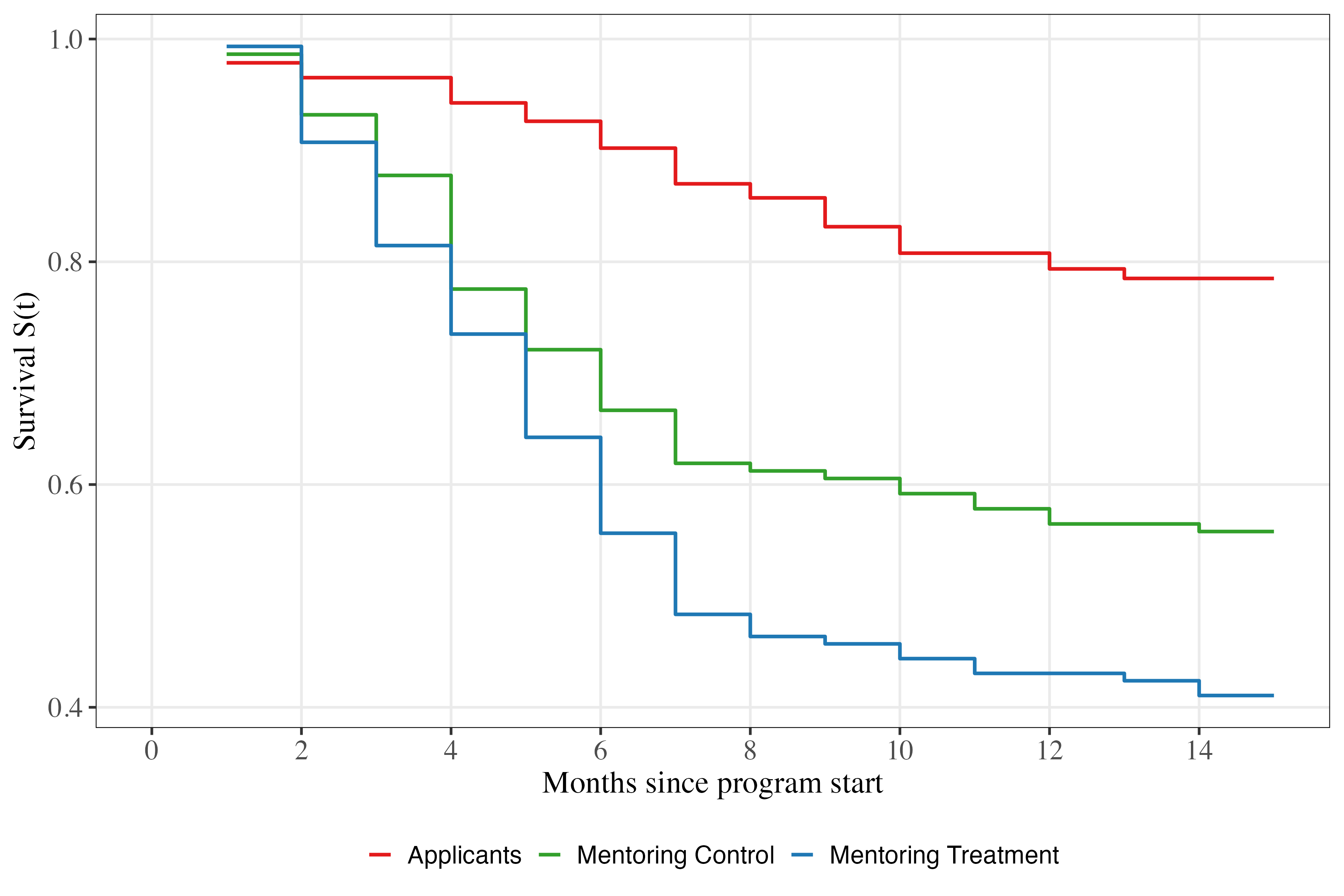}
\caption*{\footnotesize{\textit{Note: Kaplan Meier survival curves for the Mentoring program, including treated participants, control group, and applicants. A steeper decline indicates earlier job placement.}}}
    \label{fig:km_mentoring}
\end{figure}

Figure~\ref{fig:km_mentoring} shows the survival functions for the Mentoring program. 
Relative to both the control group and applicants, participants in the Mentoring treatment 
group find technology jobs more rapidly: their survival curve falls more steeply, indicating 
earlier entry into technology jobs. The difference between treated and control groups 
emerges within the first few months and grows steadily over the 15-month horizon. We can also notice that there is a substantial gap between the applicants and both experimental groups which emerges already during the first four months,

\begin{figure}[H]
    \centering
        \caption{Survival Curves Challenges Program.}
    \includegraphics[width=0.8\textwidth]{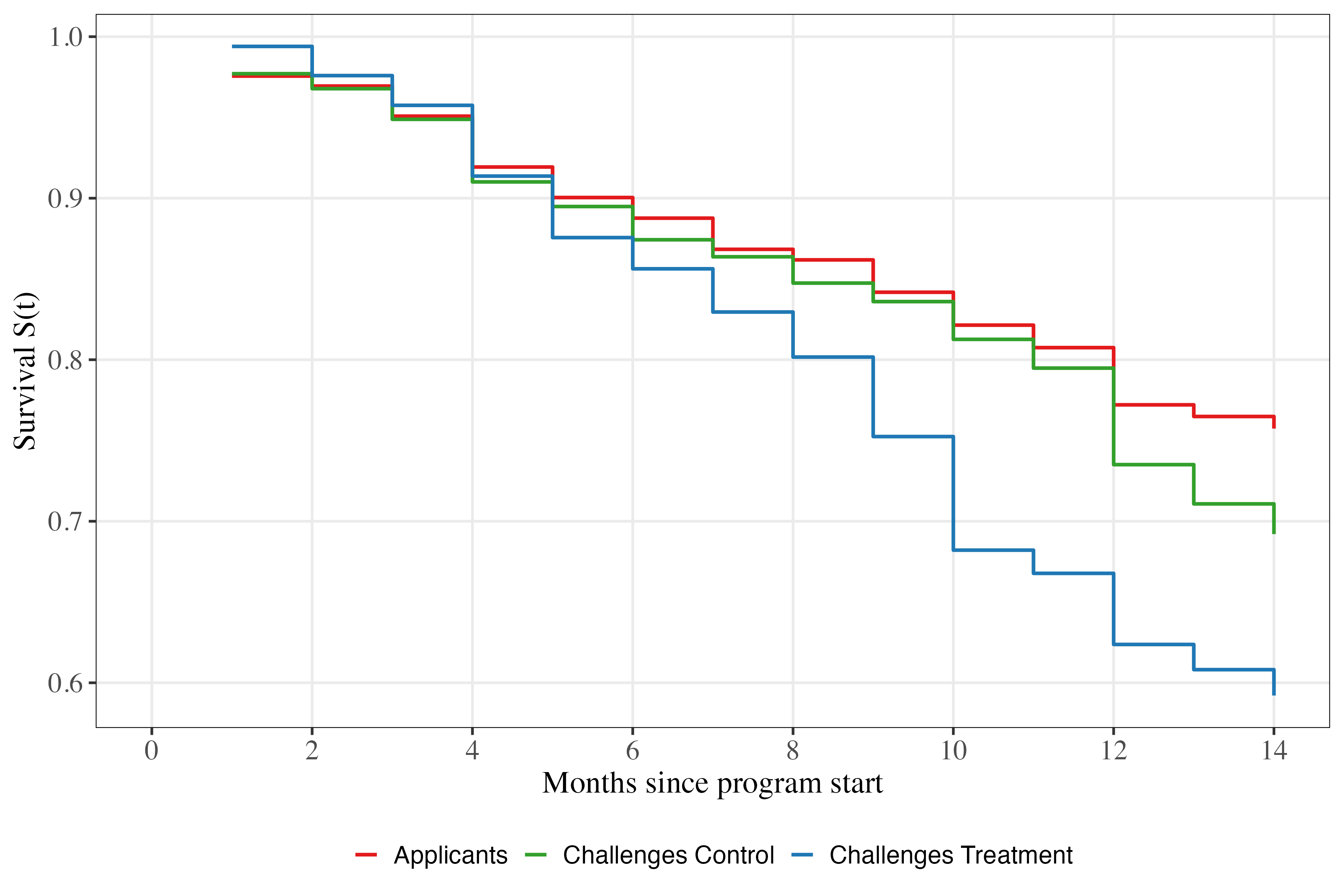}
\caption*{\footnotesize{\textit{Note: Kaplan Meier  survival curves for the Challenges program, including treated participants, control group, and applicants. Treated participants transition to jobs earlier, though the effect is smaller than in Mentoring.}}}
    \label{fig:km_challenges}
\end{figure}

Figure~\ref{fig:km_challenges} presents analogous estimates for the Challenges program. Treated participants also transition into jobs more quickly than the control group 
or applicants, though the gap is smaller in magnitude than in the Mentoring program.

\section{Comparison of salary levels from Glassdoor}\label{glassdoor}
\begin{figure}[htp]
\caption{Histograms of salary estimates from Glassdoor.}\label{hist_glass}
\centering
\includegraphics[width=.75\textwidth]{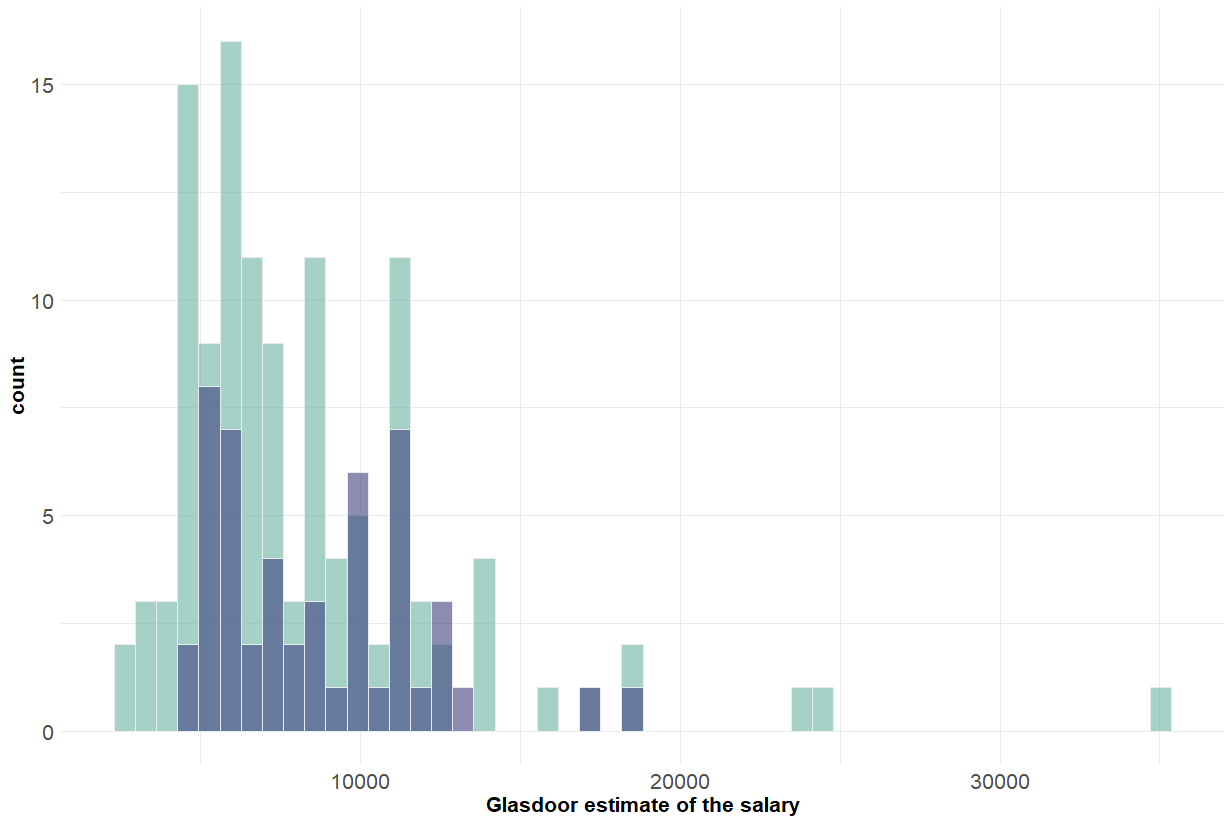}
\caption*{\footnotesize{\textit{Note: Histograms of salary estimates from Glassdoor. In green, we show tech jobs in tech firms (category 1), in blue tech jobs in non-tech firms (category 2). }}}
\end{figure}

We consider two types of jobs as a job in technology:
\begin{enumerate}
    \item All jobs in tech companies other than finance, regulatory, legal, accounting, and HR, where tech companies include firms in software development, testing, and sales; data analytics; IT services; digital marketing; and online platforms (including peer-to-peer platforms, and online sales).
    \item Jobs in non-tech companies that involve software development and testing, IT support, and data analytics. In our context, this category includes jobs in banks and management consulting agencies. 
\end{enumerate}
We analyze whether the two categories of jobs pay similar salaries. To do that we searched for salary estimates of all unique job titles that fall into the two categories using Glassdoor.com. We considered only exact matches and salary estimates for jobs in Poland. In total, we have 265 unique salary estimates. We find that the mean monthly salary for tech jobs in tech firms (category 1) is  PLN 8634 and for tech jobs in non-tech firms it is PLN 8866. The difference is statistically insignificant (p-value 0.9). Figure \ref{hist_glass} shows histograms of of salary estimates.

\section{Outcomes measured in surveys}\label{survey}

We use data from outcomes surveys to provide additional evidence of the effectiveness of the two programs that give insights into the mechanisms through which participation boosts outcomes. There were three surveys: one main survey for each program, both administered about four months after completion, and a shorter survey right after the end of the \emph{Mentoring} from which we obtained information on salaries in jobs found during or right after the mentoring program. As discussed in Section \ref{design_data}, results based on the outcomes survey suffer from a selection problem. Thus, we interpret the findings based on this data as suggestive evidence.

\paragraph{The number of job offers.}

\begin{figure}[htp]
\caption{Reported numbers of job offers: Mentoring.}\label{fig:no_offers_ment}
\centering
\includegraphics[width=.85\textwidth]{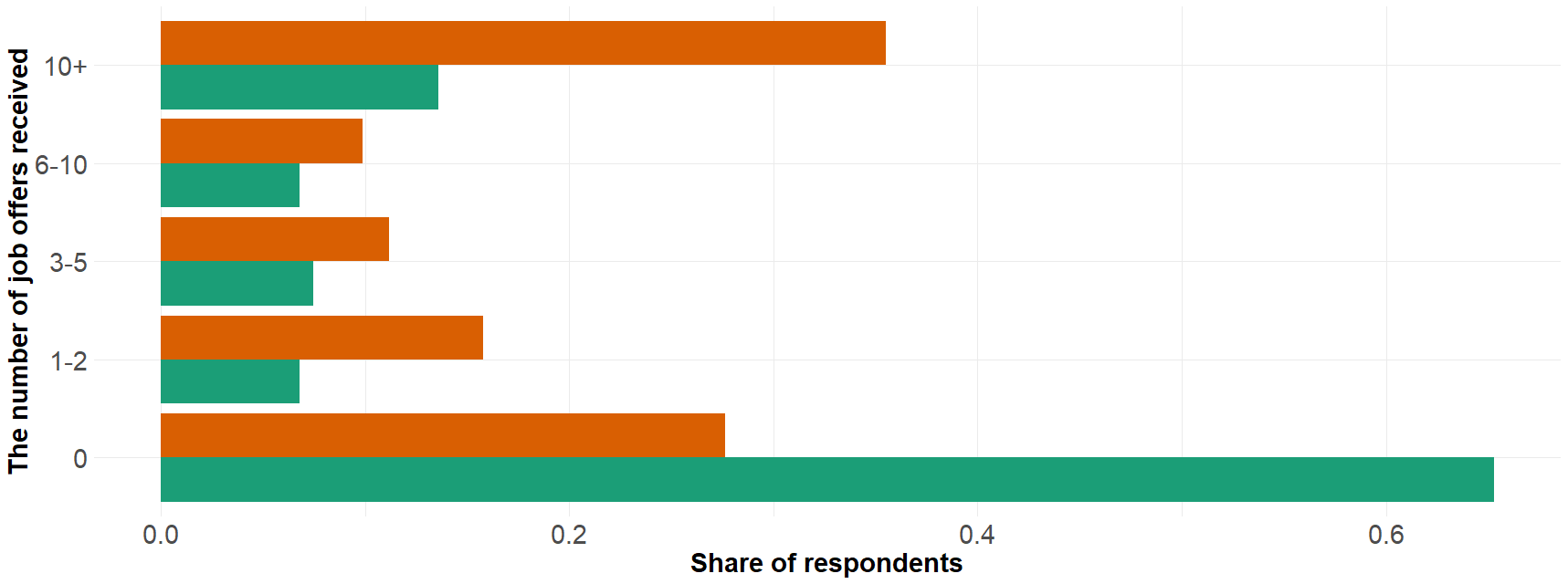}
\caption*{\footnotesize{\textit{Note: The number of job offers received by subjects in the mentoring experiment that responded to the survey.  Each bar corresponds to the share of respondents from an experimental group. Orange represents the treated group, and green represents the control group.}}}
\end{figure}

Figure \ref{fig:no_offers_ment} shows the number of job offers received by survey respondents broken by experimental groups. We find that subjects in the control group (in green) were more likely to report receiving zero job offers than the subjects in the treated group (in orange); for any non-zero range, there is a higher share of the treatment group rather than the control group. The largest difference between the experimental groups is amongst subjects reporting more than ten job offers.

\begin{figure}[htp]
\caption{Reported numbers of job offers: Challenges.}\label{fig:no_offers_challenges}
\centering
\includegraphics[width=.85\textwidth]{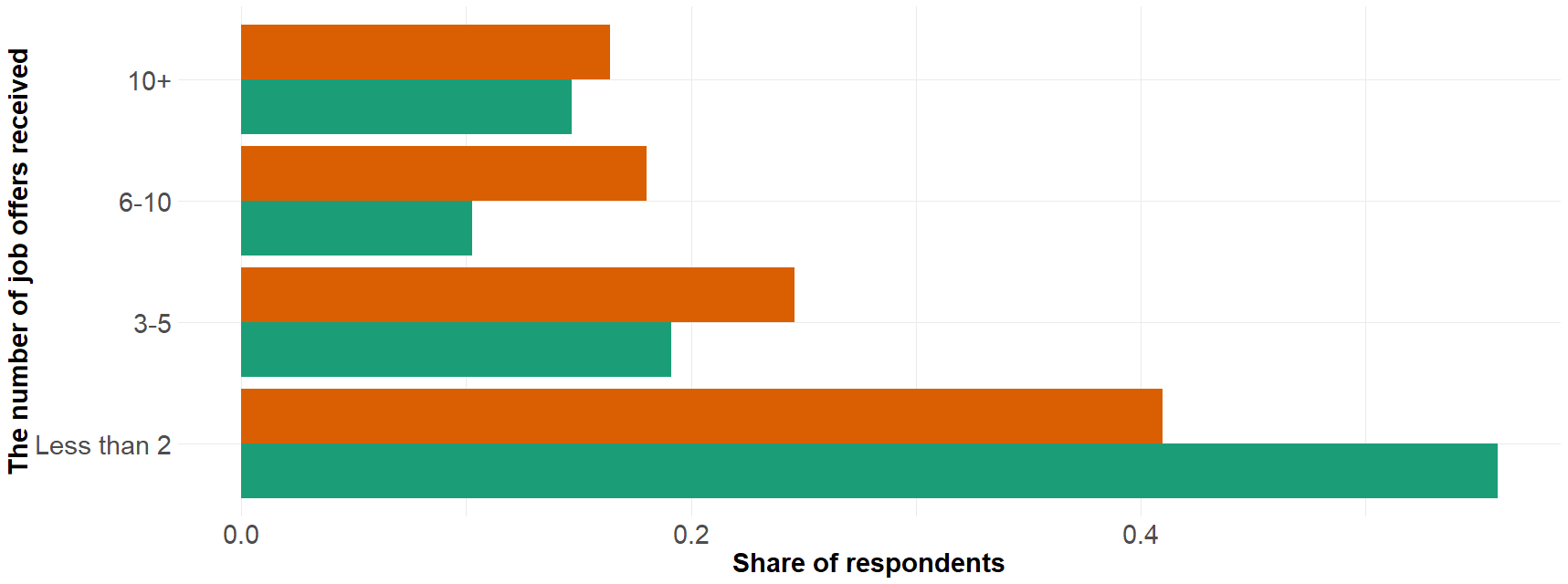}
\caption*{\footnotesize{\textit{Note: The number of job offers received by subjects in the Challenges experiment that responded to the survey.  Each bar corresponds to the share of respondents from an experimental group. Orange represents the treated group, and green represents the control group.}}}
\end{figure}

In Figure \ref{fig:no_offers_challenges}, we present evidence of the differences in the number of job offers across treatment and control groups of the Challenges experiment. We find results analogous to those from the mentoring program. The lowest category, here less than two job offers, has a higher share of control group respondents than treated group respondents. In contrast, treated subjects are more likely to report a higher number of job offers. The difference between treatment and control is more pronounced in the mentoring program.

\paragraph{Salary ranges.}

\begin{figure}[htp]
\caption{Salary ranges: Mentoring.}\label{fig:salary_ranges}
\centering
\includegraphics[width=.5\textwidth]{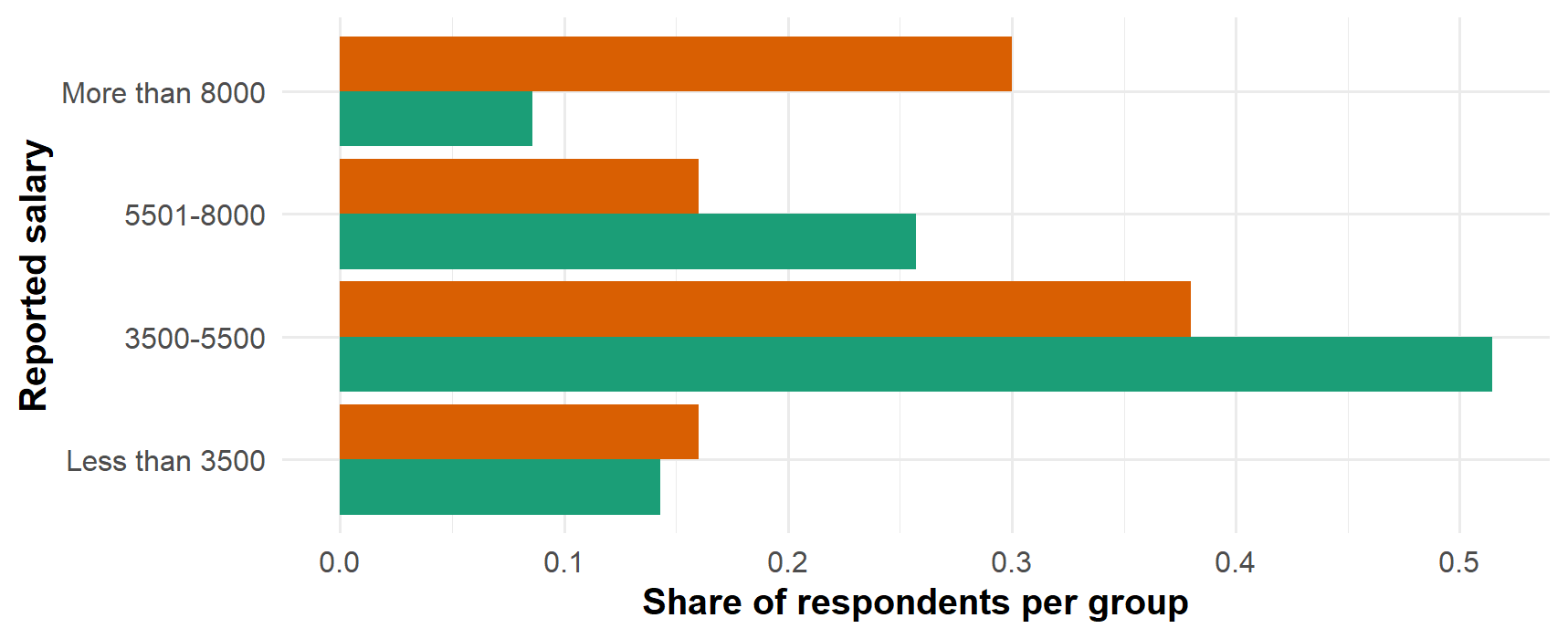}
\includegraphics[width=.5\textwidth]{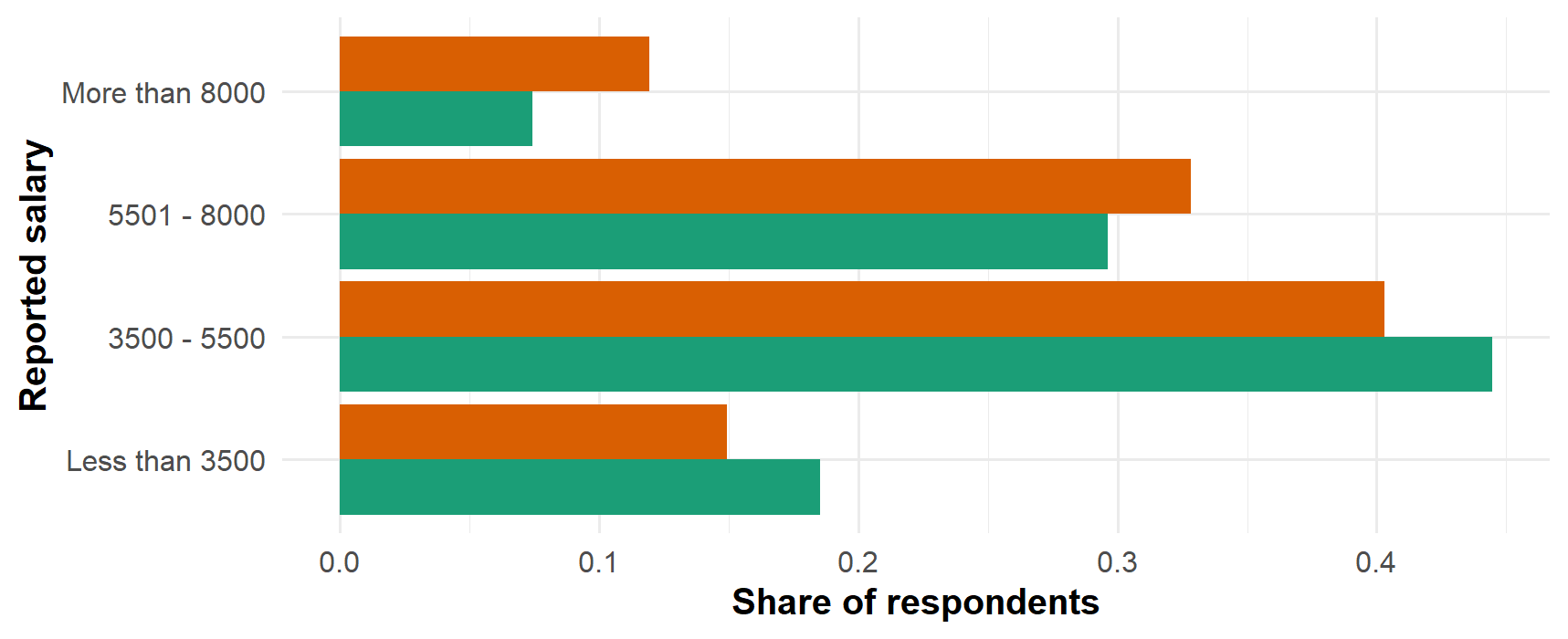}\hfill
\caption*{\footnotesize{\textit{Note: Reported salary ranges by subjects who found new jobs after the start of the mentoring program. Orange represents the treated group, and green represents the control group. The length of a bar represent the share of respondents declaring a salary within the range. The left panel survey was conducted right after the end of the program; the right panel survey was conducted four months after the program.}}}
\end{figure}

Figure \ref{fig:salary_ranges} presents the share of respondents to mentoring outcomes surveys declaring a salary within a salary range (ranges were pre-defined in the survey). Orange bars indicate the treatment group, and the green ones are the control group. The left panel is based on the survey carried out right after the end of the program, while the right panel is from the survey conducted four months after the end of the program.

Evidence presented in Figure \ref{fig:salary_ranges} suggests that subjects in the treated group of the mentoring experiment that found a new job are more likely to have a salary in a high salary range. 

\begin{figure}[htp]
\caption{Reported salary range Challenges.}\label{fig:salary_chall}
\centering
\includegraphics[width=.75\textwidth]{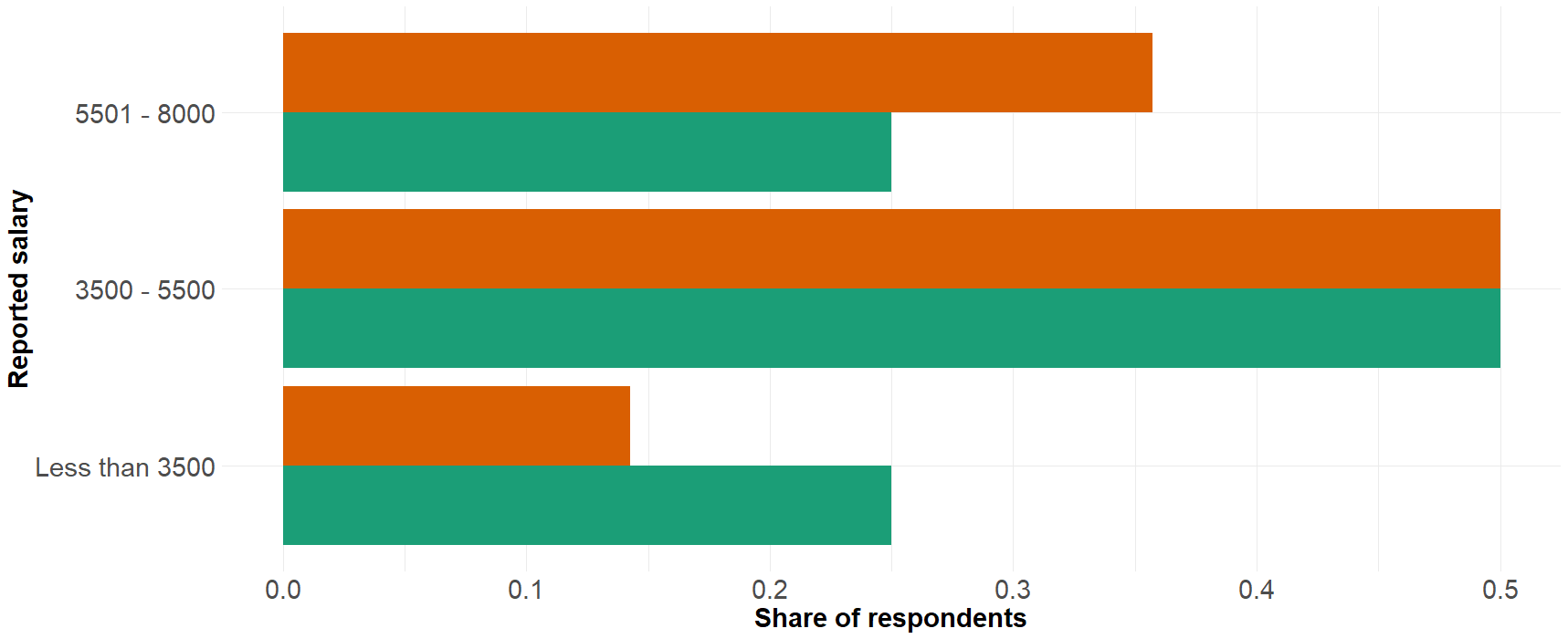}
\caption*{\footnotesize{\textit{Note:Reported salary ranges by subjects who found new jobs after the start of the Challenges experiment. The length of a bar represent the share of respondents declaring a salary within the range. Orange represents the treated group, and green represents the control group. }}}
\end{figure}

In Figure \ref{fig:salary_chall}, we report analogous results from the Challenges outcome survey. We find that a higher share of the survey respondents in the control reported salary in the lowest range as compared to the treatment group. We also find that the opposite is true for the highest salary range.

\paragraph{Negotiation.} Respondents that reported finding a job were asked whether they negotiated the conditions of their employment. Amongst the mentoring group, 33\% of treated subjects that found a new job negotiated the terms of the new job, while only 24\% of the control group did so. The difference is statistically insignificant. 

\section{Proportional hazards assumption}\label{appendix_prop_hazard}

The Cox proportional hazards model assumes the hazard rate of an event is a product of a baseline hazard and an exponential function of the covariates. Thus, in the analysis, we assume the effects of different predictors are constant over time. To test that, we correlate Schoenfeld residuals from the baseline \emph{Mentoring} and \emph{Challenges} models with months. Table \ref{prop_hazard} presents the results. We do not find any statistically significant correlations.

\begin{table}[!htbp] \centering 
  \caption{Proportional hazards assumption test} 
  \label{prop_hazard} 
  \resizebox{0.65\textwidth}{!}{%
\begin{tabular}{lcccccc} 
\hline 
\hline 
 & \multicolumn{3}{c}{Mentoring model} & \multicolumn{3}{c}{Challenges model} \\ 
\cline{2-4} \cline{5-7}
 & chisq1 & df1 & p1 & chisq2 & df2 & p2 \\ 
\hline
Treatment & 0.232 & 1 & 0.630 & 1.301 & 1 & 0.254 \\
City & 0.149 & 2 & 0.928 & 1.863 & 2 & 0.394 \\
Degree & 1.909 & 4 & 0.752 & 5.832 & 4 & 0.212 \\
STEM & 0.001 & 1 & 0.971 & 0.002 & 1 & 0.968 \\
Prof experience & 0.187 & 1 & 0.665 & 1.723 & 1 & 0.189 \\
Social science & 0.074 & 1 & 0.785 & - & - & - \\
Family friends IT & 0.074 & 1 & 0.785 & - & - & - \\
Mother & 0.164 & 1 & 0.685 & - & - & - \\
Applied before & 0.087 & 1 & 0.768 & - & - & - \\
First time mentor & 0.956 & 1 & 0.328 & - & - & - \\
Managerial experience & 0.762 & 1 & 0.383 & - & - & - \\
Career changer & 0.043 & 1 & 0.836 & - & - & - \\
Years of experience Mentor & 0.428 & 1 & 0.513 & - & - & - \\
Age & 2.716 & 3 & 0.438 & - & - & - \\
GLOBAL & 9.595 & 20 & 0.975 & 13.130 & 9 & 0.157 \\
\hline 
\hline 
\end{tabular} 
}
\caption*{\footnotesize{\textit{Note: Diagnostics of the proportional hazard model. The first three columns show results for the Mentoring model, and the last three for Challenges. City is a discrete variable with levels: Small Town, Large City, and Warsaw; degree is a discrete variable indicating the highest level of education with levels: no education mentioned, non-traditional credential, college degree, masters, and postgraduate. Prof experience is the number of years between the start date of the first employment of program participants and 2023.  }}}

\end{table}

\section{Bounds on the Average Treatment Effect Estimates}\label{ate_bounds}

In this section, we perform robustness checks on the average treatment effect estimates. First, we consider several cases in which we change which subjects are included in the analysis. Second, we estimate AT bounds following \cite{lee2021bounding}. Third, we estimate the average treatment effects at 4 months after the programs.

\paragraph{Robustness to Sample Construction and Attrition.} Our main analysis excludes participants who secured technology employment between randomization and program launch, as well as mentors who declined to participate in the experimental design. While these exclusions yield conservative estimates, we verify that our conclusions are robust to alternative sample definitions. Table \ref{tab:ate_robust_manual} presents treatment effect estimates under seven alternative specifications. For each, we report sample sizes, mean outcomes by treatment status, and the estimated average treatment effect using difference-in-means without covariate adjustment. All estimates use the twelve-month tech employment outcome.

We consider four specifications for \emph{Mentoring}. First, our baseline estimate (15.3 percentage points, SE 5.7) uses 151 treatment and 147 control participants, reflecting the 4 mentor pairs where the mentees who were initially assigned to participate found jobs before the program start and were replaced by the control group. Second, restricting to the 147 matched pairs yields a slightly smaller but still substantial effect (11.9 percentage points, SE 6.3). Third, including the 4 participants who found jobs before program launch—coding their outcomes as 1—increases the estimated effect to 16.4 percentage points. Fourth, additionally including the 9 non-participating mentors under the assumption their hypothetical mentees would not have found jobs reduces the estimate to 13.0 percentage points. All four specifications yield treatment effects between 11.9 and 16.4 percentage points, each statistically significant at the 5\% level.

For \emph{Challenges}, we examine three specifications. Our baseline estimate (11.3 percentage points, SE 4.8) excludes 17 treatment participants who secured technology jobs before program launch. Including these individuals and coding their outcomes as 1 increases the effect to 17.1 percentage points—as expected, since we add successful job-finders to the treatment group. Finally, we address the 17 treatment and 7 control participants with missing twelve-month LinkedIn data. Under the conservative assumption that all missing treatment observations had outcomes of 0 and all missing control observations had outcomes of 1—the scenario most unfavorable to finding a treatment effect—we still estimate a positive effect of 10.5 percentage points (SE 4.6, p=0.02).

\begin{table}[!h]
\centering
\caption{\label{tab:ate_robust_manual}Robustness to Sample Construction and Attrition}
\fontsize{9}{11}\selectfont
\begin{tabular}[t]{lccccc}
\toprule
\toprule
\multicolumn{1}{l}{Specification} & \multicolumn{2}{c}{Sample Size} & \multicolumn{2}{c}{Mean Outcome} & \multicolumn{1}{c}{ATE} \\
\cmidrule(lr){2-3} \cmidrule(lr){4-5} \cmidrule(lr){6-6}
 & Treated & Control & Treated & Control & (SE)\\
\midrule
\addlinespace[0.2em]
\multicolumn{6}{l}{\textbf{Panel A: Mentoring}}\\
\addlinespace[0.1em]
\hspace{1em}Main specification & 151 & 147 & 0.609 & 0.456 & 0.153***\\
 & & & & & (0.057)\\
\hspace{1em}Matched pairs only & 147 & 147 & 0.603 & 0.484 & 0.119*\\
 & & & & & (0.063)\\
\hspace{1em}Include pre-program employment & 155 & 147 & 0.619 & 0.456 & 0.164***\\
 & & & & & (0.057)\\
\hspace{1em}Include non-participants & 164 & 147 & 0.585 & 0.456 & 0.130**\\
 & & & & & (0.056)\\
\addlinespace[0.3em]
\multicolumn{6}{l}{\textbf{Panel B: Challenges}}\\
\addlinespace[0.1em]
\hspace{1em}Main specification & 166 & 218 & 0.380 & 0.266 & 0.113**\\
 & & & & & (0.048)\\
\hspace{1em}Include pre-program employment & 183 & 218 & 0.437 & 0.266 & 0.171***\\
 & & & & & (0.047)\\
\hspace{1em}Pessimistic attrition bounds & 200 & 227 & 0.400 & 0.295 & 0.105**\\
 & & & & & (0.046)\\
\bottomrule
\bottomrule
\end{tabular}
\caption*{\footnotesize{\textit{Note:Estimates of average treatment effects on twelve-month technology employment under alternative sample definitions. Main specifications correspond to Table \ref{ate_main}. Balanced pairs only (Mentoring) restricts to the 126 mentor pairs with both treatment and control observations. Include pre-program employment adds participants who secured technology jobs between randomization and program launch. Include non-participants (Mentoring) adds 9 mentors who declined experimental participation, assigning their hypothetical mentees outcomes of 0. Pessimistic attrition bounds assign outcomes of 0 to 17 missing treatment observations and 1 to 7 missing control observations. All estimates use difference-in-means without covariate adjustment. Standard errors in parentheses. *** p<0.01, ** p<0.05, * p<0.1.}}}
\end{table}

These robustness checks establish three key points. First, our main estimates are insensitive to reasonable alternative sample construction decisions, with all specifications yielding economically large and statistically significant treatment effects. Second, the range of point estimates across specifications (11.9 to 16.4 for \emph{Mentoring}, 10.5 to 17.1 for \emph{Challenges}) provides additional bounds on the true treatment effects that are consistent with our main findings. Third, even under extremely conservative assumptions about missing data—assumptions, the \emph{Challenges} program shows significant positive impacts. This addresses potential concerns that selective attrition might bias our twelve-month estimates.

\paragraph{Lee Bounds for Selective Attrition.} Our twelve-month outcome measurement experienced modest attrition in the Challenges program: 6\% had missing LinkedIn data due to profile URL changes or privacy setting modifications. We assess whether selective attrition could bias our estimated treatment effect using the bounding procedure of \cite{lee2009training}. We consider the most conservative case, in which subjects that found jobs before the start of the program are not considered; thus, we focus only on the attrition during the program, due to missing LinkedIn data.

The Lee bounding procedure addresses selective attrition by considering worst-case scenarios about which individuals are missing. Since treatment experienced higher attrition, the conservative assumption for estimating positive effects is that the best-performing treatment participants were most likely to attrite (e.g., successful job-finders who deactivated LinkedIn). To construct the lower bound, we trim 6.8\% of observations from the upper tail of the observed treatment outcome distribution, the proportion representing excess attrition relative to control, and compare the mean of this trimmed distribution to the full control group mean. The upper bound uses all observed treatment outcomes compared to all observed control outcomes. Trimmed mean is calculated over the remaining observations.

We estimate Lee bounds of [0.065, 0.114]. The upper bound of 0.114 is nearly identical to our main point estimate of 0.113, while the positive lower bound of 0.065 indicates that even under worst-case assumptions about selective attrition, the Challenges program effect remains economically meaningful. The 95\% confidence interval on the lower bound is [-0.029, 0.159]; while this marginally includes zero due to sampling variability in the observed outcomes, the point estimate is bounded away from zero.

\paragraph{Average Treatment Effect Estimates at 4 Months.}

Our main analysis uses employment outcomes measured twelve months post-program. However, we were unable to collect LinkedIn data for 17 \emph{Challenges} treatment participants and 7 control participants at the twelve-month measurement due to profile URL changes or privacy setting modifications. A natural concern is whether selective attrition biases our estimated treatment effects. To address this concern, we re-estimate average treatment effects using employment outcomes measured four months after program completion. At this earlier measurement point, we successfully collected complete LinkedIn data for all randomized participants in both programs. Any attrition-induced bias is therefore eliminated at this time horizon. Furthermore, comparing the estimates at 4 months with the main results shows persistence of the estimates over time.

Table \ref{tab:ate_robust_4} presents the results. We find substantively similar treatment effects at the four-month horizon compared to our main twelve-month estimates (reported in Table \ref{ate_main}). For \emph{Mentoring}, the four-month treatment effect is 15.3 percentage points (SE 5.7), nearly identical to the twelve-month effect of 14.5 percentage points. For \emph{Challenges}, the four-month effect is 8.9 percentage points (SE 4.3), compared to 11.2 percentage points at twelve months. The larger point estimate at twelve months suggests that the \emph{Challenges} program's impact may strengthen over time as participants leverage their portfolios in job applications.

\begin{table}[!h]
\centering
\caption{\label{tab:ate_robust_4}Average Treatment Effects of \emph{Mentoring} and \emph{Challenges} at 4 Months.}
\resizebox{0.4\textwidth}{!}{%
\begin{tabular}{lcc}
\toprule\toprule
 & \multicolumn{1}{c}{Mentoring} & \multicolumn{1}{c}{Challenges} \\
\cmidrule(lr){2-2} \cmidrule(lr){3-3}
\addlinespace[0.3em]
\hspace{1em}ATE & 0.153 & 0.089\\
\hspace{1em}SE & (0.057) & (0.043)\\
\addlinespace[0.3em]
\midrule
\hspace{1em}ATE \% baseline & 33.7 & 45.3\\
\hspace{1em}SE \% baseline & (12.6) & (21.8)\\
\addlinespace[0.3em]
\midrule
\hspace{1em}N Treatment & 151 & 183\\
\hspace{1em}N Control & 147 & 225\\
\bottomrule\bottomrule
\end{tabular}
}
\caption*{\footnotesize{\textit{Note: Estimates of the average treatment effect of Mentoring and Challenges at 4 months after the end of each program. Estimates obtained using the difference-in-means estimator.}}}
\end{table}

Importantly, these four-month estimates use the full experimental sample without any missing observations, confirming that our main findings are not an artifact of selective attrition. The consistency between four-month and twelve-month estimates also provides evidence that treatment effects do not attenuate over time—a key finding given that some labor market training programs show fadeout of impacts \citet{card2018works}. If anything, the point estimates suggest persistent or growing effects, particularly for \emph{Challenges}, though the conclusions are limited to 12 months from the start of each program.

\section{Stratified Estimation for Challenges}\label{app:stratified}

The Challenges experiment employed stratified randomization on track specialization (UX Design vs.\ Front-end Development), age category (over/under 30), and residence type (Warsaw, other large city, or small town). To account for this design, we re-estimate treatment effects using stratum-weighted estimators that properly reflect the randomization procedure.

Table \ref{ate_stratified} reports these stratum-reweighted estimates. For difference-in-means, we estimate treatment effects within each stratum and aggregate using stratum-size weights. For Cox proportional hazard models, we allow separate baseline hazards. Point estimates are similar to those in Table \ref{ate_main}, with modestly smaller standard errors reflecting the efficiency gains from accounting for stratification.

\begin{table}[!htbp] \centering 
  \caption{Average Treatment Effects on Employment Outcomes (Stratum-Reweighted)} 
  \label{ate_stratified} 
  \resizebox{0.7\textwidth}{!}{%
    \begin{tabular}{lcccc}
      \toprule      \toprule
      & \multicolumn{4}{c}{\emph{Challenges (Stratum-Reweighted)}} \\
      \cmidrule(lr){2-5}
      & \multicolumn{3}{c}{Difference-in-means} & Cox PH \\
      \cmidrule(lr){2-4} \cmidrule(lr){5-5}
      & New job & Tech job & Non-tech job & Tech job \\
      \midrule
      ATE & 0.080 & 0.120** & $-$0.040 & 0.111** \\
      & (0.051) & (0.048) & (0.042) & (0.049) \\[0.5em]
      ATE / Control mean (\%) & 16.0 & 45.2** & $-$17.2 & 41.8** \\
      & (10.2) & (18.2) & (17.8) & (18.4) \\
      \midrule
      Treated & 166 & 166 & 166 & 2,656 \\
      Control & 218 & 218 & 218 & 3,488 \\
      \bottomrule      \bottomrule
    \end{tabular}%
  }
  \caption*{\footnotesize\textit{Notes:} Average treatment effects on employment outcomes using stratum-reweighted estimators. Columns 1--3 report stratum-weighted difference-in-means estimates; column 4 reports stratified Cox proportional hazard estimates. Standard errors in parentheses. *** $p<0.01$, ** $p<0.05$, * $p<0.1$.}
\end{table}

\section{Average Treatment Effect Estimates Using Survival Forest}\label{ate_surv}

This section estimates the average treatment effects of \emph{Mentoring} and \emph{Challenges}, on employment outcomes using causal survival forests. We use experimental samples only. The outcome of interest is the time until participants secured a new job, measured over a 13-month follow-up period. This approach estimates the difference in survival probabilities between treated and control groups at a specified time horizon. For each program, we fit a causal survival forest with 2,000 trees and compute the average treatment effect at 12 months post-intervention. Results are presented in Table \ref{tab:ate_robust_12}.

\begin{table}[!h]
\centering
\caption{\label{tab:ate_robust_12}Average Treatment Effects of \emph{Mentoring} and \emph{Challenges} at 12 Months.}
\resizebox{0.4\textwidth}{!}{%
\begin{tabular}{lcc}
\toprule\toprule
 & \multicolumn{1}{c}{Mentoring} & \multicolumn{1}{c}{Challenges} \\
\cmidrule(lr){2-2} \cmidrule(lr){3-3}
\addlinespace[0.3em]
\hspace{1em}ATE & -5.180 & -2.895\\
\hspace{1em}SE & (0.291) & (0.244)\\
\addlinespace[0.3em]
\midrule
\hspace{1em}N Treatment & 151 & 183\\
\hspace{1em}N Control & 147 & 225\\
\bottomrule\bottomrule
\end{tabular}
}
\caption*{\footnotesize{\textit{Note: Estimates of the average treatment effect of Mentoring and Challenges at 12 months. Estimates obtained using causal survival forests. Estimates based on experimental samples with no control variables.}}}
\end{table}

The results indicate that the \emph{Mentoring} program reduce the time it takes to find a job by 5.18 months,, while the Challenges program reduced this time by 2.9 months. Both estimates are statistically significant. 

\begin{figure}[htp]
\caption{Treatment Effect Estimates from Survival Forest.}\label{fig:ate_over_time_surv}
\centering
\includegraphics[width=.75\textwidth]{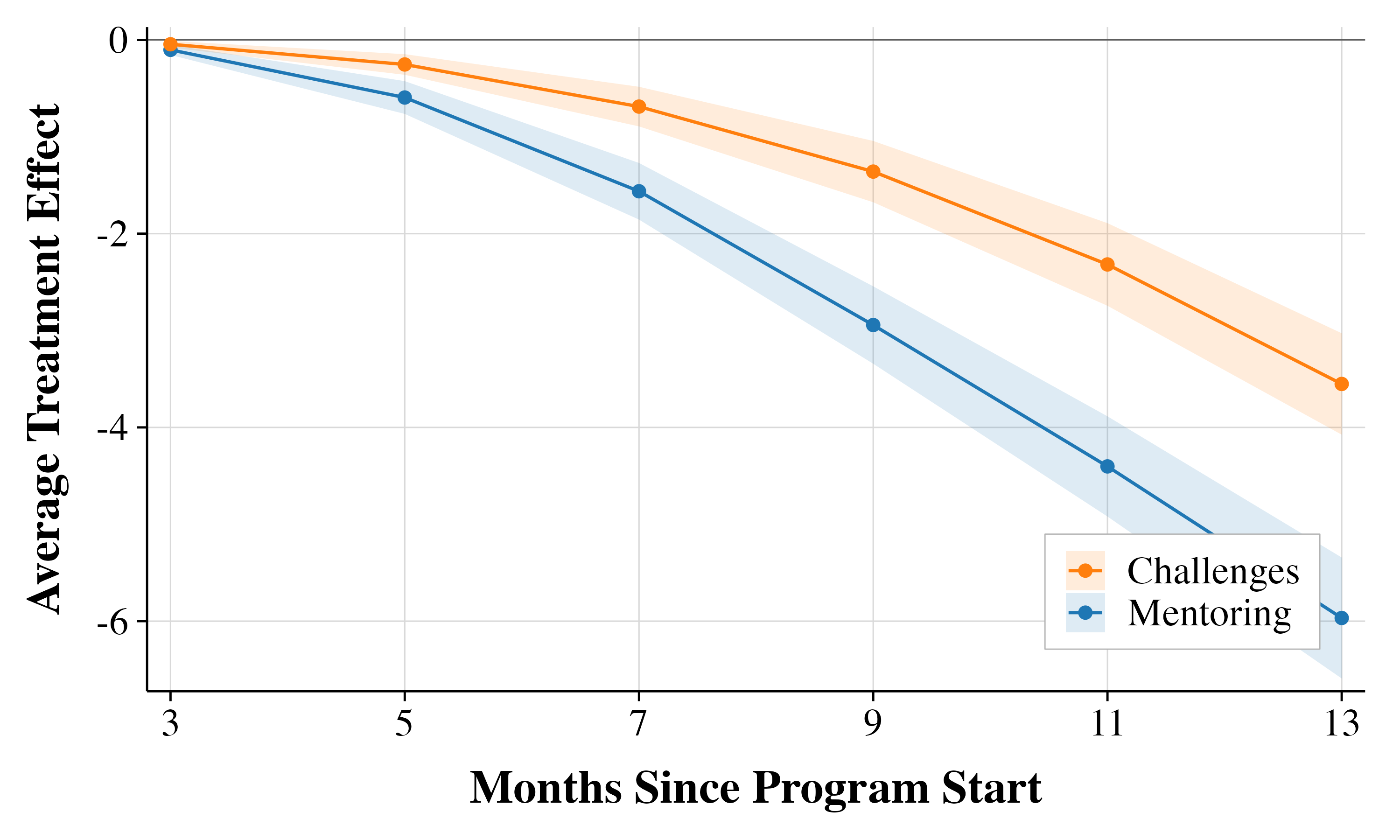}
\caption*{\footnotesize{\textit{Note: Note: Figure shows average treatment effects on employment probability at different time horizons following program completion. Treatment effects estimated using causal survival forests with 2,000 trees. Shaded regions represent 95\% confidence intervals. Horizontal line at zero indicates no treatment effect. }}}
\end{figure}

Figure \ref{fig:ate_over_time_surv} displays the average treatment effects of the \emph{Mentoring} and \emph{Challenges} programs on employment outcomes over time, measured from 3 to 13 months after program start. Both programs show negative treatment effects that grow larger in magnitude over time, indicating that participants in the treatment groups experience accelerated job finding compared to control groups. At 3 months, both programs show treatment effects near zero, suggesting minimal immediate impact. However, as time progresses, the effects become increasingly negative, meaning treated participants find jobs faster (have lower survival times). By month 13, the \emph{Mentoring} program shows a treatment effect of approximately -6 months, while the \emph{Challenges} program shows an effect of about -3.5 months.

\section{Policy estimation algorithms}\label{appendix_algorithms}

Algorithm \ref{alg:policy} describes the estimation and evaluation of \emph{optimal} policy.

\begin{algorithm}
\caption{Optimal policy}\label{alg:policy}
\begin{algorithmic}
\State 1. Considering the dataset of all applicants to both programs, randomly split the dataset into train and test with equal probability at the applicant level, denote the resulting sets as $I^{train} = \left\{1,..,\mathcal{I}^{train}\right\}$ and $I^{test}= \left\{1,..,\mathcal{I}^{test}\right\}$.
\State 2. In the training dataset, estimate outcome model $\mu_{train}$ and propensity model $e_{train}$.
\State 3. In the test set, construct predicted treatment effects using predictions from models $\mu_{train}$ and $e_{train}$. Obtain $\tau^{Op,train}_{i,t} = E[\hat{Y}(p = P)_{train} - \hat{Y}(p = Out of DareIT)_{train}|X=x_{i,t} , i \in I^{test}]$, where $\hat{Y}(p = P)_{train}$ is the predicted outcome under program $p \in $ (\emph{Mentoring}, \emph{Challenges}, \emph{Out of Dare IT}) and $\hat{Y}(p = Out of DareIT)_{train}$ is the predicted outcome under \emph{Out of Dare IT} and $t \in \left\{1,...,15\right\}$ denotes months. Compute mean treatment effects per user as: $\tau^{Op,train}_{i} = 1/T \times \sum_{t=1}^{15}\tau^{Op,train}_{i,t}$,

\State 4. Assign treatment to maximize treatment effects subject to capacity constraint. Let $Q^{p}$ be the capacity limit of program $p$ and $z_{ip}$ an indicator variable taking the value of one when applicant $i$ is assigned to program $p$ and zero otherwise. We solve the following constrained optimization problem:
\[
\max_{z_{ip}}\sum_{i\in I^{test}}\sum_{p=1}^{P}z_{ip}\tau_{i}^{Op,train} \text{ s.t. }\sum_{i\in I^{test}}z_{ip} \leq Q^{p} \forall_{p} \text{ \& }\sum_{p=1}^{P}z_{ip} = 1\forall_{i \in I^{test}}.
\]  The first constraint ensures that the capacity constraints are not violated. The second one is that every applicant is assigned to one program. There is no capacity limit on being \emph{Out of Dare IT}. We use \emph{LP Solve} algorithm to solve the problem. We obtain optimal allocation $\mathcal{A}_{X,Q}^{*} = \left\{a_{i}^{*},...,a_{\mathcal{I}^{test}}^{*}\right\}$,

\State 5. Using the test set, estimate new outcome and propensity models using cross-fitting and obtain predictions: $\hat{\mu}_{i,k}$ and $\hat{e}_{i,k}$ for all $i \in I^{test}$.  Obtain $\hat{Y}_{i,k}(a^{*})$ the AIPW estimates of the predicted outcomes using cross-fitted models trained in the test set,
    
\State 6. Obtain $\hat{V}_{X,Q}^{*} = \frac{1}{|I^{test}|}\times \sum_{i=1}^{\mathcal{I}^{test}}\hat{Y}_{i,k}(a^{*})$ as the mean of predicted outcomes under the allocation $\mathcal{A}_{X,Q}^{*}$. Estimate standard errors clustered at the applicant level: $\sigma_{X,Q}^{*}$. \Comment \\
\Return{} $(\mathcal{A}_{X,Q}^{*},\hat{V}_{X,Q}^{*},\sigma_{X,Q}^{*},\hat{Y}_{1,k}(a^{*}),...,\hat{Y}_{\mathcal{I}^{test},k}(a^{*}) )$
\end{algorithmic}
\end{algorithm}

\subsection{Assignment Models}

We pool data from \emph{Mentoring} experimental participants (N=298), \emph{Challenges} experimental participants (N=408), non-selected \emph{Mentoring} applicants (N=468), and late \emph{Challenges} applicants (N=160), yielding 1,334 total applicants. All individuals are observed over fifteen months post-application. We randomly split this dataset 50-50 at the applicant level into training (N=667) and testing (N=667) sets.

Using the training set, we estimate predicted treatment effects for each program relative to non-participation. We fit two causal survival forests \citep{athey2019generalized, cui2023estimating}: one comparing \emph{Mentoring} participants to applicants who did not participate in any program (including the control group of the \emph{Mentoring} experiment), and another comparing \emph{Challenges} participants to non-participants.

Each causal survival forest is estimated with the following parameters: 2,000 trees, honest splitting, sample fraction of 0.5 per tree, minimum node size of 5, and horizon of 15 months. Covariates include: STEM degree, residence type (Warsaw, small town), master's degree, age over 30, professional experience (continuous), and UX specialization. These settings follow standard recommendations in the generalized random forests (grf) package \citep{tibshirani2024grf}.

The causal survival forest estimates conditional average treatment effects on the restricted mean survival time (RMST) scale—the average time until technology employment over our observation horizon. More negative predicted effects indicate faster job finding under treatment. We apply the trained models to the test set to obtain predicted treatment effects 
$\hat{\tau}_i^M$ (\emph{Mentoring} vs. nothing) and $\hat{\tau}_i^C$ (\emph{Challenges} vs. nothing) for each test-set applicant.

\section{Evaluation Models}\label{eval_models}

While assignment models determine which applicants receive which program under various policies, evaluation models estimate the causal effects of these assignments. The credibility of our policy comparisons rests on these evaluation models accurately estimating treatment effects for the allocated applicants. If evaluation models correctly estimate treatment effects, our policy value estimates remain valid even if assignment models imperfectly rank applicants.

We use the held-out test set (N=667) for all policy evaluations. This sample includes experimental participants and non-participants from both programs, ensuring we can estimate counterfactual outcomes under different assignment policies. All individuals are observed over the same fifteen-month post-application horizon.

\paragraph{Model specification.}
For each policy evaluation, we estimate individual-level treatment effect predictions using causal survival forests \citep{athey2019generalized, cui2023estimating}. Within each cross-fitting fold (detailed below), we fit two causal survival forests: one comparing \emph{Mentoring} participants to applicants who did not participate in any program, and another comparing \emph{Challenges} participants to non-participants.

Each causal survival forest uses identical parameters to the assignment models: 2,000 trees, honest splitting, sample fraction of 0.5 per tree, minimum node size of 5, and horizon of 15 months. The covariate set is identical to that used in assignment models: STEM degree, residence type (Warsaw, small town), master's degree, age over 30, professional experience (continuous), and UX specialization. Model fitting follows the \texttt{grf} package implementation \citep{tibshirani2024grf}.

The causal survival forest produces augmented inverse propensity weighted (AIPW) scores $\psi_i^M$ and $\psi_i^C$ for each test-set individual, representing individual-level treatment effect estimates from \emph{Mentoring} and \emph{Challenges} relative to non-participation, respectively. The forest constructs these doubly-robust scores by combining: (1) propensity score estimates $\hat{e}(X_i)$ from a regression probability forest predicting treatment assignment, and (2) conditional mean predictions $\hat{\mu}(X_i, W)$ from honest regression forests predicting restricted mean survival time conditional on treatment status and covariates. These components are combined via the AIPW formula, yielding estimates that remain consistent if either the propensity or outcome model is correctly specified.

\paragraph{Cross-fitting procedure.}
To avoid overfitting and ensure valid inference, we employ repeated $K$-fold cross-fitting on the test set. Specifically, for each repetition $m \in \{1, \ldots, 100\}$, we randomly assign observations into $K=5$ folds. Randomization is stratified by experimental group (Mentoring treated, Mentoring control, Challenges treated, Challenges control, non-participants) to ensure each fold contains observations from all groups. For each fold $k \in \{1, \ldots, 5\}$, we:

\begin{enumerate}
\item Train causal survival forests using data from all folds except $k$ (the training sample for fold $k$).
\item Predict AIPW scores for observations in fold $k$: $\psi_{i,k}^{M,m}$ and $\psi_{i,k}^{C,m}$ for each individual $i$ in fold $k$, along with corresponding variance estimates $\hat{V}_{i,k}^{M,m}$ and $\hat{V}_{i,k}^{C,m}$.
\end{enumerate}

After completing all folds in repetition $m$, each individual has predicted treatment effects $\psi_i^{M,m}$ and $\psi_i^{C,m}$ estimated on data that did not include that individual. We then aggregate across the 100 repetitions: we take the median prediction across repetitions for each individual: $\psi_i^M = \text{Median}(\psi_i^{M,1}, \ldots, \psi_i^{M,100})$ and $\psi_i^C = \text{Median}(\psi_i^{C,1}, \ldots, \psi_i^{C,100})$. Variance estimates are similarly aggregated by taking the median across repetitions. We follow the method described in \cite{chernozhukov2017double}.

\paragraph{Policy value estimation.}
Given an allocation policy $\pi$ that assigns each test-set individual to \emph{Mentoring}, \emph{Challenges}, or neither, we estimate the policy value as:

\[
\hat{V}(\pi) = \frac{1}{n_M + n_C} \left(\sum_{i: p_{i}=M} \psi_i^M + \sum_{i: p_{i}=C} \psi_i^C\right),
\]

where $n_M$ and $n_C$ are the number of individuals assigned to \emph{Mentoring} and \emph{Challenges}, respectively. Standard errors are computed using the aggregated variance estimates from the causal survival forests, accounting for the within-arm averaging. For comparisons between policies or programs, we compute appropriate differences in AIPW scores and propagate variances assuming independence across programs.

This approach separates the assignment decision (which applicants receive which program) from the evaluation of that decision (estimating treatment effects for the assigned groups). The validity of our policy comparisons depends on the evaluation models accurately estimating treatment effects, not on the assignment models perfectly ranking applicants.

\subsection{Model Validation: Recovering Experimental Effects}

To validate our evaluation model specification, we assess whether cross-fitted AIPW scores correctly recover experimental average treatment effects. This diagnostic tests whether our models produce unbiased treatment effect estimates in the setting where we have the benchmark from the randomized experiment.

\paragraph{Validation procedure.}
We train causal survival forests following the same process as before, using 5-fold cross-fitting and generate out-of-sample AIPW predictions for each participant; we use the experimental samples. We compare these individual-level predictions to the experimental ATE by examining the distribution of deviations: $\delta_i = \hat{\tau}_i^{AIPW} - \hat{\tau}^{RCT}$. Under correct model specification, these deviations should center at zero with dispersion reflecting genuine treatment effect heterogeneity plus estimation noise.

\begin{figure}[!htbp]
\caption{Model Validation: Distribution of Deviations from RCT Estimates}\label{fig:model_validation}
\centering
\begin{tabular}{cc}
\includegraphics[width=0.48\textwidth]{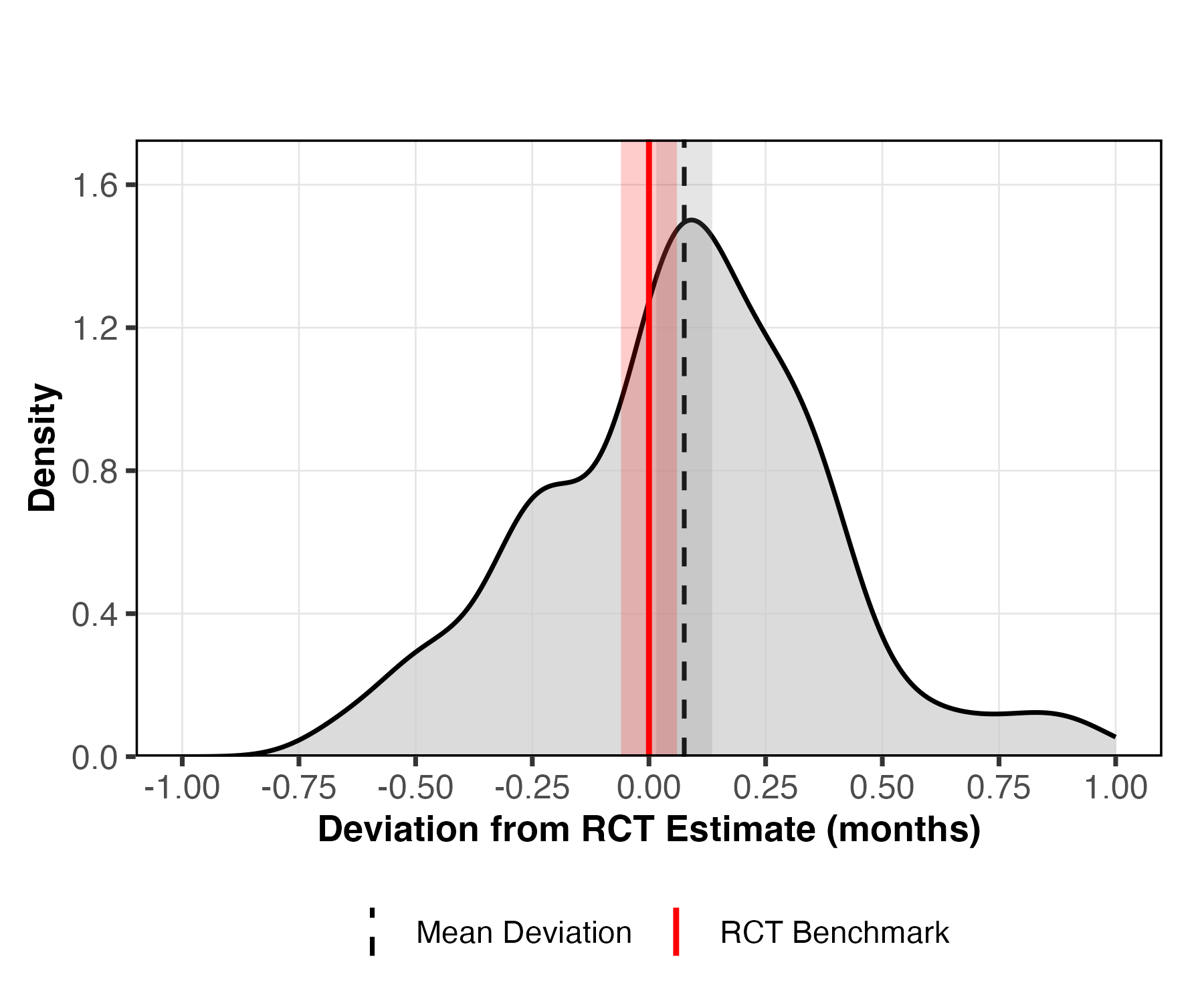} &
\includegraphics[width=0.48\textwidth]{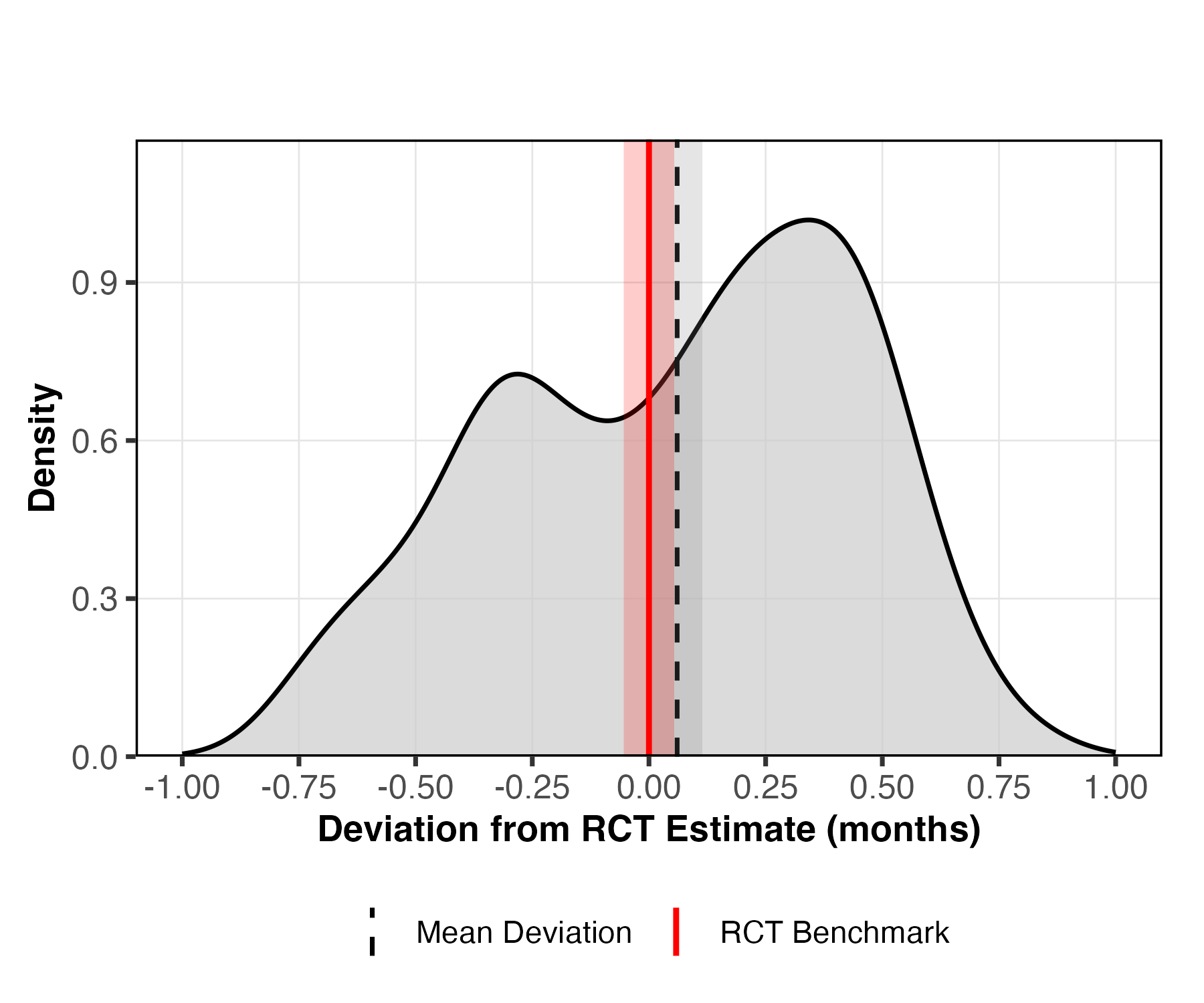} \\
\small{(a) Mentoring Program} & \small{(b) Challenges Program}
\end{tabular}

\medskip

\caption*{\footnotesize{\textit{Notes:} Distribution of deviations between cross-fitted AIPW predictions and experimental ATE benchmarks. Each panel shows $\delta_i = \hat{\tau}_i^{AIPW} - \hat{\tau}^{RCT}$ for experimental participants. The red vertical line marks zero deviation (perfect recovery), with red shading indicating the 95\% confidence interval around the RCT estimate. The black dashed line shows the mean AIPW deviation, with gray shading for its 95\% confidence interval. Causal survival forests trained exclusively on experimental samples using 5-fold cross-fitting. Mentoring: mean deviation = $0.08$ months (SE = 0.03); Challenges: mean deviation = $0.06$ months (SE = 0.04). Deviations statistically indistinguishable from zero in both programs.}}
\end{figure}

Figure~\ref{fig:model_validation} presents the results. The red solid vertical line indicates the RCT benchmark (zero deviation), with shaded regions marking 95\% confidence intervals around both the RCT estimate (red) and the mean AIPW deviation (gray). The black dashed line shows the mean deviation from the RCT estimate.

For the \emph{Mentoring} program (Panel a), AIPW predictions center slightly above the RCT benchmark (SE, with a mean deviation of $0.08$ months (SE = 0.03). The difference between RCT and the mean deviation is statistically insignificant. The distribution appears approximately symmetric and concentrated near zero, indicating our cross-fitted estimates successfully recover the experimental ATE. The modest spread in the distribution reflects both the treatment effect heterogeneity among participants and sampling variation in the predictions.

For the \emph{Challenges} program (Panel b), AIPW predictions show similarly good performance, with a mean deviation of $0.06$ months (SE = 0.04). While the distribution exhibits slightly more dispersion than \emph{Mentoring}, the mean deviation remains statistically indistinguishable from the RCT mean.

The cross-fitted causal survival forests accurately recover experimental treatment effects, demonstrating that our evaluation approach is well-calibrated and unbiased within the experimental sample. This gives us confidence that the models correctly estimate treatment effects where treatment is randomized.

\section{Alternative Policy Evaluation Using Cross-Fitting}\label{app:crossfitting}

Our main off-policy evaluation in Section~\ref{section_offpolicy} employs a sample-splitting approach in which policy estimation and evaluation occur in separate datasets. This appendix presents an alternative approach using repeated cross-fitting following \citet{fava2025}.

The procedure proceeds as follows. For each repetition $m \in \{1, \ldots, M\}$ with $M = 100$, we randomly partition the full sample of applicants into $K=5$ folds, stratified by experimental group, using a different random seed. Within each repetition, we use $K$-fold cross-fitting: for each fold $k \in \{1, \ldots, K\}$, we train causal survival forests on the remaining $K-1$ folds, obtain out-of-sample predicted treatment effects $\hat{\tau}^{M}_{i}$ and $\hat{\tau}^{C}_{i}$ for observations in fold $k$, solve the constrained optimization problem in Equation~\ref{eq_opt} to determine optimal assignments, and compute AIPW scores under these assignments. After completing all folds within a repetition, we aggregate out-of-fold AIPW scores to obtain policy value estimates and standard errors computed as the sample standard deviation of individual scores divided by $\sqrt{n}$. Following \citet{fava2025} and \citet{chernozhukov2018generic}, we aggregate across the $M$ repetitions by taking the median of point estimates and the median of confidence interval bounds.

This approach requires a uniform convergence condition on the estimated treatment effects:
\begin{equation}
    \sup_{x \in \mathcal{X}} \left| \hat{\tau}^{p}(x) - \tau^{p}(x) \right| = o_p(1) \quad \text{for } p \in \{M, C\},
\end{equation}
where $\hat{\tau}^{p}(x)$ denotes the estimated conditional average treatment effect for program $p$ at covariate value $x$ and $\tau^{p}(x)$ is the true effect. This assumption is stronger than the pointwise consistency required by the sample-splitting approach in our main specification but enables valid inference when the same data are used for both policy learning and evaluation.

\begin{table}[!htbp]
\caption{Treatment Effects Under Targeted Allocation Policies: Repeated Cross-Fitting}
\label{tab:optimal_allocation_fava}
\centering
\resizebox{0.6\textwidth}{!}{%
\begin{tabular}{lccc}
\toprule\toprule
& \emph{Mentoring} & \emph{Challenges} & Overall \\
\midrule
\multicolumn{4}{l}{\textit{Panel A: Benchmark}} \\
Random allocation & $-$2.22 (0.05) & $-$2.54 (0.02) & $-$2.38 (0.03) \\
\midrule
\multicolumn{4}{l}{\textit{Panel B: Targeted allocation policies}} \\
15\% / 15\% / 70\% & $-$5.26 (0.02) & $-$3.94 (0.03) & $-$4.60 (0.04) \\
33\% / 33\% / 34\% & $-$4.40 (0.05) & $-$3.07 (0.04) & $-$3.73 (0.04) \\
50\% / 50\% / 0\% & $-$3.41 (0.06) & $-$2.55 (0.04) & $-$2.98 (0.04) \\
\bottomrule\bottomrule
\end{tabular}%
}
\caption*{\footnotesize\textit{Notes:} Treatment effects measured as change in restricted mean survival time (months); negative values indicate faster job finding. Capacity allocations specify the share of applicants assigned to \emph{Mentoring}/\emph{Challenges}/Out. Random allocation assigns each applicant to either program with equal probability. Results are median estimates across 100 repetitions of cross-fitting with different random fold assignments. Standard errors in parentheses.}
\end{table}

Table~\ref{tab:optimal_allocation_fava} presents median treatment effects across the 100 repetitions. Point estimates are similar to those in our main specification (Table~\ref{tab:optimal_allocation}): the overall treatment effect under the most selective policy (15/15/70) is $-4.60$ months compared to $-4.34$ months in the main analysis, while effects under expanded capacity (50/50/0) are $-2.98$ versus $-2.86$ months. The qualitative conclusions are unchanged: targeted allocation substantially outperforms random assignment, with gains from targeting diminishing as capacity expands.

The principal difference between specifications lies in the estimated standard errors. Cross-fitting yields smaller standard errors, particularly for selective policies where few applicants are assigned to each program. At 15\% capacity, for example, the standard error on the \emph{Mentoring} treatment effect is 0.02 compared to 0.09 in the main specification. This difference reflects efficiency gains from using the full sample for both estimation and evaluation rather than reserving half for each task. Additionally, the median aggregation across 100 repetitions improves reproducibility: different researchers using the same data but different random seeds will obtain identical results. These smaller standard errors, however, rely on the uniform convergence assumption holding in finite samples; the sample-splitting approach provides more conservative inference under weaker conditions. The concordance of point estimates across both approaches strengthens confidence in our conclusions regarding the value of targeted program assignment.

\end{document}